\documentclass[rmp,twocolumn,superscriptaddress,showpacs,floatfix]{revtex4}

\usepackage{epsfig}
\usepackage{graphicx}
\usepackage{color}
\usepackage{longtable}

\newcommand{\be}{\begin{equation}}
\newcommand{\ee}{  \end{equation}}
\newcommand{\ba}{\begin{eqnarray}}
\newcommand{\ea}{  \end{eqnarray}}
\newcommand{\ve}{\varepsilon}

\begin{document}

\title{Random Matrices and Chaos in Nuclear Physics: Nuclear Reactions}

\author{G.~E.~Mitchell}
\email{mitchell@tunl.duke.edu}
\affiliation{North Carolina State University, Raleigh, North Carolina
27695, USA}
\affiliation{Triangle Universities Nuclear Laboratory, Durham, North
Carolina 27706, USA}

\author{A.~Richter}
\email{richter@ikp.tu-darmstadt.de}
\affiliation{Institut f{\"u}r Kernphysik, Technische Universit{\"a}t
Darmstadt, D-64289 Darmstadt, Germany}
\affiliation{$\rm ECT^*$, Villa Tambosi, I-38100 Villazzano (Trento),
Italy}

\author{H.~A.~Weidenm{\"u}ller}
\email{Hans.Weidenmueller@mpi-hd.mpg.de}
\affiliation{Max-Planck-Institut f{\"u}r Kernphysik, D-69029
Heidelberg, Germany}

\begin{abstract}
The application of random--matrix theory (RMT) to compound--nucleus
(CN) reactions is reviewed. An introduction into the basic concepts of
nuclear scattering theory is followed by a survey of phenomenological
approaches to CN scattering. The implementation of a random--matrix
approach into scattering theory leads to a statistical theory of CN
reactions. Since RMT applies generically to chaotic quantum systems,
that theory is, at the same time, a generic theory of quantum chaotic
scattering. It uses a minimum of input parameters (average $S$--matrix
and mean level spacing of the CN). Predictions of the theory are
derived with the help of field--theoretical methods adapted from
condensed--matter physics and compared with those of phenomenological
approaches. Thorough tests of the theory are reviewed, as are
applications in nuclear physics, with special attention given to
violation of symmetries (isospin, parity) and time--reversal
invariance.

\end{abstract}

\date{today}

\maketitle

\section{Introduction}
\label{int}

Random--matrix theory (RMT) as developed in the 1950s by Wigner and
Dyson (see~\cite{Wig55a,Wig57,Dys62a,Dys62b,Dys62c} and the reprint
collection~\cite{Por65}) plays an important role not only in the
analysis of nuclear spectra. Random matrices and chaos play perhaps an
even bigger role in the theory of nuclear reactions. The resulting
``Statistical Theory of Nuclear Reactions'' is the topic of the
present review. It completes the review by two of the present authors
of random matrices and chaos in nuclear structure~\cite{Wei09}.

In proposing RMT, Wigner was probably inspired by Bohr's
idea~\cite{Boh36} of the compound nucleus (CN). After the early
experimental confirmation of Bohr's idea as implemented in the
Hauser--Feshbach formula~\cite{Hau52}, the field received a boost in
1960 by Ericson's prediction~\cite{Eri60} of statistical fluctuations
in nuclear cross sections. Subsequent intense experimental work on a
number of topics in CN reactions (Ericson fluctuations, isobaric
analogue resonances, isospin mixing in nuclear reactions, tests of
time--reversal symmetry) reached saturation at the end of the 1970s,
to be followed later only by studies of parity violation in nuclear
reactions. Theoretical work extended the Hauser--Feshbach formula to
the case of direct reactions~\cite{Kaw73}. At the same time, theorists
set out to connect the statistical models of CN scattering with RMT.
That turned out to be a very challenging problem. Motivated by the
fundamental interest in CN scattering and by the need in other fields
of physics (neutron physics, shielding problems, nuclear astrophysics,
etc.) to have a viable theory of CN reactions with predictive power,
that work was carried on for a number of years and led to partial
insights into the connection between RMT and CN scattering.

Theoretical efforts at constructing a comprehensive theory of CN
reactions received a strong stimulus in the beginning of the 1980s by
developments in the theory of chaotic motion, i.e., the theory of
non--integrable classical systems and their quantum counterparts. It
is not a coincidence that the Bohigas--Giannoni--Schmit
conjecture~\cite{Boh84} (which connects properties of quantum
spectra of classically chaotic systems with RMT) and the first
papers~\cite{Ver84a,Wei84,Ver85a} establishing a firm connection
between RMT, CN scattering theory, and chaotic quantum scattering,
appeared almost simultaneously. Technically, progress became possible
by combining scattering theory based on the shell model with novel
techniques using a supersymmetric generating functional borrowed from
condensed--matter physics~\cite{Efe83}.  Physically, CN scattering was
recognized as a paradigmatic case of chaotic quantum scattering, and
the theory of CN scattering was seen to apply to chaotic scattering
processes in general. Actually the theory of CN scattering is richer
than that for most other cases of chaotic scattering, for two
reasons. First there exist conserved quantum numbers (spin and
parity). The nuclear cross section is the square of a sum of resonance
contributions each carrying these quantum numbers. Chaotic motion only
affects resonances carrying the same quantum numbers. Second, the
nucleus has internal structure. This leads to a strong increase of the
number of open channels with excitation energy. Inelastic processes
(where the masses, charges and/or excitation energies of the reaction
products differ from those of target and projectile) add complexity
and richness to the theory: Nuclear reactions can be studied versus
scattering angle, versus bombarding energy, and for different final
fragment configurations.

The statistical theory of CN reactions is generic and applies likewise
to many other cases of chaotic scattering. That fact is borne out by
applications to electron transport through disordered mesoscopic
samples~\cite{Bee97,Alh00,Imr02}, and to the passage of
electromagnetic waves through microwave cavities~\cite{Fyo05}. Some of
these cases are treated below. However, the theoretical developments
reviewed in this paper do not cover all aspects of the theory of
chaotic scattering. In systems with few degrees of freedom,
semiclassical periodic--orbit theory~\cite{Gut90,Gas91,Smi91} plays a
prominent role. That branch of scattering theory has not been much
used in nuclear many--body physics. (It has found applications, for
instance, in the scattering of two heavy ions where it applies to
relative motion). That is why it is not dealt with here. The
connection between the RMT approach and periodic--orbit theory is
discussed, for instance, in~\cite{Lew91a}.

In view of the very general applicability of the statistical theory to
chaotic scattering processes governed by RMT, we aim at a presentation
which is accessible to readers not familiar with the topic. We have in
mind, for instance, physicists working in other areas of nuclear
physics, or in chaotic scattering. Therefore, we begin this review in
Section~\ref{basi} with a summary of some basic facts and concepts of
nuclear reaction theory. We also present the central ideas and models
that were developed with the help of plausible albeit intuitive
arguments and that were used to treat CN scattering before the
connection to RMT was established. These are the Hauser--Feshbach
formula, the Weisskopf estimate for the average total width of CN
resonances, Ericson fluctuations, and modifications of the
Hauser--Feshbach formula due to direct reactions. In order to avoid
repeating our arguments later in slightly different form, we present
the arguments in modern terminology.

The modern access to the statistical theory of CN reactions is based
on RMT and presented in Sections~\ref{theo}, \ref{aver}, \ref{resu}.
While the afore--mentioned models do not all survive close scrutiny,
at least their results are vindicated and their ranges of validity are
established. Needless to say, additional results are also obtained.
Tests of the theory and applications to a number of topics are
reviewed in later sections of the paper.

The field has not been reviewed comprehensively for many years. A
review of RMT in nuclear physics~\cite{Bro81} contains sections on the
statistical theory of nuclear reactions. Shorter reviews may be found
in~\cite{Mah79,Boh88,Wei02,Fyo10}. We refer to Part I of this
review~\cite{Wei09} with the letter ``I'' so that equations, figures
or Sections in that paper are referred to, for instance, as
Eq.~(I.34), as Fig.~I.16, or as Section~I.II.A. As in part I, we have
preferred citing a review over giving a large number of references:
Readability of the article was our primary concern.

\section{Basic Facts and Concepts. Statistical Models}
\label{basi}

A stable nucleus with mass number $A$ possesses a discrete spectrum of
levels that extends from the ground state up to the lowest energy
where decay by particle emission is possible (the first ``particle
threshold'').  (Here we disregard the small widths of levels due to
beta or gamma decay). In most cases the first particle threshold
corresponds to nucleon emission ($A \to (A - 1) + n$) and typically
has an excitation energy of $6$ or $8$ MeV in nucleus $A$. The levels
above that threshold have finite widths for particle decay and appear
as resonances in the scattering of a nucleon by the nucleus with mass
number $(A - 1)$. The density $\rho(E)$ of nuclear levels increases
roughly exponentially with excitation energy $E$ (more precisely,
$\rho(E) \propto \exp \sqrt{a E}$ where $a$ is a mass-dependent
constant) and the average spacing $d = \rho^{-1}$ of resonances
decreases accordingly. The number of decay channels also grows with
$E$ since the density of states available for decay in neighboring
nuclei likewise grows nearly exponentially. As a result, the average
total decay width $\Gamma$ of the resonances grows strongly with
excitation energy. The Weisskopf estimate given in Eq.~(\ref{9}) below
shows that $\Gamma / d$ is roughly given by the number of decay
channels over $2 \pi$. Thus, we deal with isolated resonances ($\Gamma
\ll d$) at the first particle threshold and with strongly overlapping
resonances ($\Gamma \gg d$) several MeV higher. A comprehensive theory
of nuclear reactions should cover the entire range from $\Gamma \ll d$
to $\Gamma \gg d$.

\subsection{Resonances}

Resonances in the cross section play a central role in the theory. In
Part I, the empirical evidence was discussed showing that isolated
resonances measured near neutron threshold or near the Coulomb barrier
for protons display stochastic behavior: Spacings and widths of
resonances with identical quantum numbers (spin, parity) are in
agreement with predictions of RMT, more precisely, with the
predictions of the Gaussian Orthogonal Ensemble of Random Matrices
(GOE). That evidence is here taken for granted and not reviewed
again. The connection between RMT and chaotic motion was also reviewed
in Part I. We will use the term ``chaos'' as synonymous with spectral
fluctuation properties of the GOE type. The theory of nuclear
reactions makes use of the GOE properties of nuclear resonances. It is
postulated that the stochastic features found for isolated resonances
also prevail  at higher bombarding energies where resonances overlap.
The stochastic description of resonances then applies for all
bombarding energies where resonance scattering is relevant. Actually,
statistical concepts are used in one form or another to describe all
collision processes between atomic nuclei with center--of--mass
energies between $0$ and about $100$ MeV, except for reactions between
pairs of very light nuclei (where the density of resonances is too
small for a statistical approach).

A stochastic description of CN resonances is not only physically
motivated,  but a practical necessity. While typical spacings of
neighboring levels near the ground state are of the order of $100$
keV, the resonances seen in the scattering of slow neutrons have
typical spacings of $10$ eV, see Fig.~I.1. This is a consequence of
the nearly exponential growth of the average level density with $E$
mentioned above. Put differently, there are about $10^5$ to $10^6$
levels between the ground state of the CN and the isolated resonances
seen in the scattering of a slow neutron. There is no viable
theoretical approach that would allow the prediction of  spectroscopic
properties of such highly excited states. Needless to say, the
situation becomes worse as the excitation energy increases further.

Within the framework of a statistical approach, one does not predict
positions and widths of individual resonances. Rather, the GOE
predicts the distribution of spacings between resonances, and the
distribution of partial and total widths for decay into the available
channels. By the same token, the statistical theory of nuclear
reactions does not aim at predicting the precise form of some reaction
cross section versus energy or scattering angle. Rather, it aims at
predicting the average values, higher moments, and correlation
functions of cross sections obtained by averaging over some energy
interval. That interval must encompass a large number $N$ of
resonances. For isolated resonances, the resulting
finite--range--of--data error is expected to be inversely proportional
to $N$. For overlapping resonances, $N$ is replaced by the length of
the averaging interval divided by the average width of the resonances.
With the exception of the lightest nuclei, such averages are the only
theoretical predictions presently available for nuclear reactions in
the regime of resonance scattering.

\subsubsection*{Dynamic Origin of Resonances}

It is useful to have an idea how the numerous resonances dominating CN
reactions come about dynamically. Subsequent theoretical developments
then do not appear as purely formal exercises. We use the nuclear
shell model reviewed in Section~I.IV.A as the fundamental dynamical
nuclear model. In that model, two mechanisms lead to resonances. The
dominant mechanism is that of formation of bound single--particle
shell--model configurations with energies above the first particle
threshold. These states become particle--unstable when the residual
two--body interaction is taken into account. We refer to such states
as to quasibound states. \textcite{Mah69} use the term ``bound states
embedded in the continuum'' (BSEC). Details follow in the next
paragraph. A second and less important mechanism is due to barrier
effects of the shell--model potential. The angular--momentum and
Coulomb barriers cause the occurrence of more or less narrow
single--particle resonances. These occur typically within the first
MeV or so of the continuous spectrum of the single--particle
Hamiltonian. There is at most one such narrow single--particle
resonance for each angular--momentum value; resonances at higher
energies are too wide to matter in our context. It is clear that with
only one single--particle resonance for each value of angular momentum
we cannot account for the numerous CN resonances with equal spins and
average spacings of about $10$ eV observed at neutron threshold.
Therefore, the quasibound states are the main contributor to the large
number of resonances in CN scattering. Single--particle resonances can
be incorporated in the description of CN resonances as quasibound
states~\cite{Mah69} and are not mentioned explicitly again.

To describe the quasibound states in more detail, we consider nuclei
in the middle of the $sd$--shell, i.e., nuclei with mass number $28$
and with $12$ valence nucleons. That same example was extensively
discussed in Part I. Our considerations apply likewise to other
nuclei. The spacings of the energies of the single--particle states
$d_{5/2}$, $s_{1/2}$, and $d_{3/2}$ (in spectroscopic notation) in the
$sd$--shell are of the order of MeV. In $^{17}$O, for instance, the
spacing between the lowest ($d_{5/2}$) and the highest ($d_{3/2}$)
single--particle state is $5.08$ MeV~\cite{Zel96}. Thus the spectrum
of the bound $sd$--shell--model configurations of $12$ nucleons (which
typically comprises $10^3$ states for low values of total spin $J$)
extends over several $10$ MeV while the first particle threshold in
these nuclei has a typical energy of only several MeV. Inclusion of
the residual interaction within this set of bound states spreads the
spectrum further and produces chaos. But the two--body interaction
also connects the resulting bound many--body states with other
shell--model states defining the open channels. Such states are
obtained, for instance, by lifting one nucleon into the $s$--wave
continuum of the shell model and diagonalizing the residual
interaction among the remaining $11$ nucleons in the $sd$--shell. The
antisymmetrized product of the ground state or of the $n$th excited
state containing $11$ nucleons with the single--particle $s$--wave
continuum state would define channels with different threshold
energies. The coupling of the diagonalized quasibound $12$--nucleon
states to the channels causes the states above the first particle
threshold to turn into CN resonances. Lifting one or several nucleons
out of the $sd$--shell into bound states of the $pf$--shell increases
the number of bound single--particle configurations and, thus, of CN
resonances.  This shows how the (nearly exponential) increase of the
nuclear level density leads to a corresponding increase of the density
of CN resonances.

\subsection{$S$--Matrix}
\label{smat}

\begin{figure}[ht]
	\centering
	\includegraphics[width=0.45\textwidth]{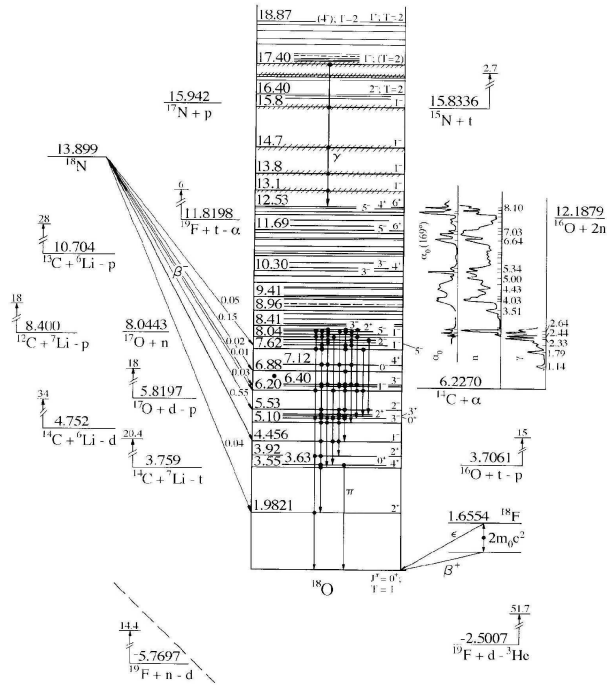}
	\caption{Level scheme of $^{18}$O. The center part of the
	figure shows the low--lying levels, beginning with the ground
	state. On the left and on the right, various thresholds are
	indicated, plus the energy dependence of some reaction cross
	sections. From~\cite{Til95}.}
	\label{fig:HW1}
\end{figure}

We need to introduce some elements of nuclear reaction theory. We
assume throughout that two--body fragmentation dominates the reaction.
(Three--body fragmentation is a rare event in the energy range under
consideration). Mass and charge and the internal states of both
fragments are jointly referred to by Greek letters $\alpha, \beta,
\ldots$. The energy of relative motion of the fragmentation $\alpha$
at asymptotic distance is denoted by $E_\alpha$ while $E$ denotes the
excitation energy of the CN. By energy conservation, $E_\alpha$ is
trivially related to $E$ and to the energy $E_\beta$ of relative
motion of any other fragmentation $\beta$ via the binding energies of
the fragments.  Channels $a, b, \ldots$ are specified by the
fragmentation $\alpha$, by the angular momentum $\ell$ of relative
motion, and by total spin $J$ and parity $\Pi$. A channel $a$
``opens'' when $E_\alpha = 0$, the corresponding value of $E$ is
called the threshold energy for fragmentation $\alpha$. A channel is
said to be open (closed) when $E_\alpha > 0$ ($E_\alpha < 0$),
respectively. We speak of elastic scattering when incident and
outgoing fragments are equal ($\alpha = \beta$) because then $E_\alpha
= E_\beta$. But elastic scattering does not imply $a = b$ (while $J$
is conserved, $\ell$ is not when the fragments carry spin). We use the
term ``strictly elastic'' for $a = b$. Some of these concepts are
illustrated in Fig.~\ref{fig:HW1}.

The central object in the theory of nuclear reactions is the
scattering matrix $S(E)$. It is a matrix in the space of open channels
and carries fixed quantum numbers $J$, $\Pi$ and, if applicable,
isospin $T$. It is defined in terms of the asymptotic behavior of the
scattering wave functions $| \Psi^+_a \rangle$. These are solutions of
the Schr{\"o}dinger equation subject to the boundary condition that
there is an incoming wave only in channel $a$. We simplify the
notation by considering only channels with neutral fragments. With
$\ell_c$ the angular momentum of relative motion in channel $c$, $r_c$
the radial coordinate of relative motion, $k_c$ the wave number, $m_c$
the reduced mass, and $h^{\pm}_\ell$ the spherical Hankel functions,
the radial part of $| \Psi^+_a \rangle$ in channel $b$ has the form
$\delta_{a b} (m_a / k^{1/2}_a) h^-_{\ell_a}(k_a r_a) - (m_b /
k^{1/2}_b)S_{b a}(E) h^+_{\ell_b}(k_b r_b)$. Thus, the element $S_{b
a}(E)$ of $S$ gives the amplitude of the asymptotic flux in channel
$b$ for unit incident flux in channel $a$ at energy $E$. The
straightforward generalization to charged fragments is obtained by
replacing the spherical Hankel functions by the Coulomb wave
functions. The definition of a channel includes the quantum numbers
$J$ and $\Pi$. In writing $S_{a b}(E)$ we omit $J$ and $\Pi$ (which
must be identical for $a$ and $b$). In general, channels $a$ and $b$
will belong to different fragmentations $\alpha$ and $\beta$,
respectively.

The nuclear Hamiltonian is invariant under time reversal. Therefore,
the amplitude of the asymptotic flux in channel $b$ for unit incident
flux in channel $a$ is equal to the amplitude of the asymptotic flux
in channel $a$ for unit incident flux in channel $b$, and $S$ is
symmetric, $S_{a b}(E) = S_{b a}(E)$.  The conservation of total flux
implies the unitarity relation
\be
\sum_c S_{a c}(E) S^*_{b c}(E) = \delta_{a b} .
\label{12a}
\ee
The dimension of $S$ is equal to the number $\Lambda$ of open channels.
With increasing $E$ that number grows nearly exponentially because so
does the number of states in the residual nuclei into which the CN may
decay. We can work with a matrix $S$ of fixed dimension $\Lambda$ only
for $E$ in an energy window with end points given by two nearest
threshold energies. The size of that window decreases nearly
exponentially with $E$. That constraint is usually neglected in
applications of the theory because the flux into channels that have
just opened is typically small due to angular--momentum and
Coulomb--barrier effects. Moreover, because of the nearly exponential
increase of the level density the number ${\cal N}$ of resonances
within each window changes only algebraically (and not exponentially)
with energy $E$, in spite of the exponentially decreasing size of the
window. At neutron threshold, where $d$ is of the order of $10$ eV and
the spacing of thresholds is of the order of several $100$ keV, we
have ${\cal N} \approx 10^4$. At $15$ MeV above neutron threshold (the
Ericson regime, see Section~\ref{eri}) ${\cal N}$ is estimated to be
one or two orders of magnitude smaller.

Given $S$, the differential cross section of any nuclear reaction is
obtained as a bilinear form in the elements $S_{a b}(E)$ of the
$S$--matrix, and similarly for other observables such as fragment
polarization. In addition to the elements of the $S$--matrix, the
formulas involve kinematical factors and angular--momentum coupling
coefficients~\cite{Bla52a} and are not reproduced here. The
statistical theory of nuclear reactions focuses on the calculation of
the statistical properties of the $S$--matrix.

For later use we introduce the eigenvalues of the unitary and
symmetric matrix $S$. These have unit magnitude and are written as
$\exp ( 2 i \delta_c(E) )$. The real phase shifts $\delta_c(E)$ are
called the eigenphase shifts of $S$. A theorem analogous to the
Wigner--von Neumann non--crossing theorem for the eigenvalues of a
Hermitean matrix also applies  in the present case~\cite{Wei67}.
Therefore, generically no two eigenphase shifts coincide, and the
matrix $S$ can be written as
\be
S_{a b}(E) = \sum_c {\cal O}_{a c}(E) \exp ( 2 i \delta_c(E) ) {\cal
O}_{b c}(E) \ .
\label{12b}
\ee
Here the real orthogonal matrix ${\cal O}(E)$ induces a transformation
from the physical channels to the ``eigenchannels'' of $S$. The
scattering wave functions in the eigenchannel representation are given
by $| \Omega_c(E)\rangle = \sum_a {\cal O}_{c a} | \Psi^+_a(E)
\rangle$. The radial part of $| \Omega_c(E) \rangle$ in the physical
channel $a$ has the form $ {\cal O}_{c a} [ h^-_{\ell_a} (k_a r_a) -
\exp ( 2 i \delta_c ) h^+_{\ell_a}(k_a r_a)]$, in keeping with the
fact that $S$ is diagonal in the eigenchannel basis. The eigenchannels
are unphysical because there is an incident wave in every physical
channel. Nevertheless the eigenchannels are helpful theoretical
constructs.

\subsection{Bohr Assumption and Weisskopf Estimate}
\label{boh}

Statistical concepts have governed the theory of CN scattering from
its inception. \textcite{Boh36} introduced the idea of the CN as an
equilibrated system of strongly interacting nucleons. The incident
nucleon shares its energy with the nucleons in the target. The system
equilibrates and ``forgets'' its mode of formation. It takes a long
time (long in comparison with the time it takes a nucleon with Fermi
velocity to traverse the nucleus) for the CN to accidentally
concentrate the available energy back onto a single nucleon which can
then be re--emitted. Therefore, formation and decay of the CN are
independent processes (``Bohr assumption''). The decay of the CN is
assumed to be governed by statistical laws (with the proviso that
energy, spin, and parity are conserved).

That intuitive picture found its first quantitative formulation in the
``Hauser--Feshbach formula''~\cite{Wol51,Hau52} for the average
differential cross section. The average is taken over an energy
interval containing a large number of resonances. We list the
assumptions that were used and defer a discussion of their validity to
Section~\ref{resu}.  Because of the presence of resonances (which
behave stochastically) the scattering matrix $S$ fluctuates randomly
in energy and is accordingly decomposed into an average part and a
fluctuating part~\cite{Fes54},
\be
S_{a b}(E) = \langle S_{a b}(E) \rangle + S^{\rm fl}_{a b}(E) \ .
\label{1}
\ee
The average over energy is indicated by angular brackets. By
definition, we have $\langle S^{\rm fl}_{a b}(E) \rangle = 0$. The
standard assumption~\cite{Fes54} on the average part is that it
vanishes for $a \neq b$,
\be
\langle S_{a b}(E) \rangle = \delta_{a b} \langle S_{a a}(E) \rangle
\ .
\label{2}
\ee
It is also assumed that in the CN, $S$--matrix elements pertaining to
different conserved quantum numbers are uncorrelated,
\be
\langle S^{\rm fl}_{a b}(E) (S^{\rm fl}_{c d}(E))^* \rangle = 0 \
{\rm for} \ J \neq J' \ {\rm and/or} \ \Pi \neq \Pi'
\label{3}
\ee
and that even when the quantum numbers are equal we have
\ba
&& \langle S^{\rm fl}_{a b}(E) (S^{\rm fl}_{c d}(E))^* \rangle = 0
\nonumber \\
&& \qquad {\rm unless} \ a = c, \ b = d \ {\rm or} \ a = d, \ b = c
\label{4}
\ea
where we have used the symmetry of $S$. Intuitively, the
assumptions~(\ref{3}) and (\ref{4}) are related to random phases for
the contributing resonances (only absolute squares survive the
averaging process). The decomposition~(\ref{1}) of the scattering
matrix implies a corresponding decomposition of the average cross
section. The average consists of a sum over terms of the form $\langle
S_{a b}(E) S^*_{c d}(E) \rangle$. We use Eq.~(\ref{1}) and focus
attention on the part which is bilinear in the elements $S^{\rm
fl}$. It is that part which contributes to the average CN cross
section $\langle \sigma^{\rm CN} \rangle$. With the
assumption~(\ref{3}), $\langle \sigma^{\rm CN} \rangle$ contains only
terms $\langle S^{\rm fl}_{a b}(E) (S^{\rm fl}_{c d}(E))^* \rangle$
where all channel indices refer to the same quantum numbers $J$ and
$\Pi$. That statement implies that average CN cross sections are
symmetric about $90$ degrees in the center--of--mass (c.m.) system.
The terms $\langle S^{\rm fl}_{a b}(E) (S^{\rm fl}_{c d}(E))^*
\rangle$ are subject to assumption~(\ref{4}) which reduces the average
CN cross section to a sum over averages of squares of $S$--matrix
elements. Each such term $\langle |S^{\rm fl}_{a b}(E)|^2
\rangle$ describes the formation of the CN from channel $a$ and its
decay into channel $b$, or vice versa. The Bohr assumption
(independence of formation and decay of the CN) is used to write
$\langle |S^{\rm fl}_{a b}(E)|^2 \rangle$ in factorized form as
\be
\langle S^{\rm fl}_{a b}(E) (S^{\rm fl}_{c d}(E))^* \rangle =
(\delta_{a c} \delta_{b d} + \delta_{a d} \delta_{b c}) \ T_a f_b \ .
\label{5}
\ee
Here $T_a$ denotes the probability of formation of the CN from
channel $a$ and is defined by
\be
T_a = 1 - |\langle S_{a a}(E) \rangle|^2 \ .
\label{6}
\ee
The factor $f_b$ gives the relative probability of CN decay into
channel $b$ and is normalized, $\sum_b f_b = 1$. Using the symmetry
and unitarity of $S$ and neglecting a term that is inversely
proportional to the number of channels, we obtain from Eqs.~(\ref{5})
and (\ref{6}) that $f_b = T_b / \sum_c T_c$. Thus,
\be
\langle S^{\rm fl}_{a b}(E) (S^{\rm fl}_{c d}(E))^* \rangle =
(\delta_{a c} \delta_{b d} + \delta_{a d} \delta_{b c}) \  T_a T_b /
\sum_c T_c
\label{7}
\ee
while the average of the CN part of the cross section takes the form
(we recall that $\alpha, \beta$ denote the fragmentation while $a, b,
\ldots$ denote the channels)
\ba
\langle {\rm d} \sigma^{\rm CN}_{\alpha \beta} / {\rm d} \omega
\rangle &=& \sum {\rm coefficients} \bigg( (1 + \delta_{a b}) T_a T_b /
\sum_c T_c \bigg) \nonumber \\
&& \times P_\ell (\cos \Theta) \ .
\label{8}
\ea
Here $\omega$ is the solid angle, $\Theta$ the scattering angle, and
$P_\ell$ the Legendre polynomial. The sum extends over all values of
$J, \ell, a, b$ and contains geometric and kinematical coefficients
not specified here, see~\textcite{Bla52a}. Except for the factor $(1 +
\delta_{a b})$, Eq.~(\ref{8}) was originally proposed
by~\textcite{Wol51} and by~\textcite{Hau52} and is commonly referred
to as ``Hauser--Feshbach formula''. The factor $(1 + \delta_{a b})$
was later shown~\cite{Vag71} to be a necessary consequence of the
symmetry of $S$ and for obvious reasons is referred to as the
``elastic enhancement factor''. In the sequel we apply the expression
``Hauser--Feshbach formula'' to both, Eq.~(\ref{7}) and Eq.~(\ref{8}).

Because of the unitarity of $S$, the average $S$--matrix is
subunitary. The ``transmission coefficients'' $T_a$ defined in
Eq.~(\ref{6}) measure the unitarity deficit of $\langle S \rangle$ (we
recall Eq.~(\ref{2})). The $T$s obey $0 \leq T_a \leq 1$. It is
natural to interpret the unitarity deficit as the probability of CN
formation (or, by detailed balance, of CN decay) from (into) channel
$a$. We speak of weak (strong) absorption in channel $a$ when $T_a$ is
close to zero (to unity), respectively. The $T$s are central elements
of the theory. In the early years of CN theory~\cite{Bla52b}, the CN
was assumed to be a black box, so that $T_a = 1$ ($T_a = 0$) for all
channels with angular momenta $\ell \leq \ell_{\rm max}$ ($\ell >
\ell_{\rm max}$, respectively). Here $\ell_{\rm max}$ is the angular
momentum corresponding to a grazing collision between both fragments.
According to Eq.~(\ref{6}), the assumption $T_c = 1$ implies $\langle
S \rangle = 0$, and the decomposition~(\ref{1}) was in fact only
introduced when Feshbach, Porter and Weisskopf~\cite{Fes54} proposed
the optical model of elastic scattering. The model was originally
formulated for neutrons but soon extended to other projectiles. The
model changed the view of CN reactions: The CN was not a black box but
was partly transparent. The transmission coefficients $T_a$ in
Eq.~(\ref{6}) were not put equal to unity but could be calculated from
the optical model which provided the first dynamical input for CN
theory (aside from the average level density that is needed to
calculate the energy dependence of CN emission products). The optical
model for nucleons and the shell model are closely connected concepts:
Both involve a single--particle central potential, see
Section~\ref{opt}.

The average time $\tau$ for decay of the CN or, equivalently, the
average width $\Gamma = \hbar / \tau$ of the CN resonances, can be
estimated using an argument due to Weisskopf~\cite{Bla52b}. For bound
levels with constant spacing $d$, the time--dependent wave function (a
linear superposition of the eigenfunctions) is, aside from an overall
phase factor, periodic with period $d / (2 \pi \hbar)$. The wave
function reappears regularly at time intervals $2 \pi \hbar / d$ at
the opening of any channel $a$ where it escapes with probability
$T_a$. For the time $\tau_a$ for escape into channel $a$ this gives
$\tau_a = 2 \pi \hbar / (d T_a)$, the partial width for decay into
channel $a$ is $\Gamma_a = \hbar / \tau_a = (d / 2 \pi) T_a$, and the
total width $\Gamma$ is
\be
\Gamma = \frac{d}{2 \pi} \sum_c T_c \ .
\label{9}
\ee
Although the derivation of Eq.~(\ref{9}) is based on equally spaced
levels, it is also used for CN resonances, with $d$ the average
resonance spacing. A precursor to Eq.~(\ref{9}) in~\textcite{Boh39},
$\Gamma = \Lambda d / (2 \pi)$, was based on the assumption of strong
absorption $T_a = 1$ in all $\Lambda$ open channels. 

\begin{figure}[ht]
	\centering
	\includegraphics[width=0.45\textwidth]{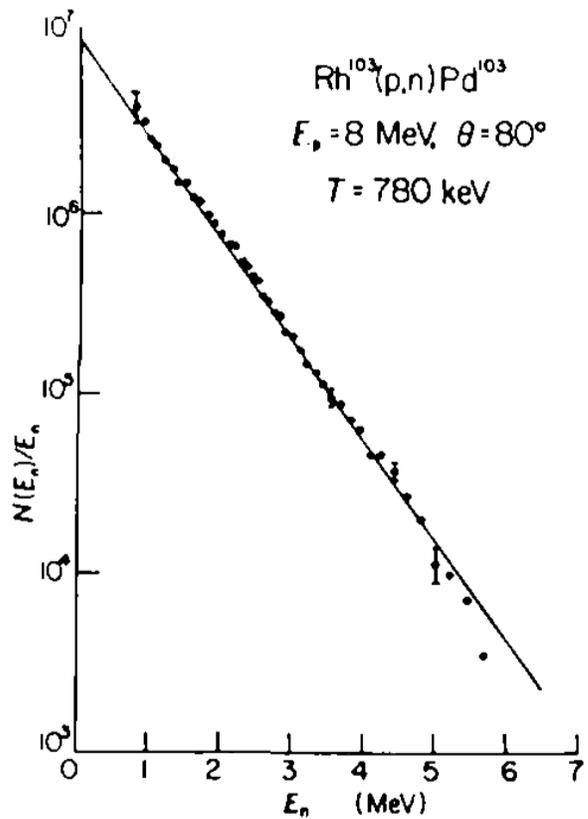}
	\caption{Spectrum of neutrons (``evaporation spectrum'')
	emitted from the CN $^{104}$Pd in the (p,n) reaction on
	$^{103}$Rh at an angle of $80$ degrees versus neutron energy
	$E_n$ (semilogarithmic plot). With all transmission
	coefficients put equal to unity (``black box'' model for the
	CN), the cross section $\propto \exp [ - E_n / T ]$ mirrors
	the level density in the residual nucleus and permits the
	determination of the nuclear temperature $T$.
	From~\cite{Hol63}.}
	\label{fig:HW2}
\end{figure}

\begin{figure}[ht]
	\centering
	\includegraphics[width=0.45\textwidth]{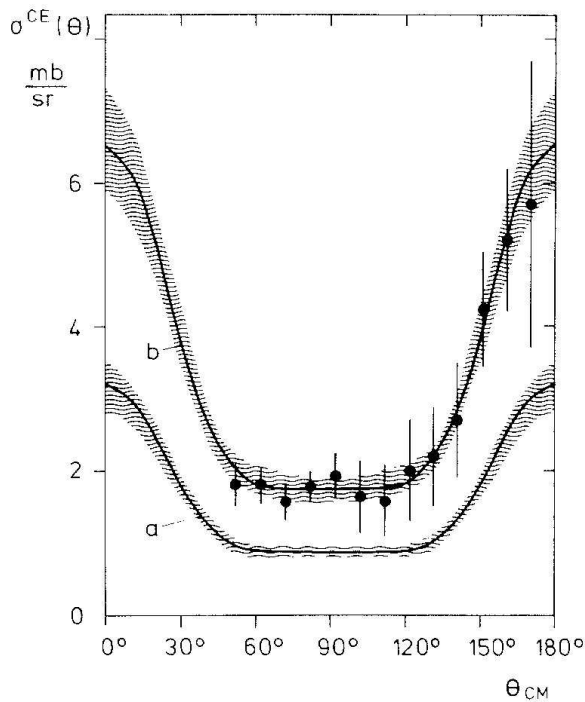}
	\caption{Angular distribution of the compound--elastic cross
	section for the $^{30}$Si(p,p) reaction at a bombarding energy
	$E_p = 9.8$ MeV. The data (dots), taken in the regime $\Gamma
	>> d$, show symmetry of the CN cross section about 90 degrees
	c.m. The solid lines (a) and (b) are predictions of the
	Hauser--Feshbach formula~(\ref{7}) without and with an elastic
	enhancement factor two, respectively. From~\cite{Kre78}.}
	\label{fig:HW3}
\end{figure}

Eq.~(\ref{3}) predicts symmetry of the CN cross section about 90
degrees c.m. while the Hauser--Feshbach formula with $T_c = 1$ in all
channels predicts that the energy distribution of CN decay products is
proportional to the density of states in the final nucleus. Early
tests of the Bohr assumption were focused on these
predictions. Examples are shown in Figs.~\ref{fig:HW2} and
\ref{fig:HW3}.

\subsection{Ericson Fluctuations}
\label{eri}

In the early days of CN theory it was widely held that in the
``continuum region'' of strongly overlapping resonances, $\Gamma \gg
d$, the cross section would be a smooth function of
energy~\cite{Bla52b}. The numerous overlapping resonances contributing
randomly at each energy were thought to yield a scattering amplitude
that varies slowly with energy. ~\textcite{Eri60} realized that this
is not the case. He predicted strong and random fluctuations of CN
cross sections that would be correlated over an energy interval of
length $\Gamma$, the average width of the CN resonances as defined in
Eq.~(\ref{9}). Ericson refined his earlier conjecture in a seminal
article~\cite{Eri63}. Together with theoretical work by
\textcite{Bri63} and \textcite{Bri64}, this paper became the basis for
a large number of experimental investigations. The almost simultaneous
advent of electrostatic Tandem van-de-Graaff accelerators that
produced ion beams with sufficiently small energy spread and with the
beam energy required to produce compound nuclei in the Ericson regime,
led to intense experimental and theoretical activity and confirmed
Ericson's conjecture. A recent example is shown in Fig.~\ref{fig:AR10}
below. ``Ericson fluctuations'', as the phenomenon came to be known,
have since been found and investigated in many areas of physics and
constitute one of the most characteristic features of chaotic
scattering in the regime of strongly overlapping resonances.

To estimate the magnitude of cross--section fluctuations, Ericson
argued that for $\Gamma \gg d$ all resonances have approximately the
same width $\Gamma$. To justify that statement he used the fact that
for each resonance, the total width can be written as the sum over
fluctuating contributions from the open channels~\cite{Eri63}. The
fluctuations of the sum are, thus, inversely proportional to
$\Lambda$, the number of open channels. From the Weisskopf
estimate~(\ref{9}) we see that $\Gamma \gg d$ is possible only for
$\Lambda \gg 1$. Thus the fluctuations of resonance widths should
become negligible for $\Gamma \gg d$ (see, however,
Section~\ref{pole}).

We simplify the presentation by excluding the case of strictly elastic
scattering, and by assuming that Eq.~(\ref{2}) applies. Taking all
resonance widths to be equal to $\Gamma$, we write the $S$--matrix in
the form
\be
S_{a b}(E) = - i \sum_\mu \frac{\gamma_{a \mu} \gamma_{\mu b}}{E -
E_\mu + (i/2) \Gamma} \ , \ (a \neq b) \ .
\label{10}
\ee
The parameters $\gamma_{a \mu}$ are the partial width amplitudes for
the decay of resonance $\mu$ into channel $a$. The fluctuations of the
resonances are due to the stochastic nature of the resonance
parameters $E_\mu$ and $\gamma_{a \mu}$. The resonance energies
$E_\mu$ and the complex partial width amplitudes $\gamma_{\mu a} =
\gamma_{a \mu}$ are assumed to be uncorrelated random variables; the
$\gamma$s are complex Gaussian random variables with mean values zero;
pairs of $\gamma$s carrying different indices are assumed to be
uncorrelated; we write $\langle |\gamma_{a \mu}|^2
\rangle = 2 \pi v^2_a$ where the angular bracket now stands for the
ensemble average. We refer to Eq.~(\ref{10}) and to these statistical
assumptions jointly as to the Ericson model.

According to Eq.~(\ref{10}), $S_{a b}(E)$ is, at fixed energy $E$ and
for $\Gamma \gg d$, a sum over many terms, each term containing the
product of two Gaussian--distributed random variables, all terms being
statistically independent of each other. We use the central limit
theorem to conclude~\cite{Bri63} that $S_{a b}(E)$ is a Gaussian
random process (i.e., the generalization of a random variable to a
random function of some parameter, here the energy $E$). A Gaussian
distribution is completely defined by its first and second
moments. Hence we need to determine only $\langle S_{a b}(E) \rangle$
and $\langle S_{a b}(E) S^*_{c d}(E + \ve) \rangle$ where $\ve$
denotes the difference between the energy arguments of $S^*_{c d}$ and
of $S_{a b}$. (We use that quite generally $\langle S_{a b} S_{c d}
\rangle = \langle S_{a b} \rangle \langle S_{c d} \rangle$, see
Section~\ref{aver}). Then all higher moments and correlation functions
of $S$ are known.

Rather than energy averages we actually calculate the ensemble
averages of $\langle S_{a b}(E) \rangle$ and $\langle S_{a b}(E)
S^*_{c d}(E + \ve) \rangle$. This is done by averaging over the
distribution of the $\gamma_{a \mu}$ and of the $E_\mu$. Energy and
ensemble averages give identical results but ensemble averaging seems
physically more transparent. It is obvious that $\langle S_{a b}(E)
\rangle = 0$, in keeping with Eq.~(\ref{2}). In calculating $\langle
S_{a b}(E) S^*_{c d}(E + \ve) \rangle$, we first carry out the
ensemble average over the $\gamma$s. In the remaining summation over
$E_\mu$ we use that for an arbitrary function $f$ we have $\sum_\mu
f(E_\mu) = \int {\rm d} E' \ f(E') \sum_\mu \delta(E' - E_\mu)$. We
also use the definition $\rho(E) = \langle \sum_\mu \delta(E - E_\mu)
\rangle$ of the average level density $\rho(E)$ of the resonances, and
we assume that $\rho(E)$ is constant over an interval of length
$\Gamma$. This yields
\ba
\langle S_{a b}(E) S^*_{c d}(E + \ve) \rangle &=& (\delta_{a c}
\delta_{b d} + \delta_{a d} \delta_{b c}) \nonumber \\
&& \times 8 \pi^3 v^2_a v^2_b \rho(E) \frac{1}{\Gamma + i \ve} \ .
\label{11}
\ea
We compare Eq.~(\ref{11}) for $\ve = 0$ with the Hauser--Feshbach
formula~(\ref{5}). Complete agreement is obtained when we use the
Weisskopf estimate Eq.~(\ref{9}), the identity $d = 1 / \rho(E)$, and
write for the transmission coefficients $T_a = 4 \pi^2 v^2_a \rho(E)$.
(In anticipation we mention that that result agrees with
Eqs.~(\ref{40}) and (\ref{29a}) if $T_a \ll 1$ for all channels). This
shows that the Ericson model yields the Bohr assumption. It does so,
however, only upon averaging. It is obvious that $|S_{a b}(E)|^2$ as
given by Eq.~(\ref{10}) does not factorize as it stands.  Calculating
the energy average (rather than the ensemble average) with the help of
a Lorentzian weight function of width $I$, we find that the averaging
interval $I$ must be large compared to $\Gamma$ for Eq.~(\ref{11}) to
hold. Thus, independence of formation and decay of the CN hold only
for cross sections averaged over an interval of length $I \gg
\Gamma$. CN cross sections measured with particle beams of sufficient
energy resolution are expected to deviate from the Hauser--Feshbach
formula. Since Eq.~(\ref{3}) involves an average, we expect the same
statement to apply to the symmetry of CN cross sections about 90
degrees c.m. In contrast to the presentation chosen in
Section~\ref{boh}, it had not always been clear prior to Ericson's
work that the Bohr assumption holds only upon averaging.

The magnitude of cross--section fluctuations is estimated by
calculating the normalized autocorrelation function of $|S_{a
b}(E)|^2$. We use the Gaussian distribution of the $S$--matrix
elements and Eq.~(\ref{11}) and obtain
\be
\frac{\langle |S_{a b}(E)|^2 |S_{a b}(E + \ve)|^2 \rangle - (\langle
|S_{a b}(E)|^2 \rangle )^2}{(\langle |S_{a b}(E)|^2 \rangle )^2} =
\frac{1}{1 + (\ve / \Gamma)^2} \ .
\label{12}
\ee
Eq.~(\ref{12}) shows that the fluctuations have the same size as the
average cross section, and a correlation width given by $\Gamma$, the
average width of the CN resonances. These results suggested that
$\Gamma$ could be measured directly using Eq.~(\ref{12}). The
Lorentzian on the right--hand side of Eq.~(\ref{12}) signals that the
CN decays exponentially in time. The lifetime is $\hbar / \Gamma$.

The Ericson model can be extended to include the case of strictly
elastic scattering~\cite{Eri63,Bri63}. Moreover, the arguments can
straightforwardly be extended to the fluctuations both in energy and
angle of the actual CN cross section as given by
Eq.~(\ref{8})~\cite{Eri63}.

In summary, in the regime of strongly overlapping resonances ($\Gamma
\gg d$) the Ericson model leads to the following conclusions: (i) The
elements $S_{a b}(E)$ of the scattering matrix are Gaussian random
processes. (ii) The Bohr assumption (i.e., the Hauser--Feshbach
formula) and the symmetry of the CN cross section about 90 degrees
c.m. hold only for the energy--averaged cross section but not for
cross sections measured with high energy resolution. The averaging
interval has to be larger than $\Gamma$. (iii) CN cross sections
(including the angular distributions) fluctuate. The fluctuations have
the same magnitude as the average cross section. The correlation
length of the fluctuations is $\Gamma$. This fact can be used to
measure $\Gamma$.

\subsection{Direct Reactions}
\label{incl}

\begin{figure}[ht]
	\centering \includegraphics[width=0.45\textwidth]{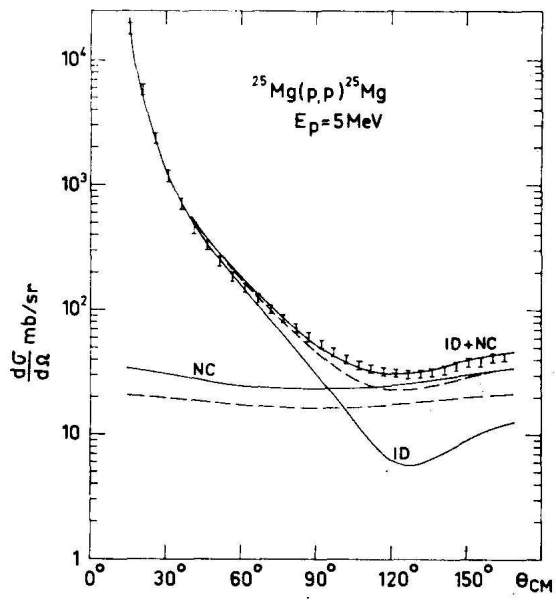}
	\caption{Elastic cross section for protons at 5 MeV scattered
	on $^{25}$Mg versus scattering angle. The data points with
	error bars are compared with predictions of the
	Hauser--Feshbach formula (NC) without (solid line) and with
	elastic enhancement factor (dashed line) and of the optical
	model (ID). From~\cite{Gal66}.}  \label{fig:HW5}
\end{figure}

Experiments in the 1950s performed with poor energy resolution showed
that CN cross sections are not always symmetric about 90 degrees c.m.
These results were confirmed with better resolution. For the case of
elastic scattering, an example is shown in Fig.~\ref{fig:HW5}.
Although the data show the cross section versus lab angle, it is clear
that the asymmetry persists also in the c.m. system. Similar results
were also oblained for inelastic processes. Such deviations were
attributed to a failure of Eq.~(\ref{2}), i.e., to inelastic
scattering processes without intermediate formation of the
CN. Theoretical efforts to develop a theory of such ``direct
reactions'' dominated nuclear reaction theory for some years and led
to explicit expressions for the non--diagonal parts of $\langle S
\rangle$. That raised the question how the Hauser--Feshbach formula
has to be modified in the presence of direct reactions, i.e., when
$\langle S \rangle$ is not diagonal.

After contributions due to~\textcite{Mol61,Mol63,Mol64}
and~\textcite{Sat63}, a definitive answer was proposed by Kawai,
Kerman, and McVoy~\cite{Kaw73}. We briefly describe the approach for
$\Gamma \gg d$. The starting point is a decomposition of the
scattering matrix of the form
\be
S_{a b} = \langle S_{a b} \rangle - i \sum_\mu \frac {g_{\mu a}
g_{\mu b}}{E - {\cal E}_\mu} \ .
\label{12c}
\ee
The first term on the right--hand side is the average $S$--matrix, and
the last term (which is identical to $S^{\rm fl}$) represents the
resonance contributions, with ${\cal E}_\mu$ denoting the complex
resonance energies. By definition, that term averages to zero. The
decomposition~(\ref{12c}) implies~\cite{Kaw73} correlations between
$g_{\mu a}$ and $g_{\mu b}$ for $a \neq b$ if $\langle S_{a b}
\rangle \neq 0$ for $a \neq b$. In that respect the approach differs
fundamentally from the Ericson model. Subsequent developments are
similar, however, to those sketched in Section~\ref{eri}. In
particular, the total widths of the resonances are effectively assumed
to be constant (independent of $\mu$). The average CN cross section is
expanded in powers of $d / \Gamma$. The term of leading order is
\be
\langle |S^{\rm fl}_{a b}|^2 \rangle = \frac{1}{{\rm Trace} P} \bigg\{
P_{a a} P_{b b} + P_{a b} P_{a b} \bigg\} \ .
\label{12d}
\ee
Here 
\be
P_{a b} = \delta_{a b} - \sum_c \langle S_{a c} \rangle \langle (S_{b
c})^* \rangle
\label{12e}
\ee
is Satchler's transmission matrix~\cite{Sat63}. It generalizes the
transmission coefficients defined in Eq.~(\ref{6}) to the case where
$\langle S \rangle$ is not diagonal. It measures the unitarity deficit
of $\langle S \rangle$. According to~\cite{Kaw73}, Eq.~(\ref{12d})
replaces the Hauser--Feshbach formula~(\ref{7}) when direct reactions
are present. We observe that if $\langle S \rangle$ is diagonal,
Eq.~(\ref{12d}) reduces to the Hauser--Feshbach formula, including the
elastic enhancement factor.

\subsection{Limitations of the Compound--Nucleus Picture}
\label{beyo}

Nuclear reaction cross sections begin to show deviations from the CN
picture some $20$ MeV or so above the first particle threshold. This
failure of the CN model is attributed to ``preequilibrium'' or
``precompound'' processes. The deviations occur because basic
assumptions on characteristic time scales of the reaction made in CN
theory do not apply any longer. In Eq.~(\ref{1}) such an assumption is
implicitly made. The average taken to calculate $\langle S \rangle$
must obviously extend over an energy interval containing many
resonances. By the uncertainty relation, a large energy interval
relates to a short time interval for the duration of the reaction, see
Section~\ref{general}. Thus, $\langle S \rangle$ describes the fast
part of the CN reaction. We estimate the minimum duration time of the
reaction by the time of passage $v_F / R$ of a nucleon with Fermi
velocity $v_F$ through a nucleus with radius $R$. In contrast, the
fluctuating part of $S$ with its rapid energy dependence describes the
slow part of the CN reaction: Decay of the CN with an average lifetime
$\hbar / \Gamma$, with $\Gamma$ given by the Weisskopf
estimate~(\ref{9}). For the decomposition~(\ref{1}) to be meaningful,
these two time scales must be well separated, $R / v_F \ll \hbar /
\Gamma$.

An even more stringent constraint on time scales is hidden in the
assumption that the CN resonances obey GOE statistics. As shown in
part I, that assumption is experimentally validated in the regime of
isolated resonances. But does it also hold for higher excitation
energies where $\Gamma \gg d$? The assumption implies that the partial
width amplitudes of the CN resonances (defined here as the
eigenfunctions of the GOE Hamiltonian introduced in Eq.~(\ref{24})
below) have a Gaussian distribution in each channel. In other words,
the couplings of all resonances to a given channel are, within
statistics, equally strong, and there does not exist a preferred state
or configuration, or a group of such states. (The coupling strengths
may differ, of course, for different channels).  This is the formal
expression of Bohr's assumption that the CN ``equilibrates''. To
display the time scale hidden in that assumption, we consider a
nucleon--induced reaction. The nucleon shares its energy with the
nucleons in the target in a series of two--body collisions.
Configurations of ever greater complexity are created. It takes
several or perhaps even many such collisions until the energy of the
incident particle is shared among many nucleons and the situation
described by assuming GOE statistics for the resonances, is attained.
The time elapsed between the first collision and the attainment of
equilibrium is the ``equilibration time'' $\tau_{\rm eq}$. The GOE
description of CN scattering holds if decay of the CN sets in after
the nucleus is equilibrated, i.e., if $\tau_{\rm eq}$ is smaller than
the average decay time $\hbar / \Gamma$. Obviously, we have $v_F / R <
\tau_{\rm eq}$ and the condition $\tau_{\rm eq} < \hbar / \Gamma$ for
applicability of RMT is usually more stringent than the condition $v_F
/ R \ll \hbar / \Gamma$ deduced from the decomposition~(\ref{1}). We
observe that $\hbar / \tau_{\rm eq}$ can be interpreted as the
spreading width of the simple configuration created in the first
collision between the incident particle and the target nucleus. The
condition $\tau_{\rm eq} < \hbar / \Gamma$ then requires that
spreading width to be large compared to the CN decay width $\Gamma$.

Precompound decay sets in when these conditions are violated. Simple
estimates show that $\tau_{\rm eq} \ll \hbar / \Gamma$ for isolated
resonances where $\Gamma \ll d$. But the complexity of CN resonances
and, therefore, $\tau_{\rm eq}$ increases with $E$. More importantly,
$\Gamma$ increases strongly with excitation energy, and $\hbar /
\Gamma$ decreases correspondingly. Thus, while $\tau_{\rm eq} \ll
\hbar / \Gamma$ for isolated resonances, $\hbar / \Gamma$ becomes
comparable with and eventually smaller than $\tau_{\rm eq}$ as $E$
increases: Particle decay is possible before the equilibrated CN is
reached in its full complexity. The associated reaction times cover
the entire range from $v_F / R$ to $\hbar / \Gamma$, and both the GOE
description of resonances and the decomposition~(\ref{1}) no longer
apply. By the same token, precompound decay does not have the
characteristic features of CN reactions: The average energy of the
emitted particles is larger than for CN decay; the emitted particles
``remember'' the incident channel(s) and emission is preferentially in
the forward direction. Pre--equilibrium reactions obviously call for a
different approach, although the large number of configurations
involved cannot be handled without statistical assumptions.
Phenomenological models designed for these reactions~\cite{Bla75} were
later followed by theories aiming at a quantum--statistical
description. The latter were compared and analyzed by~\textcite{Kon90}.
The theory of CN reactions reviewed in this paper is a closed theory
based on the concept of quantum chaos as embodied in RMT and on strong
empirical evidence reviewed in part I of this review. In its essential
aspects CN scattering is universal and occurs likewise in the
transmission of waves through disordered media. The theory of
precompound reactions is specific to nuclei. It requires additional
assumptions that go beyond chaos and RMT that cannot be tested
directly. It is not reviewed here.

Precompound processes occur in reactions induced by light particles
(mass $A < 6$ or $8$ or so) impinging on a target of mass $A > 20$ or
so. A very different situation is that of reactions between ``heavy
ions'', i.e., of two nuclei of mass $A > 50$ or $100$ each. The energy
of relative motion is usually given per nucleon. An energy of $5$ MeV
per nucleon may then easily amount to a total kinetic energy of
several $100$ MeV. In the case of a grazing collision, the reaction
transports energy and angular momentum of relative motion into
intrinsic excitations of either fragment. This is accompanied by the
transfer of nucleons between both reaction partners. Excitation
energies of several $10$ MeV are easily reached in either
fragment. But the two fragments basically keep their identity; the CN
corresponding to complete fusion of both fragments, is typically not
reached in a grazing collision. The theory of such processes uses
concepts like friction and dissipation developed in the theory of
non--equilibrium processes and methods of quantum statistical
mechanics. Naturally these are inspired by RMT and chaos but, to the
best of our knowledge, have never been strictly derived from such a
basis. This is why in the present review we will not cover that area
of the statistical theory of nuclear reactions either. In central
collisions between two heavy ions, fusion is possible even though the
density of CN resonances is limited: Quasibound states of the shell
model with too large single--particle widths do not qualify as CN
resonances~\cite{Wei64}. Such fusion processes are mainly investigated
via the gamma rays emitted in the decay of the highly excited CN; the
resulting data do yield statistically relevant information on excited
states of the CN; that information is discussed in Part I of this
review. The literature on the subject is vast because the available
energies have been much increased over the past $20$ years, and
because highly segmented gamma--ray arrays have become available. We
confine ourselves to citing two early reviews~\cite{Nor80,Wei80}.

\section{Random--Matrix Approach to Quantum Chaotic Scattering}
\label{theo}

The statistical models reviewed in Section~\ref{basi}, although
inspired by RMT, are not derived from a random--matrix description of
CN resonances. The Bohr assumption is intuitively appealing. But it is
not clear in which range of $\Gamma / d$ it applies and with what
accuracy. The Ericson model leads to interesting predictions that
agree with experiment. However, completing the model formulated in
Eq.~(\ref{10}) by an equation for the elastic case, i.e., for $S_{a
a}$, one finds that the resulting $S$--matrix violates unitarity. The
same statement applies to the inclusion of direct reactions in the
Hauser--Feshbach formula~\cite{Ker76}. As for the Weisskopf
estimate~(\ref{9}), the physical significance of the parameter
$\Gamma$ is not clear. A width parameter appears in the context of the
$S$--matrix autocorrelation function, see Eq.~(\ref{12}). We may also
define $\Gamma$ as the mean distance of the poles of $S$ from the real
axis. Are the two definitions identical? If not, which of the two (if
any) is given by the Weisskopf estimate? What is the accuracy of the
estimate?  What are the correction terms of next order? A theoretical
approach based upon a random--matrix description of CN resonances that
yields $S$--matrix distributions and $S$--matrix correlation functions
within controlled approximations, is clearly called for. That approach
is reviewed in the present and in the following two Sections. In the
present Section we formulate the approach and, at the end of the
Section, give a brief historical survey. The approach uses the average
$S$--matrix as input. Properties of and phenomenological models for
$\langle S \rangle$ are reviewed in Section~\ref{aver}. Results of the
random--matrix approach are given in Section~\ref{resu}. The approach
reviewed in this and the following Sections is not confined to CN
reactions but applies in general to a random--matrix description of
quantum chaotic scattering.

\subsection{Resonance Reactions}
\label{rere}

A statistical theory based on RMT can be formulated only on the basis
of a theory of resonance reactions. Only when the $S$--matrix is
written explicitly in terms of resonance contributions can we
implement the statistical properties of those resonances. We describe
an approach that is based upon the coupling of $N$ quasibound states
to a number of channels and yields the $S$--matrix directly in terms
of the Hamiltonian governing those quasibound states. We show why this
approach is very well suited for calculating GOE averages.

\subsubsection{Single Resonance}
\label{sing}

We begin with the simplest case, a single resonance without any
background scattering. The $S$--matrix has Breit--Wigner form,
\be
S_{a b}(E) = \delta_{a b} - i \frac{\gamma_a \gamma_b}{E - E_0 +
(i/2) \Gamma} \ .
\label{13}
\ee
The partial width amplitudes $\gamma_a$, $\gamma_b$, $\ldots$ give the
probability amplitudes for decay of the resonance into channels $a$,
$b$, $\ldots$. Factorization of the numerator in the last term in
Eq.~(\ref{13}) is implied by quantum mechanics if the resonance is
caused by a single quasibound state, see below. The matrix $S(E)$ is
obviously symmetric. We impose the unitarity condition~(\ref{12a}) for
all energies $E$ and find that all the partial width amplitudes must
be real and that
\be
\sum_a \gamma^2_a = \Gamma \ .
\label{15}
\ee

The sum of the eigenphase shifts $\sum_a \delta_a$ of $S$ increases by
$\pi$ as $E$ increases from a value far below to a value far above
$E_0$. To see this, we note that the form~(\ref{12b}) for $S$ implies
$\det S = \exp ( 2 i \sum_a \delta_a )$. To calculate $\det S$ we use
Eq.~(\ref{13}) and find in matrix notation $\det S = \det ( {\bf 1} -
i \vec{\gamma} \vec{\gamma}^T / (E - E_0 + (i/2) \Gamma) )$ where $T$
denotes the transpose. The last determinant is easily worked out and
yields $(E - E_0 - (i/2) \Gamma) / (E - E_0 + (i/2) \Gamma)$. That
term has magnitude one; the phase increases by $2 \pi$ as $E$ passes
through the resonance. The resulting increase by $\pi$ of the sum of
the eigenphase shifts of $S$ over the resonance does not imply that
one of the eigenphase shifts grows by $\pi$ while all others remain
unchanged. On the contrary, the non--crossing theorem of the
eigenphases~\cite{Wei67} referred to above implies that generically
all eigenphases increase as $E$ passes through the resonance, the
average increase of every eigenphase being $\pi / \Lambda$ where
$\Lambda$ is the number of open channels.

Eq.~(\ref{13}) can be derived by considering a single quasibound state
$| \phi \rangle$ with energy $\ve$ which interacts with a number
$\Lambda$ of continuum states $| \chi_a(E) \rangle$ (the range of the
channel index $a$ is $\Lambda$). The continuum wave functions are
orthonormal, $\langle \chi_a(E) | \chi_b(E') \rangle = \delta_{a b}
\delta(E - E')$ for all $a, b$ and orthogonal to the quasibound state,
$\langle \chi_a(E) | \phi \rangle = 0$ which in turn is normalized,
$\langle \phi | \phi \rangle = 1$. In keeping with the absence of any
background scattering in Eq.~(\ref{13}) we neglect the dynamical
coupling of the continuum states with each other and assume that the
states $| \chi_a(E) \rangle$ describe free relative motion so that the
scattering phase shift for each $| \chi_a(E) \rangle$ vanishes. We
focus attention on the coupling between the continuum states and the
quasibound state mediated by the real matrix elements $W_a(E)$. That
coupling causes $| \phi \rangle$ to become a resonance. The
Hamiltonian is
\ba
&& H = \ve | \phi \rangle \langle \phi | + \sum_a \int {\rm d} E \ E |
\chi_a(E) \rangle \langle \chi_a(E) | \nonumber \\
&& + \sum_a \int {\rm d} E \ W_a(E) ( | \phi \rangle \langle \chi_a(E)
| + | \chi_a(E) \rangle \langle \phi | ) \ . \nonumber \\
\label{16}
\ea
To determine the scattering eigenstates of $H$ we
write~(\cite{Dir58},~\cite{Fan61}, and~\cite{Mah69})
\be
| \Psi^+_a(E) \rangle = \sum_c \int {\rm d} E' \ a^{(a)}_c(E, E') |
\chi_c(E') \rangle + b^{(a)}(E) | \phi \rangle \ .
\label{17}
\ee
Here $| \Psi \rangle$ has an incoming wave in channel $a$ only. We
insert Eq.~(\ref{17}) into the Schr{\"o}dinger equation $H | \Psi
\rangle = E | \Psi \rangle$ and multiply the result from the left with
$\langle \phi |$ and $ \langle \chi_a(E) |$. That gives a set of
coupled linear equations for the coefficients $a(E, E')$ and
$b(E)$. These are solved by imposing the boundary condition for $|
\Psi^+_a(E) \rangle$. We define
\be
F(E) = \sum_c \int {\rm d} E' \ \frac{W^2_c(E')}{E^+ - E'} = \Delta -
\sum_c i \pi W^2_c(E) \ .
\label{17a}
\ee
Here $E^+$ carries an infinitesimal positive increment. That
corresponds to the boundary condition on $| \Psi^+ \rangle$. The shift
function $\Delta$ is defined in terms of a principal--value integral.
We obtain
\ba
b^{(a)}(E) &=& \frac{W_a(E)}{E - \ve - F(E)} \ , \nonumber \\
a^{(a)}_c(E, E') &=& \delta_{a c} \delta(E - E') \nonumber \\
&& + \frac{1}{E^+ - E'} \ \frac{W_a(E) W_c(E')}{(E - \ve - F(E))} \ .
\label{18}
\ea
In the second of these equations we have again used the boundary
condition: After insertion into Eq.~(\ref{17}) the delta function
yields an incoming wave in channel $a$ while the denominator $1 /
(E^+ - E')$ in the last term makes sure that all other contributions
to $| \Psi^+_a(E) \rangle$ produce only outgoing waves at infinity.
The amplitude of these waves can be easily worked out and is
\be
S^{\rm res}_{a b}(E) = \delta_{a b} - 2 i \pi \frac{W_a(E) W_b(E)}{E
- \ve - \Delta + (i/2) \Gamma} \ .
\label{19}
\ee
Here $\Gamma = 2 \pi \sum_c W^2_c(E)$. We have used the superscript
``res'' on $S$ to indicate the origin of Eq.~(\ref{19}) from a
dynamical calculation employing a quasibound state. Eq.~(\ref{19})
agrees with Eq.~(\ref{13}) if we put $\gamma_a = \sqrt{2 \pi} W_a(E)$
and $E_0 = \ve + \Delta$. The factorization of the numerator in the
second of Eqs.~(\ref{18}) causes the factorization of the resonance
numerator in Eq.~(\ref{19}) and holds for all resonances caused by a
single quasibound state.

To interpret Eqs.~(\ref{17a}) and (\ref{19}), we note that $F(E)$ is
the sum of the resolvents for free motion in the channels: The first
factor $W_c(E')$ is the amplitude for decay of the quasibound state
into channel $c$ at energy $E'$, the denominator $1 / (E^+ - E')$ is
the propagator for channel $c$, and the second factor $W_c(E')$ is the
amplitude for returning to the resonance. We have to sum over all
energies $E'$. The occurrence of $F(E) = \Delta - (i / 2) \Gamma$ in
the denominator of Eq.~(\ref{19}) signals repeated decay of and return
to the quasibound state. By that mechanism the quasibound state turns
into a resonance.  The real part $\Delta$ of $F(E)$ corresponds to
off--shell processes and gives the shift of the resonance energy
$\ve$. While $\Im F$ is due to the open channels only, $\Delta$
receives contributions from both, open and closed channels.

We turn to the general case where the background scattering is not
negligible and leads to a background scattering matrix $S^{(0)}$. The
$S$--matrix for a single resonance takes the form
\be
S_{a b}(E) = S^{(0)}_{a b} - i \frac{\tilde{\gamma}_a
\tilde{\gamma}_b}{E - E_0 + (i/2) \Gamma} \ .
\label{19a}
\ee
The matrix $S^{(0)}$ is unitary. This follows from the unitarity of
$S$ at energies far from the resonance energy $E_0$. Unitarity of $S$
at all energies $E$ implies that the complex partial width amplitudes
obey $\sum_c | \tilde{\gamma}_c |^2 = \Gamma$. The matrix $S^{(0)}$ is
also symmetric. This is implied by time--reversal invariance.
Therefore, $S^{(0)}$ possesses an eigenvalue decomposition of the
form~(\ref{12b}), $S^{(0)}_{a b} = \sum_c {\cal O}^{(0)}_{a c} \exp (
2 i \delta^{(0)}_c ) {\cal O}^{(0)}_{b c}$. We use that relation and
matrix notation to write Eq.~(\ref{19a}) in the form
\be
S_{a b}(E) = \bigg\{ {\cal O}^{(0)} \exp ( i \delta^{(0)} ) S^{\rm
res}(E) \exp ( i \delta^{(0)} ) ({\cal O}^{(0)})^T \bigg\}_{a b}
\label{14}
\ee
where the matrix $S^{\rm res}$ in the eigenchannel representation of
$S^{0}$ has exactly the form of Eq.~(\ref{13}),
\be
S^{\rm res}_{c d}(E) = \delta_{c d} - i \frac{\gamma_c \gamma_d}{E -
E_0 + (i/2) \Gamma} \ .
\label{14a}
\ee
For the unitarity condition~(\ref{12a}) to hold for all energies $E$,
the partial width amplitudes $\gamma_c = \exp ( - i \delta^{(0)}_c )
\sum_a {\cal O}_{c a} \tilde{\gamma}_a$ must be real and must obey
Eq.~(\ref{15}). It is normally assumed that $S^{(0)}$ and the
$\tilde{\gamma}$s are independent of energy. This approximation
assumes that angular--momentum and Coulomb penetration factors and all
relevant matrix elements depend smoothly on energy and is typically
justified over the width of a resonance although counterexamples
exist, see Section~\ref{fine}. The assumption does not apply for
reactions at threshold energies which are not treated here.
Analogous assumptions are made also when we later deal with many
resonances; these are not mentioned explicitly again. It is easy to
check that the sum of the eigenphases of $S$ increases by $\pi$ over
the width of the resonance.

As in the case of Eq.~(\ref{13}), it is possible to derive
Eq.~(\ref{14}) from a dynamical model. The Hamiltonian differs from
the one in Eq.~(\ref{16}) in two respects. First, the continuum states
$| \chi_a(E) \rangle$ are true scattering states with nonzero phase
shifts. Second, the continuum states carrying different channel
indices interact with each other. Taking into account these phase
shifts and continuum--continuum interactions first and neglecting the
quasibound state yields the background matrix $S^{(0)}(E)$. The actual
calculation of that matrix involves a coupled--channels problem and
may be rather involved. We show in Section~\ref{aver} how that problem
is overcome. In the eigenchannel representation Eq.~(\ref{12b}) of
$S^{(0)}$, the radial part of relative motion of the continuum wave
functions $ \exp ( - i \delta^{(0)}_c ) \Omega_c(E) \rangle$ is real.
(These functions are defined below Eq.~(\ref{12b}).  For neutral
particles the radial part in any channel $c$ is essentially given by a
spherical Bessel function $j_{\ell_c}(k_c r_c + \delta_c)$).
Therefore, the matrix elements $W_a(E)$ coupling the quasibound state
to the continuum wave functions are also real in the eigenchannel
representation. A calculation analogous to the one leading to
Eq.~(\ref{19}) then yields Eq.~(\ref{14}) with $\gamma_a =
\sqrt{2 \pi} W_a(E)$ and the resonance energy $E_0$ given by $\ve +
\Delta$ as before. We do not give that calculation here but refer
to~\cite{Mah69}. In summary, the $S$--matrix is given by
Eq.~(\ref{14}) and $S^{\rm res}(E)$ has the form of Eq.~(\ref{19})
except that the real matrix elements $W_c(E)$ now refer to the
eigenchannel representation of $S^{(0)}$. Eq.~(\ref{14}) is very
convenient as it clearly separates the effects of the
continuum--continuum interaction and the effect of resonance
scattering.

\subsubsection{$N$ Resonances}
\label{reso}

Except for the factorization of the numerator in the last term in
Eq.~(\ref{13}), it is straightforward to guess the form of the
$S$--matrix for a single resonance. In the case of several or many
overlapping resonances it is difficult to guess a form for $S$ that
fulfills the unitarity conditions~(\ref{12a}) for all values of the
resonance parameters (partial width amplitudes, resonance energies and
total widths). Therefore, it is useful to derive that form from a
dynamical model.  We describe the model but skip the derivation
because it is quite similar to the derivation leading to
Eq.~(\ref{19}).

We consider a set of $\Lambda$ orthonormal continuum states $|
\chi_a(E) \rangle$ and a set of $N$ orthonormal quasibound states $|
\phi_\mu \rangle$. The continuum states interact with each other. As
in Section~\ref{sing} that interaction gives rise to a smooth unitary
and symmetric background scattering matrix $S^{(0)}_{a b}$ for which
we can write $S^{(0)}_{a b} = $ $\sum_c {\cal O}^{(0)}_{a c} \exp (2 i
\delta^{(0)}_c) {\cal O}^{(0)}_{b c}$. The quasibound states span an
$N$--dimensional Hilbert space. The unperturbed energies of and the
interactions amongst the quasibound states are not specified in detail
but are represented jointly by the real and symmetric $N$--dimensional
matrix $H$, the projection of the total Hamiltonian onto the
$N$--dimensional Hilbert space of quasibound states. In that space, we
use an arbitrary basis of orthonormal states and write the matrix
elements of $H$ as $H_{\mu \nu}$ with $\mu, \nu = 1, \ldots, N$. As
before we define the interaction between the quasibound states and the
continuum states in the eigenchannel representation of the matrix
$S^{(0)}$ with scattering eigenfunctions $\exp ( - i
\delta^{(0)}_a) \Omega_a(E)$. Then the matrix elements $W^{(0)}_{\mu
a}(E) = W^{(0)}_{a \mu}(E)$ of that interaction are real. The upper
index zero indicates that we use the eigenchannel representation of
$S^{(0)}$. The total $S$--matrix again has the form of Eq.~(\ref{14})
but the resonance part now differs and is given by
\be
S^{\rm res}_{a b}(E) = \delta_{a b} - 2 i \pi \sum_{\mu \nu} W^{(0)
}_{a \mu}(E) (D^{-1})_{\mu \nu} W^{(0)}_{\nu b}(E) \ .
\label{20}
\ee
Here $D(E)$ is a matrix in the space of quasibound states and given by
\be
D_{\mu \nu}(E) = E \delta_{\mu \nu} - H_{\mu \nu} - F_{\mu \nu}(E) \ ,
\label{21}
\ee
with
\ba
F_{\mu \nu}(E) &=& \sum_c \int {\rm d} E' \ \frac{W^{(0)}_{\mu c}(E')
W^{(0)}_{c \nu}(E')}{E^+ - E'} \nonumber \\
&=& \Delta^{(0)}_{\mu \nu} - i \pi \sum_c W^{(0)}_{\mu c}(E)
W^{(0)}_{c \nu}(E) \ . 
\label{22}
\ea
It is easy to check that $S^{\rm res}(E)$ in Eq.~(\ref{20}) is
unitary. According to Eq.~(\ref{14}) this statement implies unitarity
of the total $S$--matrix. Also $S^{\rm res}$ is obviously symmetric,
and so is, therefore, $S(E)$. We note that the clear separation of the
influence of the continuum--continuum interaction and that of the
quasibound states expressed by Eq.~(\ref{14}) is not restricted to
the case of a single resonance but holds in general. We also observe
that Eqs.~(\ref{20}) to (\ref{22}) apply for any strength of the
coupling between resonances and channels. The equations are not
restricted to the regime of weakly overlapping resonances and will
serve as a basis for a statistical theory that applies for all values
of $\Gamma / d$.

As for the case of a single resonance, $F(E)$ given in Eq.~(\ref{22})
describes virtual (real) decay of and return to the quasibound states
via all channels by its real (imaginary) parts, respectively. This
gives rise to the real shift matrix $\Delta^{(0)}$ and the real width
matrix with elements $2 \pi \sum_c W^{(0)}_{\mu c}(E) W^{(0)}_{c
\nu}(E)$. Qualitatively speaking, the resonances are isolated (they
overlap) if the average resonance spacing is large (small) compared to
typical elements of the matrix $F(E)$. That statement is quantified in
Sections~\ref{aver} and \ref{pole} below.

To define the analogue of the total width for a single resonance, we
have to write the matrix $S^{\rm res}$ as a sum over poles in the
complex energy plane. That is straightforward only for isolated
resonances. The Hermitean matrix $H$ can be diagonalized by an
orthogonal transformation $O$. We denote the eigenvalues of $H$ by
$E_\mu$ and write for the transformed matrix elements
$\tilde{W}^{(0)}_{\mu a} = \sum_\nu O_{\mu \nu} W^{(0)}_{\nu a}$. With
a corresponding notation for $F$ the transformed matrix $D$ takes the
form $(E^+ - E_\mu) \delta_{\mu \nu} - \tilde{F}_{\mu \nu}$.  The $N$
eigenvalues $E_\mu$ are coupled to each other by the non--diagonal
elements of $\tilde{F}$. For isolated resonances, that coupling can be
neglected, and only the diagonal elements of $\tilde{F}$ are
retained. We omit $\Delta^{(0)}$ in Eq.~(\ref{20}), define $\Gamma_\mu
= 2 \pi \sum_c (\tilde{W}^{(0)}_{\mu c})^2$ and obtain for isolated
resonances the form
\ba
&&S^{\rm res}_{a b}(E) = \delta_{a b} \nonumber \\
&& - 2 i \pi \sum_\mu \tilde{W}^{(0)}_{a \mu}(E) \big( E - E_\mu +
(i/2) \Gamma_\mu \big)^{-1} \tilde{W}^{(0)}_{\mu b} \ . \nonumber \\
\label{22c}
\ea
This equation explicitly displays $N$ resonances. For the resonance
labeled $\mu$, the partial width amplitude $\tilde{W}^{(0)}_{a \mu}$
gives the probability amplitude for decay into channel $a$, and the
total width $\Gamma_\mu$ is the sum of the partial widths
$|\tilde{W}^{(0)}_{a \mu}|^2$. The scattering matrix~(\ref{22c}) is
unitary if we neglect the overlap of different resonances. The more
general Eq.~(\ref{20}) obviously describes an $S$--matrix with $N$
resonances, too, but applies also outside the regime of isolated
resonances.

The $S$--matrix in Eq.~(\ref{20}) possesses $N$ poles in the complex
energy plane. All poles of $S^{\rm res}$ are located below the real
energy axis. To see that we observe that the positions of the poles
are given by the eigenvalues of $H_{\mu \nu} + F_{\mu \nu}$. Let $z$
be one such eigenvalue. It obeys
\be
\sum_\nu (H_{\mu \nu} + F_{\mu \nu}) \Psi_\nu = z \Psi_\mu \ .
\label{31}
\ee
We multiply Eq.~(\ref{31}) by the complex conjugate eigenfunction
$(\Psi_\mu)^*$, sum over $\mu$, and take the imaginary part of the
resulting equation. That yields the ``Bell--Steinberger relation''
\be
- \pi \sum_c | \sum_\mu W_{c \mu} \Psi_\mu |^2 = \sum_\mu | \Psi_\mu
|^2 \ \Im [ z ] \ .
\label{32}
\ee
This shows that indeed $\Im [z] < 0$ unless $\sum_\mu W_{c \mu}
\Psi_\mu = \gamma_c = 0$ for all $c$. In the latter case we deal with
a bound state embedded in the continuum with vanishing partial width
amplitudes $\gamma_c$. Then $z$ is real and $S^{\rm res}$ does not
have a pole at $E = z$.
 
The scattering matrix $S^{\rm res}$ in Eq.~(\ref{20}) can be rewritten
in terms of the $K$--matrix, a matrix in channel space. The ensuing
relations~(\ref{22a}) and (\ref{22b}) are quite general and not
restricted to isolated resonances. They may be used~\cite{Mah69} to
establish the relation between the $R$--matrix approach~\cite{Wig47}
and the present framework. Using the orthogonal transformation that
leads to Eq.~(\ref{22c}) we define
\be
K_{a b}(E) = \pi \sum_\mu \frac{\tilde{W}^{(0)}_{a \mu}
\tilde{W}^{(0)}_{\mu b}} {E - E_\mu}  
\label{22a}
\ee
and find after a straightforward calculation
\be
S^{\rm res}_{a b}(E) = \bigg( \frac{1 - i K}{1 + i K} \bigg)_{a b} \ .
\label{22b}
\ee

\subsection{Stochastic Scattering Matrix}
\label{stoc}

The result of Section~\ref{reso} puts us into the position to
introduce the stochastic description of the resonances in terms of the
GOE. In a dynamical treatment of the resonances based on the shell
model we would express the matrix $H$ in Eq.~(\ref{21}) in terms of
the single--particle energies and the matrix elements of the residual
interaction. We replace such a dynamical treatment by a stochastic one
and consider $H$ as a member of the GOE. In other words,  in
Eq.~(\ref{21}) we replace the actual Hamiltonian matrix $H$ by the ensemble
$H^{\rm GOE}$. That replacement generates an ensemble of scattering
matrices $S$ each member of which is obtained by drawing $H$ from the
GOE. For the sake of completeness we recall the definition and some
properties of the GOE and refer to Part I for further details. The
independent elements $H^{\rm GOE}_{\mu
\nu}$ of the real and symmetric $N$--dimensional matrix $H^{\rm GOE}$
are uncorrelated Gaussian--distributed random variables with zero mean
values and a second moment given by
\be
\langle H^{\rm GOE}_{\mu \nu} H^{\rm GOE}_{\rho \sigma} \rangle =
\frac{\lambda^2}{N} ( \delta_{\mu \rho} \delta_{\nu \sigma} +
\delta_{\mu \sigma} \delta_{\nu \rho} ) \ .
\label{24}
\ee
Here $\lambda$ is a parameter. The average spectrum of $H^{\rm GOE}$
has semicircular shape. The radius of the semicircle is given by $2
\lambda$. Near the center of the spectrum the average level spacing
$d$ is given by $d = \pi \lambda / N$. The GOE is invariant under
orthogonal transformations of the underlying Hilbert space. The energy
argument $E$ of $S$ is taken close to (i.e., a finite number of
spacings away from) that center. All observables are calculated in the
limit $N \to \infty$. As pointed out in the Introduction, we expect
the replacement of $H$ by $H^{\rm GOE}$ to be valid at bombarding
energies of up to $10$ or $20$ MeV where the internal equilibration
time of the CN is smaller than the decay time $\hbar /
\Gamma$.

The statistical approach defined by the replacement $H \to H^{\rm
GOE}$ seems to contain a large number of parameters, i.e., the $N
\Lambda$ matrix elements $W^{(0)}_{a \mu}$ and the parameter $\lambda$
of the GOE. To appreciate the simplifications due to a statistical
treatment, it is instructive to visualize the effort required in case
we would want to determine these parameters from a dynamical model
such as the shell model. Using a shell--model basis of scattering
states we would first have to calculate the background matrix
$S^{(0)}$. We would then have to calculate the matrix elements
$W^{(0)}_{a \mu} = \exp ( - i \delta_a ) \langle \Omega_a | V |
\phi_\mu \rangle$ in the eigenchannel representation of $S^{(0)}$ in
terms of the residual interaction $V$ by constructing the quasibound
states $\phi_\mu$. For large $N$ that effort would be huge. We now
show that in the statistical approach, all this is much simplified by
the invariance properties of the GOE.

The statistical theory uses as input the elements $\langle S_{a b}
\rangle$ of the average $S$--matrix $\langle S \rangle$ defined by the
decomposition~(\ref{1}). These elements must be given in terms of some
dynamical calculation, or some suitable model. Options for a
calculation of $\langle S \rangle$ are reviewed in Section~\ref{aver}.
The statistical theory aims at predicting moments and correlation
functions (defined as ensemble averages) of $S^{\rm fl}(E)$ in terms
of $\langle S \rangle$ in the limit $N \to \infty$.  The orthogonal
invariance of the GOE implies that all such moments and correlation
functions can depend only on orthogonal invariants of the parameters
$W^{(0)}_{a \mu}$. The only such invariants are the bilinear forms
$\sum_\mu W^{(0)}_{a \mu} W^{(0)}_{\mu b}$. But the $S$--matrix is
dimensionless and can depend only on dimensionless combinations of the
parameters of the statistical approach. These are the real quantities
$\sum_\mu W^{(0)}_{a \mu} W^{(0)}_{\mu b} / \lambda$.  Their number is
$\Lambda (\Lambda + 1) / 2$, equal to the number of elements of
$\langle S \rangle$. But the elements of $\langle S \rangle$ are
complex and it even seems that the model is overdetermined. We show in
Section~\ref{aver} below that the moments and correlation functions of
$S$ actually depend, aside from overall phase factors, only on the
magnitudes $|\langle S_{a b} \rangle|$ of the elements of the average
scattering matrix. Thus, the statistical theory is well defined. We
also show below that $\langle S \rangle$ determines not only the
invariants $\sum_\mu W^{(0)}_{a \mu} W^{(0)}_{\mu b} / \lambda$ but
also the orthogonal matrix ${\cal O}^{(0)}$ and the phase shifts
$\delta^{(0)}_c$ appearing in Eq.~(\ref{14}). In other words,
knowledge of $\langle S \rangle$ suffices for a complete calculation
of CN scattering processes.

The quantities $\sum_\mu W^{(0)}_{a \mu}$ $W^{(0)}_{\mu b} / \lambda$
are the effective parameters of the statistical theory. We use this
fact to simplify further the resonance part of the scattering matrix
in Eq.~(\ref{20}). The real and symmetric matrix $\sum_\mu W^{(0)}_{a
\mu} W^{(0)}_{\mu b}$ can be diagonalized by an orthogonal
transformation ${\cal O}^{\rm CN}_{a b}$ in channel space. We define
the new real matrix elements
\be
W_{a \mu} = \sum_b {\cal O}^{\rm CN}_{b a} W^{(0)}_{b \mu}
\label{26}
\ee
which obey
\be
\sum_\mu W_{a \mu} W_{\mu b} = \delta_{a b} N v^2_a \ .
\label{29}
\ee
To express $S_{a b}$ in terms of the $W_{a \mu}$s we define the
unitary matrix
\be
U_{a b} = \bigg( {\cal O}^{(0)} \exp ( i \delta^{(0)} ) {\cal O}^{\rm
CN} \bigg)_{a b}
\label{30}
\ee
and write
\be
S_{a b}(E) = \bigg( U S^{\rm CN}(E) U^T \bigg)_{a b} \ .
\label{C1}
\ee
Here
\be
S^{\rm CN}_{a b}(E) = \delta_{a b} - 2 i \pi \sum_{\mu \nu} W_{a
\mu}(E) (D^{-1})_{\mu \nu} W_{\nu b}(E) \ ,
\label{C2}
\ee
and
\be
D_{\mu \nu}(E) = E \delta_{\mu \nu} - H^{\rm eff}_{\mu \nu}
\label{C3}
\ee
where
\be
H^{\rm eff} = H^{\rm GOE}_{\mu \nu} + F_{\mu \nu}
\label{27c}
\ee
is the non--Hermitean ``effective Hamiltonian'' that describes the
dynamics of the resonances. The non--Hermitean part of $H^{\rm eff}$
is due to the imaginary part of the matrix $F_{\mu \nu}(E)$ given by
\be
F_{\mu \nu}(E) = \sum_c \int {\rm d} E' \ \frac{W_{\mu c}(E') W_{c
\nu}(E')}{E^+ - E'} \ .
\label{28} 
\ee
That matrix describes the effect of the coupling to the channels.

The matrix $S^{\rm CN}(E)$ of Eq.~(\ref{C2}) is the $S$--matrix for CN
scattering in its purest form: Absence of all continuum--continuum
interactions and of all elastic scattering phase shifts, with real
coupling matrix elements $W_{a \mu}$ obeying Eq.~(\ref{29}). Thus,
$S^{\rm CN}$ is the central object of study of the statistical theory.
The ensemble of matrices $S^{\rm CN}(E)$ constitutes a matrix--valued
random process. (For fixed $E$, every element of $S^{\rm CN}$ is a
random variable. Because of the dependence on energy, a generalization
of the concept of random variable is called for, and one speaks of a
random process.) The moments of $S^{\rm CN}$ depend on the
dimensionless parameters
\be
x_a = \frac{\pi N v^2_a}{\lambda} = \frac{\pi^2 v^2_a}{d} \ ,
\label{29a}
\ee
see Eq.~(\ref{29}), where we have taken the average GOE level spacing
$d$ at the center of the GOE spectrum. The form of Eq.~(\ref{29a}) is
reminiscent of Fermi's golden rule. The $S$--matrix correlation
functions depend, in addition, on energy differences given in units of
$d$, see Section~\ref{resu}.

The unitary matrix $U$ in Eq.~(\ref{30}) contains all of the
non--statistical effects that connect $S^{\rm CN}$ with the physical
channels: The transformation to the eigenchannel representation, the
eigenphase shifts of $S^{(0)}$, and the transformation that
diagonalizes the bilinear form of the $W^{(0)}$s, see Eq.~(\ref{26}).
Statistical assumptions are made only with respect to the matrix
$H^{\rm GOE}$ that describes the interaction among quasibound states.
We do not impose any statistical requirements on the matrix elements
$W_{\mu c}(E)$, on $S^{(0)}$, or on $U$. This is in keeping with the
evidence on chaos in nuclei reviewed in Part I. It also corresponds to
chaotic features of quantum dots and microwave billiards.

In many cases it is possible to simplify the matrix $F_{\mu \nu}(E)$.
As in Eq.~(\ref{22}) we decompose $F_{\mu \nu}(E)$ into its real and
imaginary parts. Often the coupling matrix elements $W_{\mu a}(E)$ are
smooth functions of $E$ (so that $d \ [{\rm d} \ln W_{\mu c}(E) / {\rm
d} E] \ll 1$). Then the principal--value integral is small, the real
shift matrix can be neglected, and $F_{\mu \nu} \approx - (i/2)
\Gamma_{\mu \nu}$ where the width matrix $\Gamma_{\mu \nu}$ is given
by
\be
\Gamma_{\mu \nu} = 2 \pi \sum_c W_{\mu c}(E) W_{c \nu}(E) \ .
\label{29d}
\ee
The assumption $d \ [{\rm d} \ln W_{\mu c}(E) / {\rm d} E] \ll 1$ is
not justified near the threshold of a channel, $a$ say, where the
$W_{\mu a}$s with $\mu = 1, \ldots, N$ depend strongly upon energy. A
better approximation is obtained when we write the principal--value
integral giving the real part of $F_{\mu \nu}$ in the approximate
form~\cite{Mah69} $\sum_c W_{\mu c} \Delta_c W_{c \nu}$ where the
$\Delta_c$s are some channel--dependent constants. Then $F_{\mu \nu}
\approx \pi \sum_c W_{\mu c} (\Delta_c - i) W_{c \nu}$. That form
differs from the first approximation by the replacement of the channel
propagator $-i$ by $(\Delta_c - i)$. ~\textcite{Aga75} have shown
that such a replacement only modifies the dependence of the
transmission coefficients $T_c$ on the coupling strengths $x_c$ and
leaves the theory otherwise unchanged. But the coefficients $T_c$
serve as phenomenological input parameters anyway, and the coupling
strengths $x_c$ do not appear explicitly anywhere in the final
expressions for the moments or correlation functions of $S^{\rm
CN}$. Therefore, we use in the sequel the simple approximation $F_{\mu
\nu} \approx - i \sum_c W_{\mu c} W_{c \nu}$. With this approximation,
the effective Hamiltonian $H^{\rm eff}$ in Eq.~(\ref{27c}) takes the
form
\be
H^{\rm eff}_{\mu \nu} = H^{\rm GOE}_{\mu \nu} - i \pi \sum_c W_{\mu c}
W_{c \nu} \ . 
\label{C4}
\ee
Eqs.~(\ref{C1},\ref{C2},\ref{C3}) and (\ref{C4}) together with
Eq.~(\ref{29}) are the basic equations of the statistical model as
used in the present paper. They are used throughout except for
Section~\ref{fine}: Isobaric analogue resonances occur for proton
energies below or at the Coulomb barrier where the energy dependence
of the matrix elements $W_{c \mu}(E)$ cannot be neglected.

In some cases it is useful to write $S^{\rm CN}$ in the diagonal
representation of $H^{\rm GOE}$. We diagonalize every realization of
$H^{\rm GOE}$ with an orthogonal matrix $O$. The eigenvalues $E_\mu$,
$\mu = 1, \ldots, N$ follow the Wigner--Dyson distribution, see
Section~I.II.D. For $N \to \infty$ the elements of $O$ are
uncorrelated Gaussian--distributed random variables and so are,
therefore, the transformed matrix elements $\tilde{W}_{\mu a} = \sum_b
O_{\mu \nu} W_{\nu a}$. These are not correlated with the $E_\mu$s,
have zero mean values and have second moments given by
\be
\langle \tilde{W}_{\mu a} \tilde{W}_{\nu b} \rangle = v^2_a
\delta_{a b} \delta_{\mu \nu} \ ,
\label{27f}
\ee
see Eq.~(\ref{29}). It is assumed that the $v^2_a$s are independent of
energy. In that representation the matrix $S^{\rm CN}$ of
Eq.~(\ref{C2}) is given by
\be
S^{\rm CN}_{a b}(E) = \delta_{a b} - 2 i \pi \sum_{\mu \nu}
\tilde{W}_{a \mu}(E) (\tilde{D}^{-1})_{\mu \nu} \tilde{W}_{\nu b}(E)
\label{27g}
\ee
where
\be
\tilde{D}_{\mu \nu}(E) = (E - E_\mu) \delta_{\mu \nu} + i \pi \sum_c
\tilde{W}_{\mu c} \tilde{W}_{c \nu} \ .
\label{27h}
\ee

As done for $S^{\rm res}$ in Eq.~(\ref{22b}), $S^{\rm CN}$ in
Eq.~(\ref{C2}) can also be written in terms of the $K$--matrix. That
form was, in fact, the starting point of some work on the statistical
theory reviewed in Section~\ref{earl}. The forms~(\ref{C2}) and
(\ref{22b}), although mathematically equivalent, differ in one
essential aspect. The $S$--matrix in Eq.~(\ref{C2}) depends explicitly
on $H^{\rm GOE}$. Ensemble averages of moments of $S^{\rm CN}$ as
given by Eq.~(\ref{C2}) are calculated by integrating over the
Gaussian--distributed matrix elements of $H^{\rm GOE}$. The orthogonal
invariance of the GOE makes it possible to do the calculation
analytically for some of these averages, see Section~\ref{resu}. In
contrast, the $K$--matrix depends on the eigenvalues and eigenvectors
of $H^{\rm GOE}$. Thus, the orthogonal invariance of the GOE is not
manifest in the form~(\ref{22b}), and ensemble averages require a
separate integration over the distribution of eigenvalues and of
eigenvectors of the GOE. The calculation turns out to be prohibitively
difficult. That is why the $K$--matrix formalism has only been used
for numerical simulations.

We have derived Eqs.~(\ref{C1},\ref{C2},\ref{C3}) and (\ref{C4}) in
the framework of the shell--model approach to nuclear
reactions~\cite{Mah69}. That approach directly yields the dependence
of the scattering matrix on the Hamiltonian for the quasibound
states. Feshbach's unified theory of nuclear
reactions~\cite{Fes58,Fes62,Fes64} yields similar but more formal
expressions written in terms of the projection operators onto the
closed and the open channels. To actually implement an RMT approach
into these expressions, and to work out averages from the resulting
formulas, one has to write Feshbach's expressions explicitly in terms
of nuclear matrix elements~\cite{Lem64}. Such an approach
yields~\cite{Lew91a} formulas quite similar in structure and content
to Eqs.~(\ref{C1},\ref{C2},\ref{C3}) and (\ref{C4}). The removal of
direct reaction contributions by the matrix $U$ in Eq.~(\ref{C1}) and
the reduction of the $S$--matrix to the canonical form in
Eq.~(\ref{C2}) goes back to~\textcite{Eng73} and~\textcite{Nis85}.

Averages of observables are theoretically worked out as averages over
the ensemble of scattering matrices, i.e., over $H^{\rm GOE}$, and are
denoted by the same angular brackets as used to indicate energy
averages. The theoretical result $\langle {\cal A} \rangle$ for an
observable ${\cal A}$ is compared with the experimental running
average over energy of the same observable measured for a specific
nucleus, i.e., for a specific nuclear Hamiltonian. It is stipulated
that that nuclear Hamiltonian is a member of the GOE. Ergodicity would
then guarantee the equality of both averages, see Section~I.II.C.3.
Ergodicity holds true for ${\cal A}$ if the energy--correlation
function of ${\cal A}$ goes to zero for large values of the argument.
In the case of the scattering matrix $S^{\rm CN}$, that property is
analytically fully established for all observables formed from $S^{\rm
CN}$ only in the domain of strongly overlapping resonances. There is
no reason to doubt, however, that ergodicity applies also in the
regime of isolated and weakly overlapping resonances.

As discussed in I.II.C.2 the local spectral fluctuation measures of
the GOE do not depend on the Gaussian distribution assumed for the
matrix elements $H^{\rm GOE}_{\mu \nu}$ in Eq.~(\ref{24}) but apply
for a wide class of random--matrix ensembles. That property is
referred to as universality. The fluctuation properties of the
$S$--matrix are studied over energy intervals that are measured in
units of the mean level spacing of the resonances. The relevant
observables ($S$--matrix correlation functions) are, thus, likewise
local fluctuation measures.  The proof of universality of such
measures given by~~\textcite{Hac95} applies also to these observables.

The theoretical framework of Eqs.~(\ref{C1},\ref{C2},\ref{C3}) and
(\ref{C4}) is quite flexible and allows for extensions of the
theory. These may account for violation of isospin symmetry, of
parity, or for tests of time--reversal invariance in CN
reactions. Such applications of the statistical theory are reviewed in
Sections~\ref{isos}, \ref{pari}, and \ref{time}. Violation of isospin
symmetry or of parity is treated by replacing $H^{\rm GOE}$ in
Eq.~(\ref{C4}) by a block matrix~\cite{Ros60} as done in Eq.~(I.30)
and in Eq.~(\ref{isos1}) below. Each diagonal block refers to states
with the same isospin (or parity) quantum numbers while the
off--diagonal blocks contain the matrix elements of the
symmetry--breaking interaction, see Section~I.III.D.1. Time--reversal
invariance is broken when $H^{\rm GOE}$ is replaced by $H^{\rm GOE} +
\alpha i A$ where $\alpha$ is a real parameter and $A$ is a real and
antisymmetric random matrix, see Eqs.~(I.32) and (I.33) and
Section~\ref{time} below.

The statistical theory for $S$ formulated so far applies to every set
of fixed quantum numbers (total spin, parity) of the CN. In keeping
with Eq.~(\ref{3}) we assume that $S$--matrices referring to different
quantum numbers are statistically uncorrelated. That assumption is not
as innocent as it may look. Indeed, let us imagine that $H^{\rm eff}$
in Eq.~(\ref{C3}) is determined not from the GOE but from the shell
model. That calculation would yield the matrices $H_{\mu \nu}$ (and,
thereby, the $S$--matrices) simultaneously for all conserved quantum
numbers. A change of the residual interaction of the shell model would
cause all these matrices to change simultaneously. Different
realizations of the GOE may be thought of as corresponding to
different choices of the residual interaction. Therefore, the matrices
$H^{\rm GOE}$ appearing in $S$--matrices carrying different quantum
numbers are expected to be correlated~\cite{Mul00}. Such correlations
among Hamiltonian matrices referring to states with different quantum
numbers do indeed exist~\cite{Pap07} and were discussed in I.V.B.5.
To what extent is the assumption formulated in Eq.~(\ref{3})
invalidated by the existence of such correlations? As remarked in
Section~\ref{boh}, Eq.~(\ref{3}) implies that CN cross sections are
symmetric about 90 degrees in the c.m. system. The available
experimental evidence supports that prediction and we expect,
therefore, that deviations from Eq.~(\ref{3}) are not significant. We
return to that point in Section~\ref{corr}.

\subsection{History}
\label{earl}

The theory developed in Sections~\ref{rere} and \ref{stoc} may appear
quite natural. However, it took several decades to arrive at that
formulation.  Since the phenomenological models reviewed in
Sections~\ref{boh} and \ref{eri} led to predictions that were in good
agreement with experiment, a deeper understanding of these models was
called for from the outset. By way of justification of their
work,~\textcite{Hau52,Eri60,Bri63} themselves had referred to the
statistical properties of isolated resonances (which were then
supposed and are now known to agree with GOE predictions). But the
actual derivation of the Hauser--Feshbach formula and of Ericson
fluctuations from a GOE model for CN resonances posed a severe
challenge, especially if the aim was a comprehensive statistical
theory based on the GOE that would apply for all values of the
parameter $\Gamma / d$. That resulted in a large number of
publications in the years 1950 to 1985. Here we can give only an
outline of the main developments.

The development of the statistical theory of CN reactions depended on
the availability of a suitable theoretical framework to describe
resonance reactions. Such a framework is needed to formulate
statistical assumptions on the resonance parameters. But to a large
extent the availability of such a framework depended on the
development of nuclear--structure theory. Early attempts to formulate
statistical models mirror the development of the dynamical theory of
nuclei.

The $R$--matrix theory by~\textcite{Wig47} was formulated at a time
when virtually nothing was known about nuclear structure and is, by
necessity, a very formal theory: Except for the short range of the
nucleon--nucleon interaction, it does not refer to any specific
feature of the nuclear Hamiltonian. Resonances are constructed as
follows. The nucleus is thought to be enclosed by fictitious
boundaries (one for each two--body fragmentation). On these boundaries
fictitious boundary conditions are imposed. As a result, the spectrum
of the nuclear Hamiltonian in the ``internal region'' (the domain
enclosed by the boundaries) is discrete. The eigenvalues and
eigenfunctions depend on numerous parameters (distance of the
boundaries from the center of mass, values of the boundary
conditions). Green's theorem is used to connect these discrete levels
with the channels, and the levels become CN resonances. Even today the
resulting form of the scattering matrix is extremely useful for the
analysis of experimental data containing several partly overlapping
resonances (``multi--level $R$--matrix fit''). However, few--level
approximations to the $S$--matrix are not automatically unitary.
Moreover, the form of the $S$--matrix is rather unwieldy and depends
explicitly on the parameters just mentioned. For these reasons, the
$R$--matrix has never been used as the starting point of a statistical
approach.

\textcite{Lan57} calculated the average cross section for neutron
capture reactions in the regime of isolated resonances. The
calculation was possible without referring to a comprehensive theory
of resonance reactions. Indeed, in the regime $\Gamma \ll d$ each
resonance is independently described by a Breit--Wigner formula as in
Eqs.~(\ref{13}) or (\ref{19a}). If one assumes that the distribution
of resonance parameters of isolated resonances follows the GOE, the
partial width amplitudes in each channel are Gaussian--distributed
random variables. For each resonance the total width $\Gamma_\mu$ is
related to the partial widths $\Gamma_{a \mu}$ by the sum
rule~(\ref{15}). The resonance energies do not enter the calculation
of the energy--averaged cross section. The task consists in
calculating the ensemble average of $\Gamma_{a \mu} \Gamma_{b \mu} /
\Gamma$ where $a$ and $b$ are two specific channels (here: the neutron
and the gamma decay channel). This can be done using the GOE and as
input values for the averages of the partial widths, the strength
functions $\langle \Gamma_{a \mu} \rangle / d$, see Section~\ref{iso}.
A correction factor deduced from $R$--matrix theory accounted for the
effect of weak resonance overlap. \textcite{Mol61} extended the result
of Lane and Lynn perturbatively to weakly overlapping resonances.

The first attempt to construct a comprehensive statistical theory is
due to~\textcite{Mol61,Mol63,Mol64,Mol69,Mol75,Mol76,Mol80} (see also
references to further work therein). In order to go beyond the regime
of isolated resonances, Moldauer used an expansion of the scattering
matrix in terms of its poles in the complex energy plane. This form of
the $S$--matrix had been proposed by~\textcite{Hum61}. In contrast to
the $R$--matrix theory, the Humblet--Rosenfeld theory is a dynamical
theory: The parameters of the pole expansion are, in principle,
completely determined by the Hamiltonian, and there are no arbitrary
parameters. Unfortunately an explicit analytical connection between
the Hamiltonian and the positions of the poles and the values of the
residues, is not known. Thus, the Humblet--Rosenfeld theory is
effectively a formal theory like the $R$--matrix theory. Moreover, the
following difficulty arises. For isolated resonances there is a
one--to--one correspondence between a resonance and a pole of the
scattering matrix, see Eq.~(\ref{13}). The parameters of the pole
possess a direct physical interpretation as partial width amplitudes
and as energy and total width of the resonance. In the general case of
many overlapping resonances the Humblet--Rosenfeld expansion uses the
same parameters (residues and locations of the poles of the
$S$--matrix plus a smooth background matrix). But the simple
interpretation valid for isolated resonances does not apply. Moreover,
the theory is not manifestly unitary.  For isolated resonances the
unitarity constraint leads to the simple sum rule~(\ref{15}), while it
imposes complicated relations between the pole parameters for
overlapping resonances. These have never been untangled. Therefore,
the choice of statistical assumptions for the pole parameters was far
from obvious. At some point, Moldauer used the $K$--matrix
formulation~(\ref{22b}) of the $S$--matrix, assumed that the energies
$E_\mu$ and matrix elements $W_{a \mu}$ in Eq.~(\ref{22a}) obeyed GOE
statistics, and determined the distribution of pole parameters via a
numerical simulation. His relentless efforts met with limited success
but kept interest in the problem. Some of his results are reviewed in
Section~\ref{pole}.

Feshbach's unified theory of nuclear
reactions~\cite{Fes58,Fes62,Fes64} was the first theory of CN
reactions that expressed the scattering matrix in terms of the
Hamiltonian of the system without the help of arbitrary parameters. It
uses projection operators onto the spaces of open and of closed
channels. The projection of the nuclear Hamiltonian onto the space of
closed channels defines a self--adjoint operator with a discrete
spectrum. The bound states of that operator generate the resonances in
the full problem. The theory was used by~\textcite{Lem64} for a first
calculation of elastic neutron scattering on $^{15}$N using the
nuclear shell model. Quasibound shell--model states appeared as
neutron resonances. Although not related to the statistical theory,
that work demonstrated the possibility to account for CN resonances in
terms of a dynamical approach using the nuclear shell model. Later
work by the MIT group~\cite{Kaw73} used Feshbach's theory to formulate
an extension of the Hauser--Feshbach formula that accounts for the
presence of direct reactions, see Section~\ref{incl}. The approach
uses plausible assumptions but is not based upon a random--matrix
approach.

The general theory of nuclear resonance reactions developed in the
1960s by~\textcite{Mah69} was based on the nuclear shell model and
work by~\textcite{Dir58} and \textcite{Fan61}. It was more explicit
than Feshbach's theory of nuclear reactions. That was a natural
consequence of the fact that nuclear--structure theory had made
significant progress since the early 1950s. Resonances were shown to
be mainly due to quasibound states of the shell model. The approach
expresses the scattering matrix in terms of the Hamiltonian governing
the dynamics of the quasibound states as in Eq.~(\ref{20}). That was
the starting point of later developments. The theory is connected with
the $R$--matrix theory of~\textcite{Wig47} through relations of the
form of Eq.~(\ref{22b}).

Independently of that development,~\textcite{Eng73} showed that there
exists a unitary transformation which diagonalizes $\langle S
\rangle$. The transformation acts on channel space and, therefore,
leaves the statistical properties of the resonances
unchanged~\cite{Hof75b}. As a consequence, the transformation reduces
the problem of calculating CN cross sections in the presence of direct
reactions to the problem without direct reactions (i.e., for a
diagonal $\langle S \rangle$). That was an important simplification,
see Eq.~(\ref{C1}).

The conspicuous lack of a comprehensive theory of CN reactions with
predictive power motivated~\textcite{Tep74}
and~\textcite{Hof75a,Hof75b} to perform numerical simulations with the
aim of establishing fit formulas for average CN cross sections valid
for all values of $\Gamma / d$. The authors used the transformation
mentioned in the previous paragraph and focused attention on $S^{\rm
CN}$. They proved that aside from overall phase factors, the
distribution of $S$--matrix elements depends only on the transmission
coefficients. That fact simplified the construction of the fit
formulas. For the numerical simulations, the authors used the
$K$--matrix form~(\ref{22b}) of the stochastic scattering
matrix~(\ref{C2}). The $K$--matrix parameters were determined by the
eigenvalues and eigenfunctions of GOE matrices. Repeated random
drawings from a Gaussian distribution of the elements of GOE matrices
led to statistically meaningful averages and determined the parameters
in the fit formulas, see Section~\ref{fit}. Similar formulas were
subsequently also developed by~\textcite{Mol75}.

The first use of the explicit dependence of Eq.~(\ref{21}) on the
Hamiltonian $H$ in the calculation of CN processes was made
by~\textcite{Aga75}. CN scattering theory was extended so as to
include precompound reactions.  The authors used a stochastic model
for the matrix $H$ in Eq.~(\ref{21}) which allowed for the existence
of classes of shell--model states of increasing complexity. Averages
of cross sections were calculated in terms of an asymptotic expansion
valid for $\Gamma \gg d$. The Ericson regime was obtained as a special
case of very strong coupling between classes so that the internal
equilibration time of the CN becomes small compared to the decay time
$\hbar / \Gamma$. As a result, the Hauser--Feshbach formula and
predictions of the Ericson model were for the first time derived from
a microscopic statistical theory. That was possible because resonance
spacings were assumed to be constant (i.e., not to follow GOE
predictions). Later work showed that for $\Gamma \gg d$ that
simplification led to correct results.

In 1984, a connection was established~\cite{Ver84a} between
statistical nuclear theory and quantum field theory. More precisely,
field--theoretical concepts used in condensed-matter theory and in
statistical mechanics were applied to the RMT description of both,
nuclear spectra and nuclear reactions. In a very concrete sense, RMT
as applied to nuclei became part of the statistical mechanics of
many--body systems. Specifically, generating functionals familiar from
condensed--matter theory were used to describe fluctuations both of
nuclear spectra and of $S$--matrix elements in a common framework. In
the limit $N \gg 1$, the evaluation of the generating functionals
permitted a clear separation of average properties of observables and
of their fluctuations. The stochastic scattering matrix was defined in
the form given by Eq.~(\ref{C2}) and was written as a suitable
derivative of a generating functional $Z$. As a first application, the
$S$--matrix correlation function was calculated with the help of the
replica trick~\cite{Wei84}. That yields an asymptotic expansion in
powers of $d / \Gamma$, see Section~\ref{ericson}. Later, $Z$ was
expressed in terms of Efetov's supersymmetry
approach~\cite{Efe83}. That led~\cite{Ver84b,Ver85a} to the exact
expression for the $S$--matrix correlation function given in
Section~\ref{general}.  Eqs.~(\ref{C1},\ref{C2},\ref{C3}) and
(\ref{C4}) and the supersymmetry approach have since been established
as the main tools to study chaotic scattering based on RMT.

\section{Average $S$--Matrix}
\label{aver}

The average $S$--matrix serves as input for the statistical theory of
nuclear reactions. In the present Section we display central
properties of $\langle S \rangle$ and review ways to determine
$\langle S \rangle$ phenomenologically.

\subsection{Calculation of $\langle S^{\rm CN}(E) \rangle$}
\label{calc}

We focus attention on the only stochastic element in the scattering
matrix, i.e., on the CN part $\langle S^{\rm CN} \rangle$, see
Eq.~(\ref{C2}).  The calculation of averages of $S^{\rm CN}$ (and of
powers of $S^{\rm CN}$) is simpler than that of terms involving both
$S^{\rm CN}$ and $(S^{\rm CN})^*$ because as shown in Eq.~(\ref{32})
all poles of $S^{\rm CN}(E)$ lie below the real $E$--axis.

We calculate $\langle S^{CN}(E) \rangle$ by replacing the ensemble
average by a running average over energy with a Lorentzian weight
function of width $I$ centered at energy $E_0$~\cite{Bro59}. We choose
the center of the GOE spectrum $E_0 = 0$ as the center of the weight
function and have
\be
\langle S^{\rm CN}(E) \rangle = \frac{1}{\pi} \int_{-\infty}^{+
\infty} {\rm d} E \ S^{\rm CN}(E) \frac{I}{E^2 + (1/4) I^2} \ .
\label{33}
\ee
The average in Eq.~(\ref{33}) extends over very many resonances so
that $I \gg d$ with $d$ the average level spacing in the center of the
GOE spectrum. The integration contour can be closed in the upper half
of the energy plane. We obtain
\be
\langle S^{\rm CN}(E) \rangle = S^{\rm CN}(i I) \ .
\label{34}
\ee
We use Eqs.~(\ref{27g}) and (\ref{27h}) and expand $S^{\rm CN}(i I)$
in a Born series with respect to the imaginary part of $H^{\rm eff}$,
\be
\langle S^{\rm CN}(E) \rangle_{a b} = \delta_{a b} + 2
\sum_{n = 1}^\infty [ (- i \pi {\cal W})^n ]_{a b} \ .
\label{35}
\ee
Here ${\cal W}$ is a matrix in channel space given by
\be
{\cal W}_{a b} = \sum_\mu \tilde{W}_{a \mu} [ i I - E_\mu ]^{-1}
\tilde{W}_{\mu b} \ .
\label{36}
\ee
Since $I \gg d$, it is legitimate to neglect the fluctuations of
the $E_\mu$ in the denominator in Eq.~(\ref{36}), i.e., to assume for
the $E_\mu$ a picket--fence model with fixed nearest--neighbor spacing
$d$. The sum over $\mu$ amounts to averaging the product $\tilde{W}_{a
\mu} \tilde{W}_{\mu b}$. We have $(1/N) \sum_\mu \tilde{W}_{a \mu}
\tilde{W}_{\mu b} = (1/N) \sum W_{a \mu} W_{\mu b}$. We use
Eq.~(\ref{29}), change the summation into an energy integration, and
find
\be
{\cal W}_{a b} = \frac{- i \pi v^2_a}{d} \delta_{a b} .
\label{37}
\ee
Inserting the result into Eq.~(\ref{35}) and using the
definition~(\ref{29a}), we find
\be
\langle S^{\rm CN}_{a b} \rangle = \frac{1 - x_a}{1 + x_a}
\delta_{a b} \ .
\label{38}
\ee
Eq.~(\ref{38}) relates the average scattering matrix to the parameters
$x_a$ of the statistical theory. The result~(\ref{38}) is also
obtained in the limit $N \to \infty$ via the replica trick (see
Section~\ref{replica}) and via the supersymmetry approach (see
Section~\ref{general}). The average $S$--matrix given by
Eq.~(\ref{38}) is real because we have chosen $E$ in the center of the
GOE spectrum, $E = 0$. For a different choice of $E$, $\langle S^{\rm
CN}_{a a} \rangle$ would not be real. The additional phase caused by
such a choice is not part of the matrix $U$ constructed in
Section~\ref{stoc}.

Eq.~(\ref{38}) shows that $\langle S^{\rm CN} \rangle$ is always
diagonal. But that does not imply that $\langle S \rangle$ is
diagonal, too. In fact, Eq.~(\ref{C1}) demonstrates that $\langle S
\rangle$ is diagonal only if the matrix $U$ is diagonal. That matrix
embodies the effect of couplings between channels. That is why a
non--diagonal form of $\langle S \rangle$ is taken to be synonymous
with the presence of ``direct reactions'' (i.e., reactions that
proceed without intermediate CN formation).

The average of a product of $S$--matrix elements can be worked out in
the same way. Let $\{a_i, b_i\}$ with $i = 1, \ldots, k$ denote $k$
arbitrary pairs of channels, and consider the average of the product
of the matrix elements $S^{\rm CN}_{a_i b_i}(E_i)$ taken at energies
$E_1, E_2, \ldots, E_k$. Again using a Lorentzian averaging function,
closing the contour in the upper half of the complex energy plane, and
using Eq.~(\ref{34}) we find
\be
\langle \prod_i S^{\rm CN}_{a_i b_i}(E_i) \rangle = \prod_i \langle
S^{\rm CN}_{a_i b_i}(E_i) \rangle \ .
\label{39}
\ee
More generally, we have for any function $f ( S^{\rm CN} )$ that is
analytic in the upper half of the complex energy plane the relation
\be
\langle f ( S^{\rm CN} ) \rangle = f ( \langle S^{\rm CN} \rangle )
\ . 
\label{39a}
\ee
This is an important result. It shows that $S$--matrix elements taken
at the same or at different energies are uncorrelated. The
result~(\ref{39}) does not hold for products involving both $S^{\rm
CN}$ and $(S^{\rm CN})^*$ since these have poles on both sides of the
real energy axis, and closing the integration contour does not yield a
simple expression. The calculation of such terms poses the main
technical difficulty in the statistical theory.

\subsection{Physical Interpretation of $\langle S \rangle$}
\label{phys}

The decomposition~(\ref{1}) of the scattering matrix into an average
part and a fluctuating part applies likewise to the CN part $S^{\rm
CN}$ of $S$. The discussion in Section~\ref{beyo} of the physical
significance of that decomposition in terms of time scales for the CN
reactions then applies to $S^{\rm CN}$, too, as do the limitations of
the statistical theory established there.

Eq.~(\ref{38}) shows that the average CN $S$--matrix is subunitary,
$|\langle S^{\rm CN}_{a a} \rangle| \leq 1$. The equality sign holds
only when $x_a$ vanishes, i.e., when there is no coupling between
channel $a$ and the $N$ resonances. We emphasize that the lack of
unitarity of $\langle S^{\rm CN} \rangle$ is due to taking the average
over $N$ resonances (and, since $\langle S^{\rm CN} \rangle$ is
diagonal, not to inelastic scattering processes that would deplete the
elastic channel). To see how averaging reduces the magnitude of
$S^{\rm CN}$ we consider the simplest case of a single channel coupled
to $N$ resonances. There are no inelastic processes by definition. The
scattering amplitude has the form $S^{\rm CN}(E) = \exp ( 2 i
\delta(E) )$ and can be viewed as a point on the unit circle of the
complex plane. As $E$ increases, $\delta(E)$ increases by $\pi$ over
the width of every resonance, and $S^{\rm CN}(E)$ moves
counter--clockwise once around the origin on the unit circle. The
average of $S^{\rm CN}(E)$ over $N$ resonances must then lie in the
interior of the unit circle, i.e., be subunitary. These points are
beautifully illuminated in~\cite{Fri55}.

It is remarkable that $\langle S^{\rm CN}_{a a} \rangle$ as given in
Eq.~(\ref{38}) depends only on the coupling coefficient $x_a$ and not
on the other coupling coefficients $x_b$ with $b \neq a$. That shows
that $\langle S^{\rm CN}_{a a} \rangle$ taken all by itself describes
the loss of probability amplitude in channel $a$ due to CN formation.
Information on the manner in which the CN eventually decays back into
the various channels, is not contained in $\langle S^{\rm CN}_{a a}
\rangle$. The information as to where the lost probability eventually
reappears (in quantum mechanics, probability is conserved) is supplied
by the fluctuating part of $S^{\rm CN}$. Thus, CN scattering theory
can be viewed as a special case of quantum transport
theory~\cite{Aga75}.

In Eq.~(\ref{6}), the transmission coefficients were defined under the
assumption that $\langle S \rangle$ is diagonal. That assumption is
met by $\langle S^{\rm CN} \rangle$. We calculate $T_a$ from
Eq.~(\ref{6}) using Eq.~(\ref{38}) for $\langle S^{\rm CN} \rangle$
and find
\be
T_a = \frac{4 x_a}{(1 + x_a)^2} \ .
\label{40}
\ee
(For a non--diagonal $\langle S \rangle$ the connection with the
transmission coefficients is given in Section~\ref{satc}). The
form~(\ref{40}) for the transmission coefficients applies when we
choose the energy $E$ in the center of the GOE spectrum, $E = 0$. For
a different choice of $E$, the form of $T_a$ differs. That difference
is irrelevant in practice because the transmission coefficients serve as
phenomenological input parameters of the theory anyway.

In the theory of resonance reactions formulated in Sections~\ref{reso}
and \ref{stoc} the coupling between levels and channels seems to
increase monotonically with increasing strength of the coupling matrix
elements $W_{a \mu}$ or, equivalently, of the parameters $x_a$ defined
in Eq.~(\ref{29a}). However, Eq.~(\ref{40}) shows that this is not the
case. With increasing $x_a$, the transmission coefficient $T_a$
increases monotonically until it reaches its maximum value $T_a = 1$
at $x_a = 1$. Thereafter $T_a$ decreases monotonically with increasing
$x_a$ and tends towards zero as $x_a \to \infty$. Similarly,
Eq.~(\ref{38}) shows that $\langle S^{\rm CN}_{a a} \rangle$ decreases
from unity to zero as $x_a$ approaches unity from zero. As $x_a$
increases further, $\langle S^{\rm CN}_{a a} \rangle$ approaches minus
one. We also note that under the substitution $x_a \to 1 / x_a$, $T_a$
is invariant while $\langle S^{\rm CN}_{a a} \rangle$ changes sign,
$\langle S^{\rm CN}_{a a} \rangle(1 / x_a) = - \langle S^{\rm CN}_{a
a} \rangle(x_a)$.  All this shows that there is a maximum value of the
coupling strength, $x_a = 1$, beyond which a further increase of $x_a$
effectively reduces the coupling between channel $a$ and the
resonances.

To make that fact physically plausible we consider the matrix $D$ in
Eq.~(\ref{C3}). We follow the work
of~\textcite{Sok88,Sok89,Sok92,Sok97}. For large values of the $x_a$
(all $a$), the width matrix $\Gamma_{\mu \nu}$ (Eq.~(\ref{29d}))
dominates $H^{\rm GOE}$ in $H^{\rm eff}$, see Eq.~(\ref{C4}).
Therefore, one chooses a basis in which $\Gamma_{\mu \nu}$ is
diagonal.  Eqs.~(\ref{29}) show that $\Gamma_{\mu \nu}$ possesses
$\Lambda$ non--zero eigenvalues $2 \pi N v^2_a$ with $a = 1, \ldots,
\Lambda$. The associated orthonormal eigenvectors are given by $(1 /
\sqrt{N v^2_a}) W_{a \mu}$. Completing in some arbitrary fashion these
$\Lambda$ eigenvectors to a set of $N$ orthonormal vectors, denoting
the resulting orthogonal matrix by $G$ and transforming the matrix $D
\to \tilde{D} = G D G^T$, we obtain
\be
\tilde{D}_{\mu \nu} = (E + i \pi N v^2_\mu) \delta_{\mu \nu} -
\tilde{H}_{\mu \nu} 
\label{41}
\ee where $\tilde{H} = G H^{\rm GOE} G^T$ and where the first
$\Lambda$ diagonal entries $\pi N v^2_\mu$ are given in terms of the
$\Lambda$ non--vanishing eigenvalues $2 \pi N v^2_a$ with $a = 1,
\ldots, \Lambda$ of the width matrix $\Gamma_{\mu \nu}$ while $v^2_\mu
= 0$ for $\mu > \Lambda$. Concerning these first $\Lambda$ resonances,
we consider $\tilde{H}$ as a small perturbation. That is justified as
long as the decay widths $2 \pi N v^2_\mu$ of the resonant states $\mu
= 1, \ldots, \Lambda$ are large compared with their spreading widths
$2 \pi \langle [ \tilde{H}_{\mu \nu} ]^2 \rangle / d $ due to mixing
with the other resonances. (We recall that the concept of the
spreading width was introduced in the context of doorway states in
Section~I.II.G. The entire discussion there applies likewise to the
present case). With $d = \pi \lambda / N$ and with the help of
Eq.~(\ref{24}) that implies $v^2_a \gg d$ or $x_a \gg 1$ for all $a$.
In zeroth order in $\tilde{H}$ we then deal with $\Lambda$ poles
located at $- i \pi N v^2_\mu$, $\mu = 1, \ldots, \Lambda$ and with $N
- \Lambda$ poles located on the real axis. Taking into account
$\tilde{H}$ we obtain a ``cloud'' of $N - \Lambda$ poles located below
but close to the real axis and $\Lambda$ poles very far from that
axis. Each of the latter contributes only a smooth phase to the
$S$--matrix (smooth over a typical distance $d$). The additional phase
causes the minus sign in the relation $\langle S^{\rm CN}_{a a}
\rangle(1 / x_a) = - \langle S^{\rm CN}_{a a} \rangle(x_a)$.  Aside
from that phase, the $\Lambda$ far--away poles do not affect the CN
scattering process. On the other hand, CN scattering due to the cloud
of $N - \Lambda$ narrow resonances is, for $\Lambda$ fixed and $N \to
\infty$, indistinguishable from that of $N$ narrow CN resonances. That
is why, aside from overall phase factors, all moments and correlation
functions of $S^{\rm CN}$ do not depend on the phase shifts due to the
$\Lambda$ far--away poles and can be expressed in terms of the
transmission coefficients $T_a$. The picture of the distribution of
poles that emerges from this qualitative discussion is quantitatively
confirmed in Section~\ref{pole}: As all the coefficients $x_a$
increase monotonically from zero, the $N$ poles of $S$ move away from
the real energy axis into the complex plane. As the $x_a$ increase
beyond $x_a = 1$ (all $a$), the ``cloud'' of $N$ poles begins to
separate into a small cloud containing $\Lambda$ poles that move ever
further away from the real axis, and a big cloud of $N - \Lambda$
poles that moves back toward the real axis, see Fig.~\ref{fig:HW9}.

According to Eq.~(\ref{6}), the transmission coefficients measure the
unitarity deficit of the average $S$--matrix. As mentioned in
Section~\ref{boh}, they also measure the probability of CN formation.
It is now clear why the $T_a$s are better suited for that purpose than
the coupling coefficients $x_a$.

\subsection{Satchler's Transmission Matrix}
\label{satc}

In Section~\ref{phys} we have used $S^{\rm CN}$ (which is diagonal on
average) to express $\langle S^{\rm CN}_{a a} \rangle$ and $T_a$ in
terms of the coupling coefficients $x_a$, see eqs.~(\ref{38}) and
(\ref{40}). We now address the question how to determine the
transmission coefficients if $\langle S \rangle$ is not diagonal. That
is the generic case. In principle, the question can simply be answered
with the help of Eq.~(\ref{C1}): To find the $T_a$s, take the matrix
$S^{\rm CN} = U^\dagger S U^*$. By definition, the average of $S^{\rm
CN}$ is diagonal, and Eq.~(\ref{6}) yields the transmission
coefficients. But historically the question came up before the
form~(\ref{C1}) with $U$ defined in Eq.~(\ref{30}) was known. It
arose in conjunction with the experimental discovery of ``direct
reactions'' mentioned in Section~\ref{incl}, and with the theoretical
treatment of such reactions which are described by non--diagonal
average scattering matrices, see Section~\ref{opt}.

A general measure of the unitarity deficit of $\langle S \rangle$ is
given by Satchler's transmission matrix $P$~\cite{Sat63}, a matrix in
channel space defined in Eq.~(\ref{12e}). Since $P$ is Hermitean, it
can be diagonalized by a unitary matrix $\tilde{U}$. We denote the
eigenvalues by $p_a$ with $a = 1, \ldots, \Lambda$ and have
\be
(\tilde{U}^\dagger P \tilde{U})_{a b} = \delta_{a b} p_a \ {\rm where}
\ 0 \leq p_a \leq 1 \ .
\label{43}
\ee
Applying the transformation $\tilde{U}$ to the right--hand side of
Eq.~(\ref{12e}) and using $\tilde{U}^* \tilde{U}^T = {\bf 1}_\Lambda$
we are led to consider the matrices $A = \tilde{U}^\dagger \langle S
\rangle \tilde{U}^*$ and $A^* = \tilde{U}^T \langle S \rangle
\tilde{U}$.  Eq.~(\ref{43}) implies that $(A A^*)_{a b} = \delta_{a b}
(1 - p_a) = (A^*A)_{a b}$ which in turn shows that the symmetric
matrices $A$ and $A^*$ are normal and can be diagonalized
simultaneously~\cite{Eng73}.  The matrix product $A A^*$ being
diagonal already, we conclude that $A$ and $A^*$ are diagonal, too
(except for accidental degeneracies of the $p_a$ which we do not
consider). Thus,
\be
(\tilde{U}^\dagger \langle S \rangle \tilde{U}^*)_{a b} = \delta_{a b}
\sqrt{1 - p_a} \exp( 2 i \phi_a ) 
\label{44}
\ee
where the $\phi_a$s are real. Defining $U_{a b} = \tilde{U}_{a b} \exp
( i \phi_b )$ we conclude that the average of the matrix $U^\dagger S
U^*$ is real and diagonal, $\langle U^\dagger S U^* \rangle_{a b} =
\delta_{a b} \sqrt{1 - p_a}$. That shows that the transmission
coefficients are given by the eigenvalues $p_a$ of Satchler's
transmission matrix, $T_a = p_a$, $a = 1, \ldots, \Lambda$. Moreover,
the function $F_{\mu \nu}$ in Eq.~(\ref{28}) is invariant under an
arbitrary unitary transformation in the space of channels. Therefore,
the matrix $S^{\rm CN} = U^\dagger S U^*$ is identical with the
scattering matrix of pure CN scattering in the absence of direct
reactions. In other words, the matrix $U$ reduces the CN scattering
problem in the presence of direct reactions to the pure CN scattering
problem without direct reactions.

The transformation $U$ introduced by~\textcite{Eng73} and defined in
terms of Satchler's transmission matrix is a purely formal device.
\textcite{Nis85} have shown how $U$ is defined physically. We do not
retrace the steps of the argument here because we have used it to
construct the matrix $U$ in Eq.~(\ref{30}). The present construction
of the matrix $U$ via Satchler's transmission matrix yields exactly
the same matrix $U$ as in Eq.~(\ref{30}) if the energy $E$ is chosen
in the center of the GOE spectrum, so that $E = 0$. Otherwise, the
matrix $U$ obtained from Eqs.~(\ref{43}) and (\ref{44}) contains an
additional phase (the phase of the average $S$--matrix, see the remark
below Eq.~(\ref{38})).

\subsection{Optical Model and Strength Function. Direct Reactions}
\label{opt}

The average $S$--matrix serves as input for the statistical theory and
must be determined phenomenologically. It is intuitively clear that
the fast part $\langle S \rangle$ of the reaction amplitude $S$ can
involve only few degrees of freedom. Therefore, models that do not take
into account the full complexity of the many--body problem suffice for
an accurate determination of $\langle S \rangle$. These are the
optical model of elastic scattering and the coupled--channels
approach. The latter describes fast inelastic processes that are
referred to as direct reactions. We do not review these models in
detail but confine ourselves to those aspects that are essential for
the understanding of the statistical theory.

In the optical model~\cite{Hod63}, elastic scattering of nucleons is
described in terms of a radial Schr{\"o}dinger equation with a central
potential $V(r) + i W(r)$. The bulk of the real part $V(r)$ is due to
the shell--model potential. (The shell model is the subject of
Section~I.IV.A). The imaginary part $W(r)$ describes CN formation by
nucleon absorption. Via a dispersion relation, the imaginary part
$W(r)$ also modifies the real (shell--model) part $V(r)$. Because of
the presence of $W(r)$, the scattering amplitude calculated from the
optical model is subunitary and is identified with the average
scattering amplitude in the nucleon channel of the statistical theory.
Obvious generalizations apply to the scattering of composite
particles. Fig.~\ref{fig:HW5} shows results of an optical--model
calculation.

As mentioned in Section~\ref{earl}, the CN was originally considered a
black box. The novel aspect of the optical model introduced
by~\textcite{Fes54} was the partial transparency of the target
nucleus. That modified Bohr's picture of the CN as a system of
strongly interacting particles: Nucleons in the nucleus had a finite
mean free path. The nuclear shell model had indicated such behavior
already for the ground state and low--lying excited states. Now that
feature was extended to resonances (states above neutron threshold).
The influence of shell structure on the CN cross section is manifest
in the neutron strength function. In analogy to the spreading width
introduced in Section~I.II.G, the neutron $s$--wave strength function
$s(E)$ is defined as $s(E) = 2 \pi \langle \Gamma_{\mu a} \rangle /
d$. Here $\Gamma_{\mu a}$ is the partial width for $s$--wave neutron
emission of resonance $\mu$. With $\Gamma_{\mu a} = 2 \pi W^2_{\mu a}$
and Eq.~(\ref{29a}) that yields $s(E) = 4 x_a$ and, for $x_a \ll 1$,
$s(E) = T_a$, see Eq.~(\ref{40}).  The strength function $s(E)$
measures the intensity with which a group of resonances is coupled to
the $s$--wave neutron channel. The strength function can both be
measured (by performing a running average over a number of resonances)
and be calculated directly from the optical model. We recall that the
radius of the real part of the optical--model potential (i.e.,
essentially the shell--model potential) increases with $A$. The
strength function displays maxima versus $A$ whenever a
single--particle $s$--wave state is about to be pulled into the
potential well.

The phenomenological description of direct reactions is based upon an
extension of the optical model for elastic scattering. For a set of
channels, the radial equations describing elastic scattering with the
help of the optical model, are coupled. The term which couples two
channels is obtained as the matrix element of the interaction between
projectile and target sandwiched between the two channel wave
functions, see the reviews by~\textcite{Kon03,Aus70,Gle04}. Direct
reactions are most important for strongly coupled channels. Often
these are channels where the target nucleus is in different states of
collective excitation. Nucleon transfer between projectile and target
may also lead to strong interchannel coupling. In many cases, the
solution of the coupled--channels problem can be simplified by using
the Born approximation. The result is the distorted--wave Born
approximation: The plane waves in entrance and exit channel are
distorted by the optical potentials, the transition between both
channels is calculated perturbatively to first order.

Direct reactions induce correlations between partial--width amplitudes
relating to different channels~\cite{Huf67,Kaw73}. In the present
context, that is seen as follows. Eq.~(\ref{27f}) shows that for $a
\neq b$, the partial--width amplitudes $\sqrt{2 \pi} \tilde{W}_{\mu
a}$ and $\sqrt{2 \pi} \tilde{W}_{\mu b}$ are uncorrelated random
variables. But in the presence of direct reactions, the matrix $U$ in
Eq.~(\ref{C1}) is not diagonal, and the partial--width amplitudes
$\gamma_{\mu a}$ of the full $S$ matrix in channel $a$ are given by
$\sum_b U_{a b} \sqrt{2 \pi} \tilde{W}_{\mu b}$. For two channels $a
\neq b$ connected by a direct reaction these are, in general,
correlated. Indeed, using Eq.~(\ref{27f}) we find that $\langle
\gamma_{\mu a} \gamma^*_{\mu b} \rangle = 2 \pi \sum_c U_{a c} U^*_{b
c} v^2_c \neq 0$.

\section{Results of the Statistical Theory} 
\label{resu}

According to Eqs.~(\ref{C1},\ref{C2},\ref{C3}), $S(E)$ is a
matrix--valued random process. The goal of the statistical theory
consists in finding the joint probability distribution of $S(E)$ and
$S^*(E)$. Because of Eq.~(\ref{C1}) it would actually suffice to
determine the joint probability distribution of $S^{\rm CN}(E)$ and
$(S^{\rm CN}(E))^*$. That distribution should be given in terms of the
transmission coefficients $T_c$, of energy differences in units of the
mean GOE level spacing $d$, and overall phase factors. We are far from
that goal because averaging over the $N(N+1)/2$ random variables of
$H^{\rm GOE}$ that appear in the denominator of $S^{\rm CN}$ (see
Eqs.~(\ref{C2},\ref{C3}) and (\ref{C4})) turns out to be extremely
difficult in general. We possess only partial information, mainly on
low moments and correlation functions of $S^{\rm CN}(E)$ and $(S^{\rm
CN}(E))^*$. That information suffices for a comparison with the
available experimental data and is now reviewed, together with the
various methods that have been used to calculate the answers. We do
not go into the full complexity of some of the calculations as these
are technically quite demanding, and refer the reader to the original
literature.

In the calculation of moments and correlation functions, a
simplification arises because the coupling matrix elements obey
Eq.~(\ref{29}). It follows that all moments and correlation functions of
$S^{\rm CN}(E)$ and $(S^{\rm CN}(E))^*$ vanish unless the channel
indices are pairwise equal. For instance, using the unitary
transformation of Eq.~(\ref{C1}) to calculate $\langle |S_{a b}|^2
\rangle$, we find that only terms of the form $\langle
|S^{\rm CN}_{c d}|^2 \rangle$ and $\langle S^{\rm CN}_{c c}
(S^{\rm CN}_{c' c'})^* \rangle$ give non--vanishing contributions.

\subsection{Isolated Resonances ($\Gamma \ll d$)}
\label{iso}

Isolated resonances occur in two limiting cases, $x_a \ll 1$ and $x_a
\gg 1$ for all $a$. The two cases differ only in the phases of the
average $S$--matrix elements and lead to identical expressions for the
CN cross section. Therefore, we consider only the case $x_a \ll 1$ for
all $a$. We use the form~(\ref{27g},\ref{27h}) of $S^{\rm CN}$ and the
statistical properties listed above that equation. We replace the
ensemble average by an energy average and use contour integration.
That is possible for $\Gamma \ll d$ because the resonances are
isolated. We define $\Gamma_{\mu a} = 2 \pi \tilde{W}^2_{\mu a}$ and
$\Gamma_\mu = \sum_c \Gamma_{\mu c}$. The partial widths $\Gamma_{\mu
a}$ are squares of Gaussian--distributed random variables and follow
the Porter--Thomas distribution given in Eq.~(I.18). According to
Eq.~(I.17) that distribution depends on a single parameter, the
average width $\langle \Gamma_{\mu a} \rangle$.  We
find~\cite{Bet37,Lan57} and~\cite{Mol61} to~\cite{Mol75}
\be
\langle |S^{\rm (CN \ fl)}_{a b}|^2 \rangle = \frac{2 \pi}{d}
\bigg\langle \frac{\Gamma_{\mu a} \Gamma_{\mu b}}{\Gamma_\mu}
\bigg\rangle \ .
\label{45}
\ee
The average on the right--hand side is over the Porter--Thomas
distribution of the $\Gamma_{\mu a}$s. Since $\Gamma_\mu = \sum_a
\Gamma_{\mu a}$, the variables in numerator and denominator are not
statistically independent. We use that for $x_a \ll 1$ we have $T_a
\approx 4 x_a = (2 \pi / d) \langle \Gamma_{\mu a} \rangle$. Thus, it
is possible to express the right--hand side of Eq.~(\ref{45})
(including the Porter--Thomas distribution of the $\Gamma_{\mu a}$)
completely in terms of the transmission coefficients $T_a$, as
required by the statistical theory. We write Eq.~(\ref{45}) in the
form
\be
\langle |S^{\rm (CN \ fl)}_{a b}|^2 \rangle = \frac{T_a T_b}{\sum_c
T_c} {\bf W}_{a b} \ .
\label{46}
\ee
Here ${\bf W}_{a b}$ is referred to as the ``width fluctuation
correction'' (to the Hauser--Feshbach formula) and is defined as
\be
{\bf W}_{a b} = \bigg\langle \frac{\Gamma_{\mu a} \Gamma_{\mu b}}
{\Gamma_\mu} \bigg\rangle \frac{\langle \Gamma_\mu \rangle}{\langle
\Gamma_{\mu a} \rangle \langle \Gamma_{\mu b} \rangle} \ .
\label{47}
\ee
Eq.~(\ref{47}) shows that for the single--channel case we have ${\bf
W} = 1$. For the general case (several open channels) ${\bf W}_{a b}$
has been calculated numerically by~\textcite{Lan57} and especially
by~\textcite{Ref76}. The values range from one to three. This shows
that for isolated resonances, the Hauser--Feshbach formula is only
approximately valid. For the special case of a large number of
channels with very small transmission coefficients each (so that
$\Lambda \gg 1$ but $\sum_c T_c \ll 1$) numerator and denominator in
the first term on the right--hand side of Eq.~(\ref{47}) become
uncorrelated, we have ${\bf W}_{a b} \approx \langle \Gamma_{\mu a}
\Gamma_{\mu b} \rangle / \langle \Gamma_{\mu a} \rangle \langle
\Gamma_{\mu b} \rangle = 1 + 2 \delta_{a b}$, and the Hauser--Feshbach
formula does apply with an elastic enhancement factor of three. That
is why it is sometimes stated that the elastic enhancement factor
increases from the value $2$ for $\Gamma \gg d$ (see
Section~\ref{ericson}) to the value $3$ for $\Gamma \ll d$ although
the last value actually applies only in a special situation.

Eq.~(\ref{47}) holds for completely isolated resonances. A correction
taking into account weak resonance overlap and using $R$--matrix
theory was given by~\textcite{Lan57}. \textcite{Mol61} went beyond
Eq.~(\ref{47}) by using a perturbative expansion of the $S$--matrix in
powers of the non--diagonal elements of the width matrix. The
expressions become soon very cumbersome. Therefore,
\textcite{Mol61,Mol63,Mol64,Mol69,Mol75} (see also further references
therein) tried to go beyond the limit $\Gamma \ll d$ by using the pole
expansion of the $S$--matrix as formulated by~\textcite{Hum61}. But
the pole parameters (locations of poles and values of the residues)
are linked by unitarity in a complicated way and, thus, not
statistically independent. That difficulty has never been resolved. We
return to some aspects of Moldauer's work in Section~\ref{pole}.

Eq.~(\ref{46}) predicts average cross sections for elastic and
inelastic scattering as functions of the transmission coefficients.
It would be of interest to calculate perturbatively also the
cross--section autocorrelation function $\langle |S^{\rm (CN \ fl)}_{a
b}(E_1)|^2 |S^{\rm (CN \ fl)}_{a b}(E_2)|^2 \rangle$. That seems not
to have been done yet. Likewise, the quantity $\langle S^{\rm CN \
fl}_{a a} (S^{\rm CN \ fl}_{b b})^* \rangle$ seems not to have been
worked out explicitly for $\Gamma \ll d$. As remarked in the
introduction to Section~\ref{resu}, that quantity is needed for the
average CN cross section in the presence of direct reactions. The
reason for the neglect is probably that $\Gamma \ll d$ is strictly
realized only for a single open channel.

\subsection{Ericson Regime ($\Gamma \gg d$)}
\label{ericson}

The Ericson model reviewed in Section~\ref{eri} leads to interesting
predictions that agree with experiment but has the shortcomings listed
at the beginning of Section~\ref{theo}. While it is probably not
possible to derive the Ericson model as such from the RMT approach
developed in Section~\ref{theo} we now show that it is indeed possible
to derive both, the Hauser--Feshbach formula and all results of the
Ericson model, from that approach. That is done with the help of an
asymptotic expansion in powers of $d / \Gamma$. Two methods have been
used, a diagrammatic approach~\cite{Aga75} and the replica
trick~\cite{Wei84}. These are reviewed in turn. Combining the results
of both, one obtains a complete theory of CN reactions in the regime
$\Gamma \gg d$.

\subsubsection{Diagrammatic Expansion}
\label{diag}

We use the form~(\ref{27g},\ref{27h}) of $S^{\rm CN}$ and the
statistical properties that come with that representation. We note
that with $\tilde{F}_{\mu \nu}(E) = - i \pi \sum_c \tilde{W}_{\mu c}
\tilde{W}_{c \nu}$, we have 
\be
\langle \tilde{F}_{\mu \nu} \rangle = - i \pi \delta_{\mu \nu} \sum_c
v^2_c  = - i \delta_{\mu \nu} f \ .
\label{50}
\ee
The last equation defines $f$. With $\delta F_{\mu \nu} = F_{\mu \nu}
- \delta_{\mu \nu} f$, the matrix $\tilde{D}$ is written as
$\tilde{D}_{\mu \nu} = (E - E_\mu - i f) \delta_{\mu \nu} - \delta
F_{\mu \nu} = \tilde{D}^{(0)}_\mu \delta_{\mu \nu} - \delta F_{\mu
\nu}$. In the approach of~\textcite{Aga75}, each of the two
$S$--matrix elements in the product $S^{\rm CN}_{a b} (E_1) (S^{\rm
CN}_{c d}(E_2))^*$ is expanded in a Born series with respect to
$\delta F_{\mu \nu}$. To calculate $\langle S^{\rm CN}_{a b}(E_1)
(S^{\rm CN}_{c d}(E_2))^* \rangle$, one first performs the ensemble
average over the $\tilde{W}$s. That average is taken separately for
each term of the double Born series. The matrix $\delta F_{\mu \nu}$
is bilinear in the Gaussian--distributed $\tilde{W}$s. Each term of
the double Born series therefore contains a product of
$\tilde{W}$s. The ensemble average over such a product is taken by
Wick contraction: The $\tilde{W}$s are grouped in pairs, each pair is
replaced by its ensemble average, see Eq.~(\ref{27f}). All possible
ways of pairing the $\tilde{W}$s must be taken into account, including
pairs where one member stems from the Born series of $S^{\rm CN}$ and
the other, from that of $(S^{\rm CN})^*$. The number of ways of
pairing the $\tilde{W}$s increases dramatically with the order of the
double Born series. Among these, only those which contribute to
leading order in $d / \Gamma$, are kept. These are identified with the
help of two rules. (i) Contraction patterns are neglected that yield
terms of the form $\sum_\mu (\tilde{D}^{(0)}_\mu)^{n_1}
((\tilde{D}^{(0)}_\mu)^*)^{n_2}$ with $n_1 \geq 2$ and/or $n_2 \geq
2$. That is because when the summation over $\mu$ is changed into an
energy integration and the integrals are done via contour integration,
the result vanishes. (ii) The second rule is illustrated by the
following example. We compare two simple contraction patterns both of
which occur as parts in the Born series and yield a term proportional
to $(\tilde{D}^{(0)}_\mu)^3$. The first one occurs in the term
$\tilde{D}^{(0)}_\mu [ ( \tilde{W} i \tilde{W} \tilde{D}^{(0)} )^2
]_{\mu \nu}$ and involves the contraction of the first with the third
and of the second with the fourth factor $\tilde{W}$. From
Eq.~(\ref{27f}) the result is $- \delta_{\mu \rho}
(\tilde{D}^{(0)}_\mu)^3 \sum_a (v^2_a)^2$. The second pattern occurs
in the term $\tilde{D}^{(0)}_\mu [(\tilde{W} i \tilde{W}
\tilde{D}^{(0)})^4]_{\mu \nu}$ and involves the contraction of the
first with the fourth, of the second with the third, of the fifth with
the eighth, and of the sixth with the seventh factor $\tilde{W}$. The
result is $\delta_{\mu \nu} (\tilde{D}^{(0)}_\mu)^3 ((-i \pi / d)
\sum_a (v^2_a)^2)^2$. The first (second) pattern yields a single
(double) sum over channels. Thus, the first pattern is small of order
$d / \Gamma$ compared to the second and is neglected. Denoting the
contraction of a pair of $\tilde{W}$s with an overbar, one finds that
only those contractions survive in leading order in $d / \Gamma$ where
no two contraction lines intersect (nested contributions).

In evaluating these patterns we have replaced the summation over
eigenvalues $E_\mu$ by an integration over energy. That yields the
factors $1 / d$ in the second pattern. Doing so corresponds to the
neglect of correlations between eigenvalues (the actual eigenvalue
distribution is replaced by one with constant spacings). Wigner--Dyson
eigenvalue correlations have a typical range given by $d$, and for $d
\ll \Gamma$ one expects that the neglect is justified. With the help
of that same approximation the remaining terms in the double Born
series can be resummed. With $\ve = E_2 - E_1$ that yields
\ba
&& \langle S^{(\rm CN \ fl)}_{a b}(E_1) (S^{(\rm CN \ fl)}_{c d}(E_2))^*
\rangle = (\delta_{a c} \delta_{b d} + \delta_{a d} \delta_{b c})
\nonumber \\
&& \qquad \qquad \times \frac{T_a T_b}{\sum_e T_e + 2 i \pi \ve / d} \ .
\label{52}
\ea
For $\ve = 0$ that coincides with the Hauser--Feshbach
formula~(\ref{7}) and for $\ve \neq 0$ agrees with Ericson's
prediction~(\ref{11}) if we identify in the latter the factors $4
\pi^2 v^2_a / d$ with the transmission coefficients $T_a$, see the
text following Eq.~(\ref{11}). Moreover, in the framework of the
diagrammatic approach, the Weisskopf estimate~(\ref{9}) is seen to
yield the exact expression for the width of the $S$--matrix
correlation function in the Ericson regime. Eq.~(\ref{52}) implies
that the elastic ($a = b$) CN cross section is enhanced over the
inelastic one by a factor of two (``elasic enhancement factor'').
This result is beautifully confirmed experimentally, see
Fig.~\ref{fig:HW3}.  In~\cite{Aga75} the terms of next order in $d /
\Gamma$ were also calculated. We return to that point in
Section~\ref{replica}.

The diagrammatic expansion can also be used to calculate higher--order
$S$--matrix correlation functions. It is found~\cite{Aga75} that to
leading order in $d / \Gamma$, such correlations can be expressed
completely in terms of correlations of pairs of $S$--matrix elements.
That implies that in the Ericson regime, the $S$--matrix elements
possess a Gaussian distribution.

In summary, the diagrammatic approach shows that for $\Gamma \gg d$,
the $S$--matrix elements are Gaussian--distributed random processes,
with first and second moments given by Eqs.~(\ref{38}) and (\ref{52}),
respectively, and with a correlation width given by the Weisskopf
estimate~(\ref{9}). All of this agrees with the predictions of the
Ericson model. We conclude that within the diagrammatic approach
(i.e., under the neglect of eigenvalue correlations), the distribution
of the $S$--matrix elements is completely known and is determined in
terms of the average $S$--matrix elements~(\ref{38}) via the
transmission coefficients~(\ref{6}).

\subsubsection{Replica Trick}
\label{replica}

Calculation of the average of $|S^{\rm CN}_{a b}|^2$ is difficult
because $H^{\rm GOE}$ appears in the denominator of $S^{\rm CN}$. A
way to overcome that problem consists in the use of a generating
functional $Z$~\cite{Wei84}. The observable (here: $|S^{\rm
CN}_{a b}|^2$) is given in terms of a suitable derivative of $Z$, and
$H^{\rm GOE}$ appears in $Z$ as the argument of an exponential. That
fact would greatly simplify the integration over the random variables
(the matrix elements of $H^{\rm GOE}$) were it not for the need to
normalize $Z$. The normalization problem is overcome by the replica
trick originally developed in condensed--matter physics~\cite{Edw75}.

We introduce $N$ real integration variables $\psi_\mu$, $\mu = 1,
\ldots, N$ and define the generating functional
\ba
&&Z(E, J) = \bigg( \prod_{\mu = 1}^N \int_{- \infty}^{+ \infty}
{\rm d} \psi_\mu \bigg) \exp \bigg\{ (i/2) \sum_{\mu} \psi_\mu
E \psi_\mu \bigg\} \nonumber \\
&& \times \exp \bigg\{ (i/2) \sum_{\mu \nu} \psi_\mu( - H_{\mu \nu}
+ i \pi \sum_c \hat{W}_{\mu c} \hat{W}_{c \nu} ) \psi_\nu \bigg\} \ .
\nonumber \\
\label{53}
\ea
Here the matrix elements $\hat{W}$ include the ``source terms'' $J$
and are defined as
\be
\hat{W}_{\mu c} = W_{\mu c} + J \delta_{c a} W_{\mu b} + J
\delta_{c b} W_{\mu a} \ .
\label{54}
\ee
The CN scattering matrix is given by
\be
S^{\rm CN}_{a b} = \delta_{a b} + \frac{\partial}{\partial J} \ln
Z(E, J) \bigg|_{J = 0} \ .
\label{55}
\ee
Unfortunately, little has been gained because calculating $\langle
S^{\rm CN}_{a b} \rangle$ or $\langle S^{\rm CN}_{a b} (S^{\rm CN}_{c
d})^* \rangle$ from Eq.~(\ref{55}) involves averaging the logarithmic
derivative of $Z$ or the product of two logarithmic derivatives of $Z$
and is next to impossible to do. The difficulty is that at $J = 0$,
$Z$ is not and cannot easily be normalized to unity. The problem is
overcome by using the identity
\be
\ln Z = \lim_{n \to 0} \frac{1}{n} ( Z^n - 1 ) \ .
\label{56}
\ee
For integer values of $n$, the average of $Z^n$ (or of a product of
such terms) can be calculated, and Eq.~(\ref{56}) is then used to
calculate $\langle \ln Z \rangle$ or of $\langle ( \ln Z )^2 \rangle$.
Instead of $\ln Z$ we use $n$ replicas of $Z$ to calculate averages,
hence the name of the method. The calculation must be done
analytically as otherwise the limit $n \to 0$ cannot be taken. That is
possible only for positive integer values of $n$.  However,
Eq.~(\ref{56}) is strictly valid only when $n$ is not restricted to
integer values. Otherwise one may miss non--zero contributions that
happen to vanish for all positive integer $n$. That is why the method
is not guaranteed to be exact and is referred to as a ``trick''. Later
investigations have shown, however, that when used for an asymptotic
expansion as is done below, the replica trick gives correct
answers~\cite{Ver85b}.

A detailed description of the calculation~\cite{Wei84} exceeds the
frame of this review. We only sketch the essential steps. In the case
of $\langle Z^n \rangle$, we deal with $n \times N$ real integration
variables $\psi^k_\mu$ with $k = 1, \ldots, n$ and $\mu = 1, \ldots,
N$. With the help of Eq.~(\ref{24}), averaging over $H^{\rm GOE}$
yields a quartic term in the $\psi^k_\mu$s. That term can be written
as the trace of the square of the symmetric real $n
\times n$ matrix $A_{k k'} = \sum_\mu \psi^k_\mu \psi^{k'}_\mu$. The
form of $A$ reflects the orthogonal invariance of the GOE. The
bilinear term in $A$ is removed by means of a Hubbard--Stratonovich
transformation,
\ba
&&\exp \bigg\{ - \frac{\lambda^2}{4 N} {\rm Trace} \ (A^2) 
\bigg\} \propto \nonumber \\
&&\prod_{k \leq k'} \int_{- \infty}^{+ \infty} {\rm d} \sigma_{k k'}
\ \exp \bigg\{ - \frac{N}{4} {\rm Trace} \ (\sigma^2) \bigg\}
\nonumber \\
&& \qquad \times \exp \bigg\{ - i \lambda {\rm Trace} (\sigma A)
\bigg\} \ .
\label{57}
\ea
In Eq.~(\ref{57}) we have introduced a set of new integration
variables which appear in the form of the symmetric real
$n$--dimensional matrix $\sigma_{k k'}$. After removal of the quartic
term, the remaining integral over the $\psi^k_\mu$s is Gaussian and
can be done. The integral over $\sigma_{k k'}$ is performed with the
help of the saddle--point approximation. For $N \gg 1$ that
approximation is excellent. For $\langle S^{\rm CN} \rangle$, it
yields an asymptotic expansion in inverse powers of $N$ with the
right--hand side of Eq.~(\ref{38}) as the leading term. Proceeding in
the same way for $\langle S^{({\rm CN \ fl})}_{a b}(E_1) (S^{({\rm CN
\ fl})}_{c d}(E_2))^* \rangle$ one finds instead of a single saddle
point a continuum of saddle points. In a seminal
paper,~\textcite{Sch80} have shown in a different context how to deal
with that manifold. That is done with the help of a suitable
parametrization of the matrix $\sigma_{k k'}$ which displays both, the
massive integration variables (that describe integration points
outside the saddle--point manifold) and the Goldstone mode (which
defines the coordinates within the saddle--point manifold). After
these steps, one finds that the energy difference $E_2 - E_1 = \ve$
appears in the integrand only in the combination $\sum_c T_c + (2 i
\pi \ve / d)$. That shows that the correlation width $\Gamma$ of the
correlation function is given by the Weisskopf estimate~(\ref{9}). The
integrals over the saddle--point manifold cannot be done exactly. For
the calculation of $\langle S^{({\rm CN \ fl})}_{a b}(E_1) (S^{({\rm
CN \ fl})}_{c d}(E_2))^* \rangle$, the replica trick can only be used
to calculate an asymptotic expansion in inverse powers of $\sum_c
T_c$. The leading term agrees with Eq.~(\ref{52}). Terms of higher
order also agree with the result given by~\textcite{Aga75}. Moreover,
these terms can be used to confirm unitarity for each order of the
expansion and to show that the leading term gives reliable answers for
$\sum_c T_c \geq 10$ or so, see also Section~\ref{general}.

In the case of direct reactions, we use Eq.~(\ref{C1}) and obtain
from Eq.~(\ref{52}), the definition~(\ref{12e}), and with $T_a = p_a$
as a result Eq.~(\ref{12d}). That shows that to leading order in $d /
\Gamma$, the result of~\textcite{Kaw73} agrees with that of the
statistical theory.

\subsubsection{Summary}

The replica trick yields an asymptotic expansion in inverse powers of
$\sum_c T_c$ for the $S$--matrix correlation function. The leading
term is given by Eq.~(\ref{52}). That form implies that the
correlation width agrees exactly with the Weisskopf
estimate~(\ref{9}). The diagrammatic approach neglects eigenvalue
correlations of the GOE Hamiltonian in favor of a constant--spacing
model but leads likewise to Eq.~(\ref{52}). That shows that for
$\Gamma \gg d$ such correlations are indeed negligible, and that the
results of the diagrammatic approach are trustworthy. Investigating
higher--order correlation functions within the diagrammatic approach
(which seems prohibitively difficulty in the framework of the replica
trick), one finds that the $S$--matrix elements are
Gaussian--distributed random variables with first and second moments
given by Eqs.~(\ref{38}) and (\ref{52}), respectively.  We see that
combining the replica trick and the diagrammatic approach we obtain a
complete theoretical understanding of the distribution of the elements
of $S^{\rm CN}$ in the Ericson regime. The results agree with
predictions of the Ericson model and, in the presence of direct
reactions, with those of~\textcite{Kaw73}.

\subsection{$S$--Matrix Correlation Function}
\label{general}

The replica trick yields only an asymptotic expansion but not the full
$S$--matrix correlation function. That function can be obtained
exactly using ``supersymmetry''~\cite{Efe83,Ver85b}. We briefly
motivate the use of that method.

The use of the generating functional defined in Eq.~(\ref{53}) would
greatly simplify if it were possible to normalize $Z$ so that $Z(E,0)
= 1$. Then in Eq.~(\ref{55}) we would have $[\partial \ln Z(E, J) /
\partial J]|_{J = 0} = [\partial Z(E, J) / \partial J]|_{J = 0}$.
Averaging $S$ would amount to averaging $Z$ (and not $\ln Z$), and it
would not be necessary to use the replica trick. Since in $Z$ the
random variables appear in the exponent, the calculation of $\langle Z
\rangle$ would be straightforward.

That goal is achieved by defining the normalized generating functional
as the product of two factors. The first factor is $Z^2$ with $Z$ as
defined in Eq.~(\ref{53}). The second factor has the same form as $Z$
except that the integration variables anticommute. Integration with
anticommuting variables is a well--known mathematical
technique~\cite{Ber86}. In the present context, we use the following
property. The normalization factor of a Gaussian integral involving
$N$ commuting real integration variables $\psi_\mu$, $\mu = 1, \ldots,
N$ and a symmetric matrix $A$ is given by
\ba
&& \prod_\mu \int_{- \infty}^{+ \infty} {\rm d} \psi_\mu \ \exp
\bigg\{(i/2) \sum_{\mu \nu} \psi_\mu A_{\mu \nu} \psi_\nu \bigg\} =
\nonumber \\
&& \qquad \qquad [\det (A / (2 i \pi))]^{- 1/2} \ .
\label{58}
\ea
The same integral with the commuting variables $\psi_\mu$ replaced by
anticommuting variables $\chi_\mu$ and $\chi^*_\mu$ is given by
\be
\prod_\mu \int {\rm d} \chi^*_\mu {\rm d} \chi_\mu \ \exp \bigg\{ i
\sum_{\mu \nu} \chi^*_\mu A_{\mu \nu} \chi_\nu \bigg\} = [\det (A
/ (2 i \pi))] \ .
\label{59}
\ee
Combining two factors of the form of the left--hand side of
Eq.~(\ref{58}) with one factor of the form of the left--hand side of
Eq.~(\ref{59}) we obtain a normalized generating functional involving
both commuting and anticommuting integration variables. The term
``supersymmetry'' commonly used in that context is somewhat
inappropriate. It stems from relativistic quantum field
theory~\cite{Wes74a,Wes74b} where fermionic (i.e., anticommuting) and
bosonic (i.e., commuting) fields are connected by a
``supersymmetry''. That specific symmetry does not occur in the
present context.

In a formal sense, the calculation of $\langle S^{\rm CN} \rangle$ and
of the correlation function runs in parallel to the one for the
replica trick. In content, the steps differ because of the
simultaneous use of commuting and anticommuting variables. The steps
are: Averaging the generating functional over the GOE, replacing the
quartic terms in the integration variables generated that way with the
help of a Hubbard--Stratonovich transformation, execution of the
remaining Gaussian integrals over the original integration variables,
use of the saddle--point approximation for the remaining integrals
over the supermatrix $\sigma$. That matrix is introduced via the
Hubbard--Stratonovich transformation and has both commuting and
anticommuting elements. For $N \gg 1$ the saddle--point approximation
is excellent and yields for $\langle S^{\rm CN} \rangle$ an asymptotic
expansion in inverse powers of $N$, the leading term being given by
the right--hand side of Eq.~(\ref{38}) when $E$ is taken in the center
of the GOE spectrum. When the same formalism is used for the
$S$--matrix correlation function instead of $S$ itself, the
saddle--point changes into a saddle--point manifold. With the help of
a suitable parametrization~\cite{Sch80} of $\sigma$, the integration
over that manifold can be done exactly. The result is again valid to
leading order in $(1 / N)$. With $\ve = E_2 - E_1$, one
obtains~\cite{Ver85a}
\ba
&& \langle S^{(\rm CN \ fl)}_{a b}(E_1) (S^{(\rm CN \ fl)}_{c d}(E_2))^*
\rangle = \prod_{i = 1}^2 \int_{0}^{+ \infty} {\rm d} \lambda_i
\int_{0}^{1} {\rm d} \lambda \nonumber \\
&& \qquad \times \frac{1}{8} \ \mu(\lambda_1, \lambda_2, \lambda) \ 
\exp \bigg\{ - \frac{i \pi \ve}{d} (\lambda_1 + \lambda_2 + 2 \lambda)
\bigg\} \nonumber \\
&& \qquad \times \prod_e \frac{(1 - T_e \lambda)}{(1 + T_e
\lambda_1)^{1/2} (1 + T_e \lambda_2)^{1/2}} \nonumber \\
&& \qquad  \times J_{a b c d} (\lambda_1, \lambda_2, \lambda) \ . 
\label{60}
\ea
The factor $\mu(\lambda_1, \lambda_2, \lambda)$ is an integration
measure and given by
\be
\mu(\lambda_1, \lambda_2, \lambda) = \frac{(1 - \lambda) \lambda
|\lambda_1 - \lambda_2|}{\prod_{i = 1}^2 [((1 + \lambda_i)
\lambda_i)^{1/2} (\lambda + \lambda_i)^2]} \ ,  
\label{61}
\ee
while
\ba
&& J_{a b c d}(\lambda_1, \lambda_2, \lambda) = (\delta_{a c}
\delta_{b d} + \delta_{a d} \delta_{b c}) T_a T_b \nonumber \\
&& \times \bigg( \sum_{i = 1}^2 \frac{\lambda_i(1 + \lambda_i)}{(1 +
T_a \lambda_i)(1 + T_b \lambda_i)} + \frac{2 \lambda (1 - \lambda)}
{(1 - T_a \lambda) (1 - T_b \lambda)} \bigg) \nonumber \\
&& \qquad + \ \delta_{a b} \delta_{c d} T_a T_c \langle
S^{\rm CN}_{a a} \rangle \langle (S^{\rm CN}_{c c})^* \rangle
\nonumber \\
&& \qquad \times \bigg( \sum_{i = 1}^2 \frac{\lambda_i}{1+ T_a
\lambda_i)} + \frac{2 \lambda}{1 - T_a \lambda} \bigg) \nonumber \\ 
&& \qquad \times \bigg( \sum_{i = 1}^2 \frac{\lambda_i}{1+ T_c
\lambda_i)} + \frac{2 \lambda}{1 - T_c \lambda} \bigg) 
\label{62}
\ea
describes the dependence of the correlation function on those channels
which appear explicitly on the left--hand side of Eq.~(\ref{60}).

Eqs.~(\ref{60}) to (\ref{62}) give the $S$--matrix correlation
function in closed form, i.e., in terms of an integral representation.
It does not seem possible to perform the remaining integrations
analytically for an arbitrary number of channels and for arbitrary
values of the transmission coefficients. In the way it is written,
Eq.~(\ref{60}) is not suited very well for a numerical evaluation
because there seem to be singularities as the integration variables
tend to zero. Moreover, the exponential function oscillates strongly
for large values of the $\lambda$s. These difficulties are overcome by
choosing another set of integration variables. Details are given
in~\cite{Ver86}. For the case of unitary symmetry, formulas
corresponding to Eqs.~(\ref{60}) to (\ref{62}) were given
in~\cite{Fyo05}.

We turn to the physical content of Eqs.~(\ref{60}) to (\ref{62}). The
unitarity condition~(\ref{12a}) must also hold  after averaging for
$S^{\rm CN}$. Using a Ward identity one finds that unitarity is indeed
obeyed~\cite{Ver85a}. Except for the overall phase factors of the
average $S$--matrices appearing on the right--hand side of
Eq.~(\ref{62}), the $S$--matrix correlation function depends only on
the transmission coefficients, as expected. Eqs.~(\ref{60}) to
(\ref{62}) apply over the entire GOE spectrum if $d$ is taken to be
the average GOE level spacing at $E = (1/2) (E_1 + E_2)$. That
stationarity property enhances confidence in the result. When $E$ is
chosen in the center of the GOE spectrum, $E = 0$, then $\langle
S^{\rm CN} \rangle$ is real, see Eq.~(\ref{38}), and the complex
conjugate sign in the last term in Eq.~(\ref{62}) is redundant.
Writing the product over channels in Eq.~(\ref{60}) as the exponential
of a logarithm, expanding the latter in powers of $\lambda_1,
\lambda_2$, and $\lambda$, and collecting terms, one finds that $\ve$
appears only in the combination $[(2 i \pi \ve / d) + \sum_c T_c]$.
This fact was also established in the framework of the replica
trick. It shows that $\Gamma$ as defined by the Weisskopf
estimate~(\ref{9}) defines the scale for the dependence of the
correlation function on $\ve$. Put differently, the energy difference
$\ve$ appears universally in the dimensionless form $\ve / \Gamma$
(and not $\ve / d$), and for $\{a,b\} = \{c,d\}$ the correlation
function~(\ref{60}) has the form $f(1 + i \ve / \Gamma; T_a, T_b; T_1,
T_2, \ldots, T_\Lambda)$. The right--hand side of Eq.~(\ref{62}) is
the sum of two terms. These correspond, respectively, to $\langle
S^{({\rm CN \ fl})}_{a b}(E_1)$ $(S^{({\rm CN \ fl})}_{a b}(E_2))^*
\rangle$ and to $\langle S^{({\rm CN \ fl})}_{a a}(E_1)$ $(S^{({\rm CN
\ fl})}_{b b}(E_2))^* \rangle$ and are exactly the terms expected, see
the introduction to Section~\ref{resu}.

The limiting cases ($\Gamma \ll d$ and $\Gamma \gg d$) of
Eqs.~(\ref{60}) to (\ref{62}) were studied and compared with previous
results~\cite{Ver86}. The case $\Gamma \ll d$ is obtained by
expanding the result in Eqs.~(\ref{60}) to (\ref{62}) in powers of the
transmission coefficients. The result agrees with Eq.~(\ref{45}).
Conversely, expanding the result in Eqs.~(\ref{60}) to (\ref{62}) in
inverse powers of $\sum_c T_c$ one generates the same series as
obtained from the replica trick, see Section~\ref{replica}. That shows
that the Hauser--Feshbach formula~(\ref{7}), the extension~(\ref{12d})
of that formula to the case of direct reactions, and the Ericson
result~(\ref{52}) all are the leading terms in an asymptotic expansion
of the relevant expressions in powers of $d / \Gamma$. Moreover, in
the limit $\Gamma \gg d$ the Weisskopf estimate~(\ref{9}) gives the
exact expression for the correlation width. All of these facts were
also obtained previously with the help of the replica trick, see
Section~\ref{replica}. A more detailed investigation using
supersymmetry~\cite{Dav88,Dav89} has shown that the elastic
$S$--matrix elements $S_{a a}$ possess the Gaussian distribution
assumed by Ericson only for strong absorption ($T_a \approx 1$). If
that condition is not met, deviations from the Gaussian are caused by
unitarity, see Section~\ref{csf}.

Eqs.~(\ref{60}) to (\ref{62}) contain the central result of the
statistical theory. The $S$--matrix correlation function is given
analytically for all values of $\Gamma / d$. These equations supersede
both the perturbative approach (Section~\ref{iso}) and the replica
trick (Section~\ref{replica}) because the results of both these
approaches turn out to be special cases of the general result. It
would be highly desirable to extend the supersymmetry approach to the
calculation of moments and correlation functions involving higher
powers of $S^{\rm CN}$ and $(S^{\rm CN})^*$ than the first. For the
correlation function~(\ref{62}), the supermatrix $\sigma$ has
dimension eight. For the cross--section correlation function, that
matrix has dimension sixteen. Integration over all matrix elements
must be done analytically. While that is feasible for the
function~(\ref{62}), it seems beyond reach for the cross--section
correlation. That is why the results of the diagrammatic approach
(Section~\ref{diag}) are needed to complete the theory in the Ericson
regime: They show that the $S$--matrix elements have a Gaussian
distribution. There is no analytical information about cross--section
fluctuations versus energy outside the Ericson regime.

We note that in both limits $\Gamma \ll d$ and $\Gamma \gg d$ the
fluctuation properties of the $S$--matrix are completely determined by
the Gaussian distribution of the eigenvectors of $H^{\rm GOE}$ and are
independent of the fluctuation properties of the eigenvalues of that
matrix. Indeed, for $\Gamma \ll d$ the value of $\langle | S^{\rm CN
fl}_{a b} |^2 \rangle$ in Eqs.~(\ref{46}) and (\ref{47}) is
determined entirely by the Porter--Thomas distribution of the partial
widths (which in turn is a consequence of the Gaussian distribution of
the eigenvectors of $H^{\rm GOE}$). And for $\Gamma \gg d$ (Ericson
limit) the asymptotic expansion~\cite{Aga75} based on a
picket--fence model for the eigenvalues gives the same result as the
calculation that takes fully into account the eigenvalue correlations
of $H^{\rm GOE}$. A significant dependence of the $S$--matrix
fluctuations on the distribution of the eigenvalues of $H^{\rm GOE}$
can, thus, occur only in the intermediate domain $\Gamma \approx d$.

A dynamical model for an interacting fermionic many--body system
coupled to a number of open channles was investigated
by~\textcite{Cel07,Cel08}. Spinless fermions are distributed over a
number of single--particle states with Poissonian level statistics and
interact via a two--body interaction of the EGOE(2) type, see
Section~I.V.A. The Hilbert space is spanned by Slater determinants
labeled $\mu = 1, \ldots, N$. These are coupled to the open channels
labeled $a, b, \ldots$ by amplitudes $A^a_\mu$ which are assumed to
have a Gaussian distribution with mean value zero and a second moment
given by $\langle A^a_\mu A^b_\nu \rangle = \delta_{a b} \delta_{\mu
\nu} \gamma^a / N$. The parameters $\gamma^a$ determine the strength
of the coupling to the channels. One focus of these papers is on the
way the $S$--matrix fluctuations change when the intrinsic dynamics is
changed from regular (Poisson statistics) to chaotic (Wigner--Dyson
statistics) by increasing the strength of the two--body interaction.
Many of the features discussed in this review are illustrated by the
numerical simulations by~\textcite{Cel07,Cel08}. The assumption that
the amplitudes $A^a_\mu$ have a Gaussian distribution with a second
moment that is proportional to the unit matrix in the space of Slater
determinants, puts a constraint on the calculations as it implies that
the distribution of the $A^a_\mu$ is invariant under orthogonal
transformations in the space of Slater determinants. Put differently,
together with the intrinsic Hamiltonian $H_{\mu \nu}$ all Hamiltonian
matrices obtained from $H_{\mu \nu}$ by orthogonal transformations
yield the same distribution for the $S$--matrix elements. This induced
orthogonal invariance of $H$ implies that the eigenvectors of $H$ are
Gaussian--distributed random variables, completely independently of
the detailed form of $H$, and that only the distribution of the
eigenvalues of $H$ depends on whether the intrinsic dynamics is
regular or chaotic. The statistical assumption on the amplitudes
$A^a_\mu$ is physically justified when many channels are open and when
the channel wave functions describe states that are chaotic
themselves. But the assumption allows for a partial test only of the
transition from regular to chaotic motion.

Eqs.~(\ref{60}) to (\ref{62}) have been derived in the limit $N \to
\infty$ and $\Lambda$ fixed. \textcite{Leh95b} pointed out that in the
extreme Ericson regime $\sum_c T_c \gg 1$, another limit (first
considered by \textcite{Sok88,Sok89,Sok92}) may be more appropriate:
$\Lambda$ and $N$ tend jointly to $\infty$ while the ratio $m =
\Lambda / N$ is kept fixed. The authors calculated the $S$--matrix
correlation function in that limit. Using supersymmetry, they find
corrections of order $m$ to the saddle--point equation.  These modify
the $S$--matrix correlation function but keep the correlation width
(given by the Weisskopf estimate) unchanged. The modifications are due
to the fact that in the limit considered, the range of the correlation
function becomes comparable to the range of the GOE spectrum, and the
universality that characterizes the regime $\Gamma \ll \lambda$ (where
$2 \lambda$ is the radius of the GOE semicircle) is lost. Further
details are given in Section~\ref{pole}.

\subsubsection*{Decay in Time of the Compound Nucleus}

Rather than studying the dependence of the correlation
function~(\ref{60}) on the energy difference $\ve$, we investigate the
Fourier transform of that function. We confine ourselves to pure CN
scattering (so that the matrix $U$ in Eq.~(\ref{C1}) is the unit
matrix). We first show that for a very short wave packet incident in
channel $a$, the time dependence of the CN decay feeding channel $b$
is given by the Fourier transform of the correlation
function~(\ref{60})~\cite{Lyu78a,Lyu78b,Dit92,Har92}. We then use
the explicit form of that function to determine the time dependence of
CN decay.

Let $\int {\rm d} E$ $\exp( - i E t) \ g(E)$ $| \Psi_a(E) \rangle$
with $\int {\rm d} E$ $|g(E)|^2$ $= 1$ be a normalized wave packet
incident in channel $a$. The functions $| \Psi_a(E) \rangle$ are
solutions of the Schr{\"o}dinger equation for a CN system with
$\Lambda$ channels and $N \gg 1$ quasibound states subject to the
boundary condition that there is an incident wave with unit flux in
channel $a$ only. We put $\hbar = 1$. The wave packet is short in time
if $g(E)$ is very broad, i.e., covers many resonances. The outgoing
flux in any channel $b$ is asymptotically (large distance) given by
$\int {\rm d} E_1 \ \int {\rm d} E_2$ $\exp [ i (E_2 - E_1) t ]$
$g(E_1) g^*(E_2)$ $S^{\rm CN}_{a b}(E_1) (S^{\rm CN}_{a b}(E_2))^*$.
We introduce new integration variables $E = (1/2) (E_1 + E_2)$ and
$\ve = E_2 - E_1$ and obtain $\int {\rm d} E \ \int {\rm d}
\ve$ $\exp [ i \ve t ]$ $g(E - (1/2) \ve) g^*(E + (1/2) \ve)$ $S^{\rm
CN}_{a b}(E - (1/2) \ve) (S^{\rm CN}_{a b}(E + (1/2) \ve))^*$. Under
the assumption that $g(E)$ changes very slowly over a scale of order
$d$ or $\Gamma$ (whichever is bigger), we may write that expression as
$\int {\rm d} \ve$ $\exp [ i \ve t ]$ $\int {\rm d} E \ |g(E)|^2
S^{\rm CN}_{a b}(E - (1/2) \ve) (S^{\rm CN}_{a b}(E + (1/2) \ve))^*$
$= \int {\rm d} \ve$ $\exp [ i \ve t ]$ $\langle S^{\rm CN}_{a b}(E -
(1/2) \ve) (S^{\rm CN}_{a b}(E + (1/2) \ve))^* \rangle$. We use the
decomposition~(\ref{1}) and obtain a sum of two contributions. The
first, $\delta(t) |\langle S^{\rm CN}_{a b} \rangle|^2$, shows that
the average $S$--matrix describes the fast part of the reaction as
claimed in Section~\ref{aver}. The second has the form
\ba
p_{a b}(t) &=& \int {\rm d} \ve \exp [ i \ve t ] \langle S^{({\rm CN
\ fl})}_{a b}(E - (1/2) \ve) \nonumber \\
&& \times (S^{({\rm CN \ fl})}_{a b}(E + (1/2) \ve))^* \rangle \ .
\label{62a}
\ea
That shows that the Fourier transform (FT) of the correlation
function~(\ref{60}) gives the time dependence of the CN decay. More
precisely, the FT is the derivative of the function which describes
the decay in time of the CN in channel $b$ within the statistical
theory~\cite{Lyu78a,Lyu78b,Dit92,Har92}. The form~(\ref{C3}) of the
matrix $D$ implies that $p_{a b}(t) = 0$ for $t < 0$ as
expected~\cite{Dit92,Har92}.

The function $p_{a b}(t)$ cannot be worked out analytically in its
full generality. Some limiting cases are of
interest~\cite{Dit92,Har92}. For a single open channel, the decay is
not exponential. Rather, it is asymptotically ($t \to \infty$) given
by the power--law dependence $t^{- 3/2}$. For $\Lambda$ channels
weakly coupled to the resonances (all $T_a \ll 1$), the decay has
asymptotically the form $t^{- 1 - \Lambda / 2}$. In both cases, the
characteristic mean decay time is given by the average of the coupling
strengths of the resonances to the channels. Exponential decay given
by $\exp [ - \Gamma t ]$ with $\Gamma$ given by the Weisskopf
estimate~(\ref{9}) is realized only in the Ericson regime $\Gamma \gg
d$. The deviations from the exponential decay law are due to the
Porter--Thomas distribution of the partial widths, see
Section~I.II.D.1. The decay in time of an isolated resonance is
exponential, of course, except for very small and very large times.
But superposing many such resonances with different widths causes
deviations from the exponential distribution.  Such deviations have
been observed experimentally, see Section~\ref{wocr}. In~\cite{Har09},
a somewhat different conclusion was drawn. The decay in time of
excitations in a chaotic microwave cavity was investigated. The cavity
mimicks a chaotic quantum system. Therefore, the general RMT results
should apply. The authors find that in the limit $t \to \infty$, the
ensemble average considered in~\cite{Dit92,Har92} differs from the
behavior of (every) single realization. For the latter, the
longest--living resonances finally dominate, these are well separated
in energy, and the decay, therefore, becomes eventually
exponential. The transition from algebraic to exponential decay
follows a universal law if time is properly
normalized~\cite{Har09}. The apparent discrepancy with the result
of~\cite{Dit92,Har92} is due~\cite{Ott09} to a finite--size
effect:~\textcite{Har09} considered the case of a finite number $N$ of
resonances, while the result of~\cite{Dit92,Har92} holds in the limit
$N \to \infty$.  The turning point in time where in~\cite{Har09}
power--law decay changes into exponential decay increases with $N$ and
is moved to infinity for $N \to \infty$.

Another approach to time delay uses the Wigner--Smith time--delay
matrix~\cite{Wig55b,Smi60a,Smi60b}
\ba
Q_{a b} &=& i \bigg\{ \frac{\rm d}{{\rm d} \ve} \sum_c \langle
S^{({\rm CN \ fl})}_{a c}(E - (1/2) \ve) \nonumber \\
&& \times (S^{({\rm CN \ fl})}_{b c}(E + (1/2) \ve))^* \rangle
\bigg\}_{\ve = 0} \ .
\label{62b}
\ea
That matrix is the matrix of average time delays. Indeed, it is easy
to see that $Q_{a a} = (2 \pi)^{-1} \sum_b \int {\rm d} t \ t \ p_{a
b}(t)$. The eigenvalues of the Hermitean matrix $Q$ are called the
proper delay times. For $\Lambda$ channels with $T_c = 1$ for all $c$,
the joint probability distribution and the density of the proper delay
times have been worked out~\cite{Bro97,Bro99}.~\textcite{Sav01}
generalized that approach to $\Lambda$ equivalent channels with $T_c
\neq 1$. Analytical formulas valid for an arbitrary number of channels
and arbitrary values of the transmission coefficients are given
in~\cite{Leh95b}. For time--reversal non--invariant systems these
issues were treated in~\cite{Fyo97a,Fyo97b}. The single--channel case
was studied in~\cite{Oss05}.

\subsection{Distribution of $S$--Matrix Elements}
\label{sme}

Except for the Ericson regime, correlation functions that relate to
physical observables and involve higher powers of $S$ than in
Eq.~(\ref{60}), are not known. Is the situation better when we ask for
the distribution of $S$--matrix elements all taken at the same energy? 
As we shall see, the answer lies strangely between yes and no.

\subsubsection{Fit Formulas}
\label{fit}

Fit formulas for the second moments of $S$ based upon an RMT
simulation were developed prior to the derivation of the general
result in Eqs.~(\ref{60}) to (\ref{62}) and, in principle, have been
superseded by that development. We give these formulas here because
they are still frequently used in applications.~\textcite{Tep74}
and~\textcite{Hof75a,Hof75b} used an ansatz for $\langle |S^{({\rm CN
\ fl})}_{a b}|^2 \rangle$ that was inspired by the Hauser--Feshbach
formula. It reads
\be
\langle |S^{({\rm CN / fl})}_{a b}|^2 \rangle = \frac{V_a V_b}{\sum_c
V_c} (1 + ({\bf W}_a - 1) \delta_{a b} ) \ .
\label{63}
\ee
Unitarity relates the expressions on the right--hand side of
Eq.~(\ref{63}) with the transmission coefficients $T_a$. The resulting
equations possess unique solutions for the $V_a$s provided the ${\bf
W}_a$s are known. That leaves the latter as the only parameters to be
determined by fits to a numerical simulation. For numerous sets of
transmission coefficients, the simulation was done using the
$K$--matrix form~(\ref{22b}) of $S^{\rm CN}$, taking in
Eq.~(\ref{22a}) the $E_\mu$s as eigenvalues and determining the
$\tilde{W}^{(0)}_{a \mu}$s in terms of the eigenfunctions of a GOE
matrix, with a strength defined by $T_a$. The result is a fit formula
for ${\bf W}_a$,
\be
{\bf W}_a - 1 = \frac{2}{1 + (T_a)^{0.3 + 1.5 (T_a / \sum_c T_c)}}
+ 2 \bigg[ \frac{T_a - T}{\sum_c T_c} \bigg]^2 \ . 
\label{64}
\ee
Here $T$ is the arithmetic mean of the $T_a$s. The other
non--vanishing bilinear form $\langle S^{({\rm CN \ fl})}_{a a}
(S^{({\rm CN \ fl})}_{b b})^* \rangle$ was similarly assumed to
factorize for $a \neq b$. Fit formulas for that function based on
simulations are likewise given in~\cite{Hof75b}. Numerical
evaluation~\cite{Ver86} of Eqs.~(\ref{60}) to (\ref{62}) showed good
agreement with Eqs.~(\ref{63}) and (\ref{64}) within the statistical
errors expected.  Similar formulas were subsequently developed
by~\textcite{Mol76}.

\subsubsection{Many Open Channels}

\textcite{Dys62a} has defined the orthogonal ensemble of unitary
symmetric matrices $S$ of dimension $N$ (the ``circular orthogonal
ensemble'') and has studied the distribution of its eigenvalues for $N
\to \infty$. The members of that ensemble may be interpreted as
scattering matrices. By definition, these have average value zero and,
thus, transmission coefficients $T_a = 1$ in all channels. Moreover,
the limit $N \to \infty$ implies the (unrealistic) limit of infinitely
many channels. It is perhaps for these reasons that the orthogonal
circular ensemble has not found notable applications in nuclear
physics.

The distribution of $S$--matrix elements is analytically accessible
for systems with absorption. Absorption occurs, for instance, in
microwave resonators where it is due to Ohmic losses. Absorption is
described by introducing ficticious channels and associated
transmission coefficients. If the resulting total widths of the levels
are dominated by absorption, they become statistically independent of
the partial width amplitudes of the physical channels. The calculation
of the distribution of $S$--matrix elements is then much
simplified~\cite{Fyo05, Sav06}.

The approach may be used to obtain partial information on the
distribution of $S$--matrix elements in CN reactions. One must focus
attention on a distinct pair $(a, b)$ of channels and assume that many
channels are open (these may all have small transmission coefficients
so that we do not necessarily work in the Ericson regime). All
channels different from $(a, b)$ then play the same role as the
ficticious channels in the case of absorption, and the results
obtained by~\textcite{Fyo05} and~\textcite{Sav06} may be used to
determine the distribution of $S_{a b}$. The method obviously does not
yield the joint distribution function of all elements of the
scattering matrix. We are not aware of applications of that approach
to CN reactions and do not reproduce the relevant formulas here, see,
however, Section~\ref{csf}.

\subsubsection{Exact Results for Low Moments}

The supersymmetry approach can be used to calculate the third and
fourth moments of the scattering matrix~\cite{Dav88,Dav89}. The need
to introduce supermatrices of dimension larger than eight (unavoidable
if one wishes to calculate higher--order correlation functions) is
circumvented by writing these moments as higher--order derivatives of
the very same generating functional that is used to calculate the
result~(\ref{60}). The resulting analytical formulas are valid for all
values of $\Gamma / d$. For more details, see Section~\ref{csf}. In
view of the complexity of the calculations, an extension of that
approach to higher moments than the fourth seems very difficult.

\subsubsection{Maximum--Entropy Approach}

The approach developed by the Mexican group~\cite{Mel85} is based on
an appealing idea. If $S^{\rm CN}$ is determined by an RMT approach
(as done in Section~\ref{stoc}), then $S^{\rm CN}$ itself should be as
random as is consistent with basic properties of that matrix. These
properties are unitarity, symmetry, and the property Eq.~(\ref{39}) to
which the authors refer as ``analyticity--ergodicity''. In addition,
it is required that $\langle S^{\rm CN}_{a a} \rangle$ should have the
value given by the $S$--matrix $S^{\rm opt}_{a a}$ of the optical
model. With these requirements used as constraints, expressions for
the distribution $F(S)$ of $S^{\rm CN}$ are then obtained from either
an analytical approach or a variational principle. The results
agree. In the latter case, the probability density for $S^{\rm CN}$ is
determined by maximizing the entropy $- \int F(S) \ln F(S) {\rm d}
\mu(S)$ under the said constraints. Here $\mu(S)$ is the Haar measure
for unitary and symmetric matrices, see Section~I.II.B. With $\Lambda$
the number of channels, ${\bf 1}_\Lambda$ the unit matrix in $\Lambda$
dimensions, and $V$ a $\Lambda$--dependent normalization factor, the
result is
\be
F(S^{\rm CN}) = \frac{1}{V} \frac{[\det ({\bf 1}_\Lambda -
(S^{\rm opt})^* S^{\rm opt}) ]^{(\Lambda + 1)/2}}{|\det
({\bf 1}_\Lambda - (S^{\rm opt})^* S^{\rm CN} )|^{\Lambda + 1} } \ . 
\label{65}
\ee
The function $F$ is the most likely distribution function for $S^{\rm
CN}$ under the constraints mentioned. Do results calculated from
$F(S)$ agree with those based on the RMT approach in
Section~\ref{stoc}? The answer is a (conditional) yes. First, the
distribution~(\ref{65}) can be derived~\cite{Bro95} from the
stochastic $S$--matrix in Eq.~(\ref{C2}) under the assumption that the
Hamiltonian is a member of the ``Lorentzian ensemble'' (rather than of
the GOE). The Lorentzian ensemble and the GOE have the same
eigenvector distribution and the same level--correlation functions in
the large $N$ limit~\cite{Bro95}. It is, therefore, extremely likely
that the distribution~(\ref{65}) also holds for the stochastic
$S$--matrix in Eq.~(\ref{C2}) with the Hamiltonian taken from the
GOE. That view is supported further by the following facts. For strong
absorption ($\langle S_{a b} \rangle = 0$ for all $\{a, b\}$)
Eq.~(\ref{65}) reduces to $F(S^{\rm CN}) =$ constant. In other words,
the distribution of $S$--matrix elements is determined entirely by the
Haar measure. In that limit the ensemble~(\ref{65}) agrees, therefore,
with Dyson's circular ensemble~\cite{Dys62a}. In the Ericson regime,
Eq.~(\ref{65}) yields the Hauser--Feshbach formula with an elastic
enhancement factor of two~\cite{Fri85}. Moreover, for $\Lambda = 2$
and $\ve = 0$ Eq.~(\ref{65}) agrees~\cite{Ver86} with Eqs.~(\ref{60})
to (\ref{62}). Further support comes from results for the unitary case
\cite{Fyo97b}.

All these facts make it seem highly probable that Eq.~(\ref{65})
correctly describes the distribution of $S$--matrix elements for
chaotic scattering. Unfortunately, in the general case of several open
channels the expression on the right--hand side of Eq.~(\ref{65}) is
so unwieldy that it has not been possible so far to evaluate it. That
puts us into the strange situation that we do seem to know the
distribution of $S$--matrix elements without being able to use it. It
must also be remembered that the maximum--entropy approach does not
yield information on $S$--matrix correlation functions.

\subsection{Cross--Section Fluctuations}
\label{csf}

For the analysis or prediction of cross--section fluctuations, the
theoretical results reviewed so far in this Section do not suffice.
While the third and fourth moments of $S^{\rm fl}(E)$ at fixed energy
$E$ are known analytically~\cite{Dav88,Dav89}, information on the
corresponding correlation functions of $S^{\rm fl}(E)$ does not
exist. The problem was addressed by~\textcite{Die09c}. The authors
used the available analytical results and information obtained
numerically and/or experimentally from microwave billiards (see
Section~\ref{wocr}) to investigate cross--section fluctuations, and to
identify the range of parameters where predictions can safely be made.

We use Eqs.~(\ref{C2}) to (\ref{27c}) for $S^{\rm CN}$. The
cross--section autocorrelation function ${\cal C}_{a b}(\ve)$ is
defined as
\begin{eqnarray}
	{\cal C}_{a b}(\ve) &=& \langle |S^{\rm CN}_{a b}(E+\ve/2)|^2
	|S^{\rm CN}_{a b}(E - \ve/2)|^2 \rangle \nonumber \\
        && - \bigg[\langle |S^{\rm CN}_{a b}(E)|^2 \rangle \bigg]^2 \ .
	\label{cs1}
\end{eqnarray}
Using Eq.~(\ref{1}) we write $S^{\rm CN}(E) = \langle S^{\rm CN}
\rangle + S^{({\rm CN \ fl})}$. We use the fact (see Eq.~(\ref{38}))
that $\langle S^{\rm CN} \rangle$ is real and obtain
\begin{eqnarray}
	&& {\cal C}_{a b}(\ve) = 2 \delta_{a b} \bigg\{| \langle
	S^{\rm CN}_{a a} \rangle |^2 \ \Re [ C^{(2)}_{a a}(\ve) ]
	\nonumber \\ &+& \langle S^{\rm CN}_{a a} \rangle \ \Re [
	\langle S^{({\rm CN \ fl}) *}_{a a}(E + \ve/2) |S^{({\rm CN \
	fl})}_{a a}(E - \ve/2)|^2 \rangle ] \nonumber \\ &+& \langle
	S^{\rm CN}_{a a} \rangle \ \Re [ \langle S^{({\rm CN \ fl})
	*}_{a a}(E - \ve/2) |S^{({\rm CN \ fl})}_{a a}(E + \ve/2)|^2
	\rangle ] \bigg\} \nonumber \\ &+& \bigg\{ \langle |S^{({\rm
	CN \ fl})}_{a b}(E + \ve/2)|^2 |S^{({\rm CN \ fl})}_{a b}(E -
	\ve/2)|^2 \rangle \nonumber \\ && \qquad - \bigg[ \langle
	|S^{({\rm CN \ fl})}_{a b}|^2 \rangle \bigg]^2 \bigg\} \, .
	\label{cs2}
\end{eqnarray}
Here $C^{(2)}_{a b}(\ve)$ is the $S$--matrix autocorrelation function
in Eq.~(\ref{60}) taken at $c = a$, $d = b$. Eq.~(\ref{cs2}) shows
that there is a substantial difference between the elastic case ($a =
b$) and the inelastic one ($a \neq b$), caused by the fact that
$\langle S^{CN} \rangle$ is diagonal. We address the inelastic case
first. In the Ericson regime, $S^{({\rm CN \ fl})}$ is Gaussian, and
the autocorrelation function of $|S^{({\rm CN \ fl})}|^2$ is,
therefore, given by the square of the $S$--matrix correlation function
in Eq.~(\ref{60}). However, that relation cannot be expected to hold
much outside the Ericson regime because we must expect the
cross--section fluctuations (in units of the average cross section) to
increase significantly as $\Gamma / d$ decreases. This expectation is
quantitatively confirmed in~\cite{Die09c}: The autocorrelation
function of $|S^{({\rm CN \ fl})}|^2$ is, at least approximately,
given by the square of the $S$--matrix correlation function in
Eq.~(\ref{60}) whenever $S^{({\rm CN \ fl})}$ posseses a bivariate
Gaussian distribution, and that is essentially the case when $\Gamma >
d$ or so.  For the elastic case ($a = b$) the situation is difficult
even in the Ericson regime unless $|\langle S^{CN}_{a a} \rangle| \ll
1$. Indeed, whenever that constraint is violated, the distribution of
$S^{({\rm CN \ fl})}_{a a}$ cannot be Gaussian: Combined with the
decomposition $S^{\rm CN}_{a a} = \langle S^{CN}_{a a} \rangle +
S^{({\rm CN \ fl})}_{a a}$, the constraint $|S^{\rm CN}_{a a}| < 1$
implied by unitarity forces the distribution of $S^{({\rm CN \
fl})}_{a a}$ to be skewed. The distortion of the Gaussian distribution
grows with decreasing $\Gamma / d$ and is strongest when the coupling
to the channels becomes very small. (Then $S^{\rm CN}_{a a}$ is
dominated by the first term on the right--hand side of
Eq.~(\ref{C2}).) The terms in the second and third line of
Eq.~(\ref{cs2}) vanish only if the distribution of $S^{({\rm CN \
fl})}_{a a}$ is Gaussian so that the phase of $S^{({\rm CN \ fl})}_{a
a}$ is uniformly distributed in the interval $\{0, 2 \pi \}$. That
condition is found to be violated~\cite{Die09c} already when $\Gamma /
d < 4$, and a full evaluation of all terms on the right--hand side of
Eq.~(\ref{cs2}) is then necessary.

This is possible with the help of the results of
~\textcite{Dav88,Dav89} who calculated analytically the functions
\begin{eqnarray}
F^{(4)}_{ab}(\ve)&=&\langle \left[S^{\rm fl *}_{a b}(E+\ve/2)\right]^2
\left[S^{\rm fl}_{a b}(E - \ve/2)\right]^2 \rangle \ , \nonumber\\
F^{(3)}_{a b}(\ve)&=&\langle S^{\rm fl *}_{a b}(E+\ve/2)
\left[S^{\rm fl }_{a b}(E - \ve/2)\right]^2 \rangle \, .
\label{cs3}
\end{eqnarray}
We note that for $\ve \neq 0$, these functions differ from the
expressions appearing on the right--hand side of Eq.~(\ref{cs2}).
However, numerical simulations and experimental data
show~\cite{Die09c} that for all values of $\Gamma / d$ the last curly
bracket in Eq.~(\ref{cs2}) (denoted by $C^{(4)}_{a b}(\ve)$) is well
approximated by $C^{(4)}_{ab}(0) F^{(4)}_{ab}(\ve) / F^{(4)}_{ab}(0)$,
and that similarly we have $\Re [ \langle S^{({\rm CN \ fl}) *}_{a
a}(E + \ve/2) |S^{({\rm CN \ fl})}_{a a}(E - \ve/2)|^2 \rangle ]
\approx F^{(3)}_{a a}(\ve)$. The last relation holds with good
accuracy except for the regime $\Gamma \approx d$ of weakly
overlapping resonances. With these results, Eq.~(\ref{cs2}) takes
the form
\begin{eqnarray}
{\cal C}_{a b}(\ve) &\approx& 2 \delta_{a b} \bigg\{ | \langle
S^{CN}_{a a} \rangle |^2 \ \Re [ C^{(2)}_{a a}(\ve) ] \nonumber \\
&+& \bigg( \langle S^{CN} _{a a} \rangle \ \Re [ F^{(3)}_{a b}(\ve)
+ F^{(3)}_{a b}(-\ve) ] \bigg) \bigg\} \nonumber \\
&+& \frac{C^{(4)}_{a b}(0)}{F^{(4)}_{a b}(0)} F^{(4)}_{a
b}(\ve) \ .  \label{cs4}
\end{eqnarray}
All terms in Eq.~(\ref{cs4}) are known analytically. Expressions
useful for a numerical computation are given in the Appendix
of~\cite{Die09c}. Thus, from a practical point of view theoretical
expressions for cross--section fluctuations are available for all
values of $\Gamma / d$ except for the elastic case where the
relation~(\ref{cs4}) does not hold for $\Gamma \approx d$.

\subsection{Poles of the $S$--Matrix}
\label{pole}

Resonances correspond to poles of the $S$--matrix. The distribution of
the poles of the stochastic $S$--matrix defined in Eq.~(\ref{C2})
has, therefore, attracted theoretical attention from the beginning.
Obvious questions are: How is the correlation width~(\ref{9}) related
to the distance of the poles from the real axis? Is it possible to
verify quantitatively the picture drawn of the pole distribution in
Section~\ref{phys}?

It was mentioned in Section~\ref{earl}
that~\textcite{Mol61,Mol63,Mol64,Mol69,Mol75,Mol76,Mol80} based his
approach to CN scattering on the pole expansion of the $S$--matrix. He
seems to have been the first author to determine the distribution of
poles numerically~\cite{Mol64}. He made a number of important
discoveries that stimulated later research. (i) There exists a gap
separating the poles from the real axis. (ii) For strong coupling to
the channels, some poles occur far away from the real energy
axis~\cite{Mol75}. (iii) A ``sum rule'' for resonance reactions
(see~\cite{Mol69} and further references therein) relates the
transmission coefficients and the mean distance of the poles from the
real axis.

Later work by various authors has led to a deeper understanding of
these results. We begin with the ``Moldauer--Simonius sum rule for
resonance reactions'' and follow the derivation due
to~\textcite{Sim74}. We assume that the unitary $S$--matrix defined in
Eq.~(\ref{C2}) has $N$ simple poles ${\cal E}_\mu$ only.  (Coincidence
of two poles is considered fortuitous and actually excluded by
quadratic repulsion of poles, see below). Then
\be
\det S^{\rm CN} = \exp \{ 2 i \phi \} \prod_{\mu = 1}^N \frac{E -
{\cal E}^*_\mu}{E - {\cal E}_\mu} \ .
\label{66}
\ee
The denominator on the right--hand side represents the $N$ poles. The
form of the numerator follows from the unitarity of $S^{\rm CN}$:
$\det [ (S^{\rm CN})^* ]$ is inverse to $\det [ S^{\rm CN} ]$. The
only energy dependence is due to the poles of $S^{\rm CN}$, and the
phase $\phi$ is, therefore, constant. With $\Im {\cal E}_\mu = - (1/2)
\Gamma_\mu$ we have
\be
\ln \det S^{\rm CN} - 2 i \phi = \sum_\mu \ln \bigg( 1 - i
\frac{\Gamma_\mu}{E - {\cal E}_\mu} \bigg) \ .
\label{67}
\ee
To average Eq.~(\ref{67}) over energy, we use a Lorentzian averaging
function with width $I$, see Section~\ref{calc}, and expand the
logarithm in powers of $\Gamma_\mu / (E - {\cal E}_\mu)$. Contour
integration shows that only the linear term gives a non--vanishing
contribution. With $I \gg \Gamma_\mu$ for all $\mu$ we find
\be
\bigg\langle \ln \bigg( 1 - i \frac{\Gamma_\mu}{E - {\cal E}_\mu}
\bigg) \bigg\rangle = - i \frac{\Gamma_\mu}{E - \Re({\cal E}_\mu) +
i I} \ .
\label{68}
\ee
This yields
\ba
&& \langle \ln \det S^{\rm CN} \rangle + \langle \ln (\det S^{\rm
CN})^* \rangle \nonumber \\
&& = - 2 \pi \sum_\mu \frac{\Gamma_\mu (I / \pi)}{(E -
\Re({\cal E}_\mu))^2 + I^2} = - 2 \pi \frac{\langle \Gamma_\mu
\rangle}{d} \ .
\label{69}
\ea
We use Eq.~(\ref{39a}) (i.e., the equality of $\langle \ln \det
S^{\rm CN} \rangle$ and of $\ln \det \langle S^{\rm CN} \rangle$)
and the definition~(\ref{6}) and obtain the Moldauer--Simonius sum
rule
\be
\langle \Gamma_\mu \rangle = - \frac{d}{2 \pi} \sum_c \ln (1 - T_c) \ .
\label{70}
\ee
Eq.~(\ref{70}) implies $\langle \Gamma_\mu \rangle \geq \Gamma$ where
$\Gamma$ is the correlation width given by the Weisskopf
estimate~(\ref{9}). Moreover, $\langle \Gamma_\mu \rangle$ diverges
whenever a single (or several) transmission coefficients approach
unity. The divergence is caused by the fact that one (or several)
pole(s) of $S^{\rm CN}$ is (are) shifted far below the real $E$--axis,
see Section~\ref{phys} and the text below. In deriving Eq.~(\ref{70})
we have assumed that the averaging interval $I$ is large compared to
$\Gamma_\mu$ for all $\mu$. It is not clear from the derivation
whether the Moldauer--Simonius sum rule applies as one of the $T$s
approaches unity. However, work on the unitary case (GUE)~\cite{Fyo96}
and on the single--channel case for the GOE~\cite{Som99} has shown
that the distribution of the poles acquires a tail that causes the
divergence as $T_c \to 1$.

The locations of the poles of $S$ are given by the eigenvalues of the
effective Hamiltonian~(\ref{29d}).~\textcite{Sok88,Sok89,Sok92} used
Eq.~(\ref{29d}) in an effort to extend statistical spectroscopy (as
based on the properties of $H^{\rm GOE}$) to resonances. Numerical
work of~\textcite{Kle85} had amplified Moldauer's observation that for
large coupling to the channels, one or several poles is (are) located
far from the real energy axis. Sokolov and Zelevinsky gave a
semiquantitative analytical explanation of that observation summarized
in Section~\ref{phys} and showed that at the same time, the majority
of poles moves back towards the real axis (``trapped states''). For
the single--channel case they derived the distribution of poles in the
complex energy plane. They connected the existence of one or several
poles far from the real energy axis with the phenomenon of
superradiance~\cite{Dic54} in quantum optics. They showed that the
non--Hermitean part of $H^{\rm eff}$ causes quadratic repulsion of the
poles in the complex plane, and that finding the density of poles of
$S$ in the complex energy plane is equivalent to the reconstruction of
the two--dimensional charge density from a given electrostatic field.

To determine analytically the joint probability density of the poles
in the complex energy plane for the $S$--matrix defined in
Eq.~(\ref{C2}), two approaches have been taken. If the limit $N \to
\infty$ is taken with $\Lambda$ fixed, terms of order $m = \Lambda /
N$ do not contribute to the saddle--point equation, neither in the
replica approach (see Section~\ref{replica}) nor in the supersymmetry
approach (see Section~\ref{general}). A model different from but
related to Eqs.~(\ref{C2}) introduced by~\textcite{Sok88,Sok89,Sok92}
makes it possible to overcome that limitation, and to discuss the pole
distribution in the framework of the saddle--point approximation. The
number $\Lambda$ of channels is assumed to be large and to go with $N$
to infinity while the ratio $m = \Lambda / N$ is held fixed. The
parameters $W_{a \mu}$ describing the coupling of level $\mu$ with
channel $a$ are taken to be Gaussian--distributed random variables
with mean value zero and common second moment
\be
\langle W_{a \mu} W_{b \nu} \rangle = \frac{d \gamma}{\pi^2} \delta_{a
b} \delta_{\mu \nu} \ .
\label{71}
\ee
The $\Lambda$ channels are all equivalent, and the resulting ensemble
of $S$--matrices is invariant with respect to orthogonal
transformations of the channel space. The dimensionless strength
$\gamma$ of the coupling is the only parameter. In contrast to the use
of $H^{\rm GOE}$ in Eq.~(\ref{29d}), the invariance of the
distribution of the $W$s under orthogonal transformations of the
channels cannot be deduced from quantum chaos and is, therefore,
somewhat arbitrary.  In~\cite{Leh95a} it was shown, however, that the
pole distribution obtained from the model~(\ref{71}) is quite similar
to the one where the $W$s are fixed. The limit $\Lambda \to \infty$
corresponds to the Ericson regime. However, by choosing $\gamma \ll
1$, one can approach the limit of weakly overlapping resonances.

The model~(\ref{71}) was used for an extensive discussion of the
distribution of poles of the $S$--matrix in~\cite{Haa92} (where the
replica trick was used) and in~\cite{Leh95a} (where supersymmetry was
applied). In both cases, the saddle--point equations differ from those
obtained for fixed channel number by terms of order $m = \Lambda /
N$. These equations are used to determine the average pole
distribution in the complex energy plane with the help of an
electrostatic analogy similar to the one mentioned above. Analytic
expressions are obtained for the boundary curve separating the area of
non--vanishing pole density from the empty one. Typical results are
shown below.

As mentioned above, Moldauer discovered a gap separating the poles of
$S$ from the real energy axis. Later,~\textcite{Gas89a,Gas89b} deduced
the existence of a gap in the framework of the semiclassical
approximation for chaotic systems with few degrees of freedom. That
suggests that the gap is a universal feature of chaotic scattering.
For the model of Eq.~(\ref{71}), the gap was shown to exist and the
gap parameters were worked out analytically in~\cite{Haa92,Leh95a}. In
the center of the GOE semicircle, the width $\Gamma_{\rm gap} / 2$ of
the gap separating the cloud of poles and the real axis is given
by~\cite{Leh95a} \be
\Gamma_{\rm gap} = \frac{d}{2 \pi} \Lambda \frac{4 \gamma}{1 +
\gamma^2} \ .
\label{72}
\ee
Using Eq.(\ref{40}) and $x_a = \gamma$ we can write the right--hand
side as $(d / (2 \pi)) \sum_c T_c$. This is the Weisskopf estimate.
In~\cite{Leh95a}, the $S$--matrix correlation width for the model of
Eq.~(\ref{71}) was also worked out and found to coincide with the one
found in the framework of Eqs.~(\ref{C2}) (fixed number of channels),
i.e., with the Weisskopf estimate. For the model of Eq.~(\ref{71}),
the equality of gap width and Weisskopf estimate holds unless $m$
approaches unity. Then, the correlation width becomes comparable to
and is modified by the range of the GOE spectrum. Put differently and
positively, the result implies that $S$--matrix fluctuations are
universal as long as there is a clear separation of the two energy
scales, the gap width and the range of the GOE spectrum.

\begin{figure}[ht]
	\centering
	\includegraphics[width=0.45\textwidth]{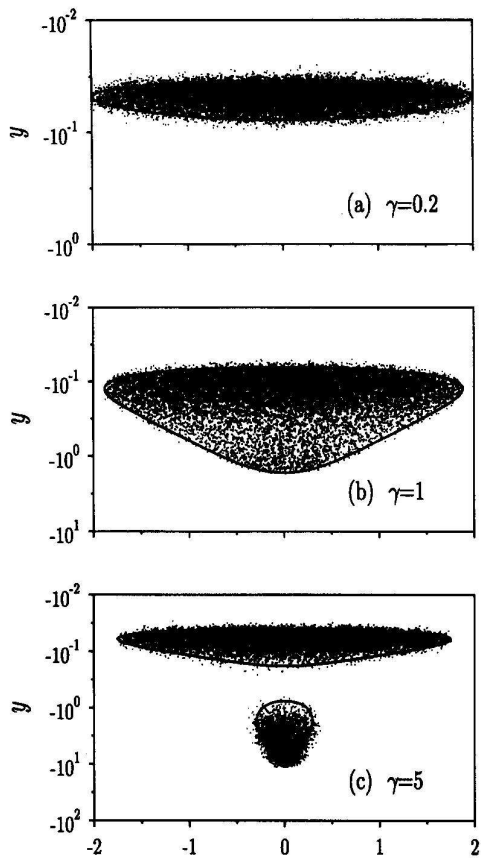}
	\caption{Distribution of the poles of the $S$--matrix in the
	complex energy plane. The distribution is shown for several
	values of the strength $\gamma$ of the average coupling to the
	channels. The abscissa spans the entire range of the spectrum.
	From~\textcite{Leh95a}.}
	\label{fig:HW9}
\end{figure}

The joint probability density of the poles of $S$ in the complex plane
can be obtained from the above--mentioned algebraic equations. With
$\Re (E) / \lambda = x$, $\Im (E) / \lambda = y$, Fig.~\ref{fig:HW9}
shows the distribution below the real $E$--axis. The single cloud seen
for $\gamma = 0.2$ and for $\gamma = 1$ splits into two as $\gamma$ is
increased further. The separation begins to develop at $\gamma =
1$. From the Moldauer--Simonius sum rule, we would expect that many
poles move to $- i \infty$ as $\gamma$ approaches unity. That is not
seen in the figure.  We ascribe the discrepancy to the assumption
(used in the derivation of the sum rule) that the entire energy
dependence of $S^{\rm CN}$ is due to the poles, see
Eq.~(\ref{66}). That assumption fails when the widths of the
resonances become comparable with the range of the GOE spectrum.

The second approach~\cite{Fyo96a,Fyo03} takes the limit $N \to \infty$
for fixed channel number $\Lambda$ and arbitrary values of the
transmission coefficients $T_c$ (``almost Hermitian matrices''). For
the unitary case (GUE Hamiltonian) it yields the complete joint
probability density of the eigenvalues of the effective Hamiltonian in
Eq.~(\ref{27c}).  Corresponding results for the GOE are not known
except for the single-channel case $\Lambda = 1$~\cite{Ull69,Sok89}.

The first experimental study~\cite{Kuh08} of the distribution of poles
of the scattering matrix in the complex energy plane employed a
microwave resonator and the method of harmonic analysis, see also
Section~\ref{sec:EricFluc}. The results agree with theoretical
predictions~\cite{Som99}.

In summary: The distribution of the poles of $S^{\rm CN}$ in the
complex energy plane gives valuable insight into the scattering
mechanism, even though that information is not sufficient to construct
the scattering amplitude(s). The Moldauer--Simonius sum rule shows
that in general the average distance of the poles from the real axis
is bigger than would be concluded from the Weisskopf
estimate~(\ref{9}). With increasing coupling to the channels, up to
$\Lambda$ poles are moved ever further away from the real axis, a fact
related to superradiance in quantum optics. An example for the actual
distribution of poles in the regime of many strongly coupled channels
($N \to \infty$, $\Lambda \to \infty$ and $m = \Lambda / N$ fixed) is
shown in Fig~\ref{fig:HW9}. A gap separates the cloud of poles from
the real $E$--axis. The width of the gap is given by the Weisskopf
estimate. The regime of almost Hermitean matrices has also been worked
out.

\subsection{Correlations of $S$--Matrix Elements carrying different
Quantum Numbers}
\label{corr}

We return to the question raised in the last paragraph of
Section~\ref{stoc}: Is the assumption~(\ref{3}) justified that
$S$--matrix elements carrying different quantum numbers are
uncorrelated?

While RMT {\it per se} is obviously not in a position to give an
answer to that question, a realistic large--scale shell--model
calculation would. We have in mind a calculation using the two--body
random ensemble (TBRE) of the nuclear shell model. The TBRE was
introduced in Section~I.V.B where details and references to the
original papers may be found. In the TBRE, several neutrons and
protons occupy the single--particle states of a major shell of the
shell--model and interact via a two--body interaction. The matrix
elements of that interaction are taken to be Gaussian--distributed
random variables. Spin is a good quantum number, and the need to
couple nucleon angular momenta and spins to good total spin creates
considerable complexity. Matrix elements of $S$ carrying different
spin quantum numbers would have to be calculated numerically for
different realizations of the TBRE, and their correlations worked
out. For reasons given below, the dimensions of the underlying
shell--model spaces would have to be very large. We are not aware of
any such calculation.

However, in the conclusions of~\cite{Pap08} the authors argue that for
sufficiently large shell--model spaces, the correlations between
$S$--matrix elements carrying different quantum numbers might be very
weak. The arguments are based on the study of a model~\cite{Pap08}
that is conceptually close to but technically simpler than the TBRE
and, therefore, analytically accessible. In the model parity is the
only quantum number, and $m$ spinless fermions occupy $\ell_1$
($\ell_2$) degenerate single-particle states with positive (negative)
parity, respectively. The fermions interact via a parity--conserving
two--body interaction with random Gaussian--distributed uncorrelated
two--body matrix elements. The $n$th moments ${\cal M}_n(\pm)$ of the
Hamiltonian H of the model (defined as normalized traces of $H^n$ with
positive integer $n$) can be worked out analytically for the
many--body states of both positive ($+$) and negative ($-$)
parity. The case of large matrix dimension is attained in the ``dilute
limit'' defined by $\ell_1, \ell_2,m \to
\infty$, $m / \ell_1 \to 0$, $m / \ell_2 \to 0$. It is shown that
${\cal M}_n(+) = {\cal M}_n(-)$ for all $n$ up to a maximum value that
is bounded from above by $m$ but tends to $\infty$ in the dilute
limit, and that the two moments differ ever more strongly when $n$
grows beyond that bound. That result shows that the spectra of states
with positive and negative parity are strongly correlated. In
particular, the shapes of the two average spectra are extremely
similar. At the same time, the result suggests that the local spectral
fluctuations of the two spectra are uncorrelated. Indeed, in their
work on the use of moments for nuclear spectroscopy, French and
collaborators concluded that such fluctuations are determined by the
very highest moments of the Hamiltonian~\cite{Bro81}.

The results just stated apply in the dilute limit only. They do not
contradict earlier findings~\cite{Pap07} for the TBRE on correlations
between spectra carrying different quantum numbers. These calculations
involved Hilbert spaces of small dimension only.

Assuming that a result similar to the one just stated holds in the
limit of large matrix dimension for the TBRE and observing that the
fluctuation properties of the $S$--matrix are caused by the local
spectral fluctuation properties of the underlying Hamiltonian (see
Eqs.~(\ref{20},\ref{21})), we conclude that within the framework of
the nuclear shell model, $S$--matrix elements carrying different
quantum numbers are likely to be uncorrelated, in agreement with
Eq.~(\ref{3}).

\section{Tests and Applications of the Statistical Theory}

The statistical theory reviewed in Sections~\ref{aver} and \ref{resu}
has been tested thoroughly. Moreover, it has found numerous
applications, both within the realm of nuclear physics and beyond. In
this Section we review some recent such tests and applications.

\subsection{Isolated and Weakly Overlapping CN resonances}
\label{wocr}

In the regime of isolated resonances, thorough tests of the
statistical theory of nuclear reactions ($\Gamma \ll d$) were
undertaken already many years ago. Especially for neutron resonances
there exists a comprehensive review~\cite{Lyn68}. In the regime of
weakly overlapping resonances, tests have so far not been performed in
nuclei. Here we do not summarize the early works but rather focus
attention on recent data and tests of the theory. These have become
possible in microwave billiards~\cite{Die08}. Such devices simulate
the CN and its resonances or, for that matter, any other chaotic
quantum scattering system. Indeed, in sufficiently flat microwave
resonators and for sufficiently low values of the radio frequency (rf)
--- so--called quantum billiards --- only one vertical mode of the
electric field is excited, and the Helmholtz equation is
mathematically equivalent to the Schr{\"o}dinger equation for a
two--dimensional quantum billiard~\cite{Sto90,Sri91,Gra92,So95,Ric99,
Sto00}. If the classical dynamics (free motion within the microwave
resonator and elastic scattering by its boundary) is chaotic, the
statistical properties of the eigenvalues and eigenfunctions of the
closed resonator in the quantum case follow RMT
predictions~\cite{Boh84}, and the scattering of rf amplitudes by the
resonator corresponds to quantum chaotic scattering.

\begin{figure}[ht]
	\centering
	\includegraphics[width=0.45\textwidth]{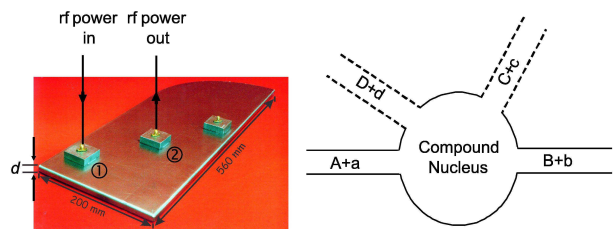}
	\caption{Flat microwave resonator (left--hand side) as a model
	for the compound nucleus (right--hand side). The height $d =
	0.84~{\rm cm}$ of the flat resonator makes it a chaotic
	quantum billiard up to a frequency of $18.75~{\rm GHz}$.}
	\label{fig:AR1}
\end{figure}

The left--hand side of Fig.~\ref{fig:AR1} shows a typical quantum
billiard realized in the form of a flat microwave resonator. For the
measurement of the spectrum, rf power is coupled via an antenna
labeled 1 into the resonator, thereby exciting an electric field mode
within the resonator, and the reflected output signal at the same
antenna (or the transmitted one at the antenna labeled 2) is
determined in magnitude and phase in relation to the input signal.
Hence, the resonator is an open scattering system where the antennas
act as single scattering channels. The scattering process is analogous
to that of a CN reaction as indicated schematically on the right--hand
side of Fig.~\ref{fig:AR1}. The incident channel $A + a$ consists of a
target nucleus $A$ bombarded by a projectile $a$ leading to a compound
nucleus which eventually decays after some time into the channel with
the residual nucleus $B$ and the outgoing particle $b$. (We disregard
angular momentum and spin). Attaching more antennas to the resonator
(or dissipating microwave power in its walls) corresponds to more open
channels $C + c$, $D + d$, \dots\ of the compound nucleus. The
resonator in Fig.~\ref{fig:AR1} has the shape of a so--called
Bunimovich stadium billiard which is known to be fully chaotic in the
classical limit~\cite{Bun85}.

\begin{figure}[ht]
	\centering
	\includegraphics[width=0.45\textwidth]{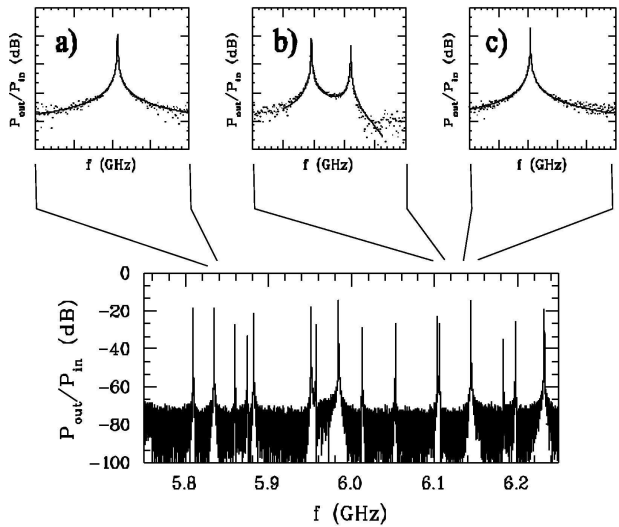}
	\caption{Part of the transmission spectrum of a
	superconducting microwave billiard. We note the extremely high
	resolution of the resonances. Two singlets (a) and (c) and a
	doublet (b) of resonances are magnified in the upper part. For
	all three cases $R$--matrix resonance formulas~\cite{Bec03}
	based on expressions from \textcite{Lan58} were fitted to
	the data. From~\cite{Dem05}.}
	\label{fig:AR2}
\end{figure}

The resonator in Fig.~\ref{fig:AR1} was made of niobium and operated
in a superconducting mode. This strongly increases the quality factor
of the resonator and yields very high resolution. The resonator was
used for both, the study of spectral properties, i.e., the statistics
of the resonances in the cavity~\cite{Gra92}, and a measurement of
their decay widths~\cite{Alt95}. The latter are proportional to the
square of the billiard eigenfunctions at the locations of the
antennas. Figure~\ref{fig:AR2} shows a transmission spectrum. Spectra
of that quality can be typically obtained in superconducting billiards
in the regime of isolated and weakly overlapping resonances. The
measured ratio of $P_{\rm out, b}$, the rf power signal transmitted
into antenna $b$, and of $P_{\rm in, a}$, the incoming rf power signal
at antenna $a$, is shown on a semilogarithmic plot. The ratio is equal
to $|S_{a b}(f)|^2$ where $S_{ab}(f)$ with $a,b = 1,2$ are the
elements of the complex--valued frequency--dependent $2 \times 2$
scattering matrix $S$. More generally, measurements of the modulus and
phase of the outgoing and the incoming signals performed with a
network vector analyzer determine magnitude and phase of all elements
$S_{ab}(f)$ of $S$. Such detailed information is not usually available
for other chaotic scattering systems where in general one can only
measure intensities. For a number of isolated resonances labeled $\mu$
with $\mu = 1, \dots, N$ without any background scattering the
$S$--matrix is a sum of Breit--Wigner terms,
\begin{equation}
	S_{ab} = \delta_{ab} - i \sum_\mu \frac{\Gamma_{\mu a}^{1/2}\,
	\Gamma_{\mu b}^{1/2} }{f - f_\mu + (i/2)\, \Gamma_\mu}\, .
	\label{eqn:AR1}
\end{equation}
Here $f_\mu$ and $\Gamma_\mu$ are the real and imaginary parts of the
eigenvalues of an effective Hamiltonian $\hat{H}_{\rm eff} = \hat{H} -
i\,\pi\,\hat{W} \hat{W}^\dagger$ for the microwave billiard in which
the $\hat{W}$s denote the coupling of the resonator states to the
antenna states and to the walls of the billiard, see
Section~\ref{reso}. With $\tilde{W}^{(0)}_{a \mu} \to \Gamma^{1/2}_{a
\mu}$, Eq.~(\ref{eqn:AR1}) is completely equivalent to Eq.~(\ref{22c})
which describes the coupling of isolated nuclear quasi--bound states
to the channels.

\begin{figure}[ht]
	\centering
	\includegraphics[width=0.45\textwidth]{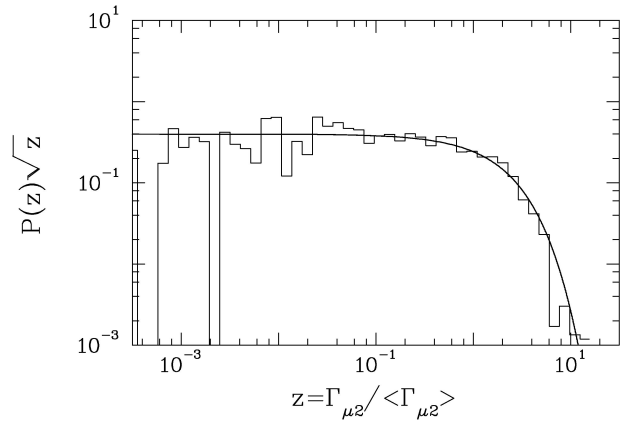}
	\caption{The experimental distribution of the the partial
	widths $\Gamma_{\mu 2}$ in units of their mean value
	$\langle \Gamma_{\mu 2} \rangle$. The solid line corresponds
	to a Porter--Thomas distribution. From~\cite{Alt95}.}
	\label{fig:AR4}
\end{figure}

As discussed below Eq.~(\ref{22c}) the partial width amplitude
$\Gamma_{\mu a}^{1/2}$ is the probability amplitude for decay of
resonance $\mu$ into channel $a$, and $\Gamma_\mu = \sum_a \Gamma_{\mu
a}$ is the total width. In an experiment with the superconducting
quantum billiard (shown on the left--hand side of Fig.~\ref{fig:AR1})
coupled to three antennas $c = 1,2,3$ (corresponding to a CN reaction
with three open channels), the complete set of resonance parameters
(resonance energies, partial widths, total widths) for 950 resonances
was measured~\cite{Alt95}. The total widths $\Gamma_\mu$ and the
partial widths $\Gamma_{\mu 2}$ are found to fluctuate randomly about
a slow secular variation with $\mu$, i.e., with frequency. The GOE
predicts a Gaussian distribution for the decay amplitudes $\Gamma_{\mu
c}^{1/2}$ or, equivalently, a $\chi^2$ distribution with one degree of
freedom (the Porter--Thomas distribution~\cite{Por56}, see Eq.~(I.18)).
The distribution of $\Gamma_{\mu 2}$ in Fig.~\ref{fig:AR4} exhibits
this behavior impressively.

\begin{figure}[ht]
	\centering
	\includegraphics[width=0.45\textwidth]{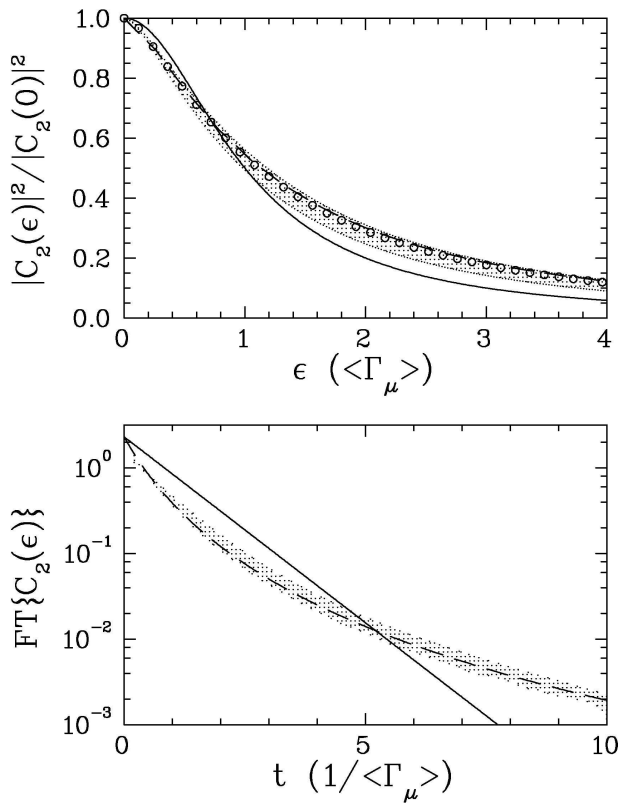}
	\caption{Upper part: The experimental autocorrelation function
	$| C_2(\varepsilon) |^2$ (circles within the shaded band of
	errors), the prediction of the statistical theory (dashed
	line), and a Lorentzian (solid line). Lower part: Fourier
	coefficients of the autocorrelation function (dots) with
	errors indicated by the shaded band together with the
	prediction of the statistical theory (dashed line) and an
	exponential (Fourier transform of a Lorentzian). The
	non--exponential decay in time of the Fourier transform of the
	$S$--matrix autocorrelation functions is clearly
	visible. From~\cite{Alt95}.}
	\label{fig:AR5}
\end{figure}

As a further test of the statistical theory, the autocorrelation
function of the $S$--matrix of the cavity, defined as
$C_c(\varepsilon) = \langle S_{cc}(f) \, S_{cc}^\ast(\varepsilon+f)
\rangle - | \langle S_{cc}(f) \rangle |^2$, was determined for
channels $c = 1, 2$, and $3$. The average is taken over frequency. For
$c = 2$ the result is plotted as circles in the upper part of
Fig.~\ref{fig:AR5}. The shaded band is a measure of the experimental
uncertainty of $C_2(\varepsilon)$. The shape of the shaded band
differs markedly but not unexpectedly~\cite{Lew91a} from that of a
Lorentzian with width $\langle \Gamma_\mu \rangle$ shown as a solid
line in the upper part of Fig.~\ref{fig:AR5}. The contribution of each
of the over 900 individual resonances in Fig.~\ref{fig:AR5} is, of
course, Lorentzian in shape with width $\Gamma_\mu$. However,
different resonances have different widths, and the average over all
resonances is not a Lorentzian. This happens only when the total
widths $\Gamma_\mu$ fluctuate strongly with $\mu$. Thus, both the
number of open channels and the absorption in the walls of the
microwave billiard (also contained in $\Gamma_\mu$) must be
small. Both conditions can be satisfied in experiments with
superconducting microwave resonators, but generally not in experiments
at room temperature. (In the microwave experiment of~\textcite{Dor90},
for instance, absorption was strong and consequently the data did not
display a non--Lorentzian line shape~\cite{Lew91}.) Clearly, the
observed non--Lorentzian shape is a quantum phenomenon: In the
semiclassical approximation, i.e., for many open channels, we have
purely exponential decay. The result displayed in the upper part of
Fig.~\ref{fig:AR5} is in quantitative agreement with the statistical
theory, i.e., with Eqs.~(\ref{60}) to (\ref{62}).

The Fourier transform (FT) of the autocorrelation function
$C_2(\varepsilon)$ shown in the lower part of Fig.~\ref{fig:AR5}
decays non--exponentially as a function of time. This feature, too,
reflects the non--Lorentzian shape of the autocorrelation function
shown in the upper part of Fig.~\ref{fig:AR5} and is in agreement with
the statistical model. We refer to the discussion in the paragraph
following Eq.~(\ref{62a}). One may say that in the experiment
non--exponential decay of a quantum system with chaotic dynamics has
been ``observed'' for the first time.

\begin{figure}[ht]
	\centering
	\includegraphics[width=0.45\textwidth]{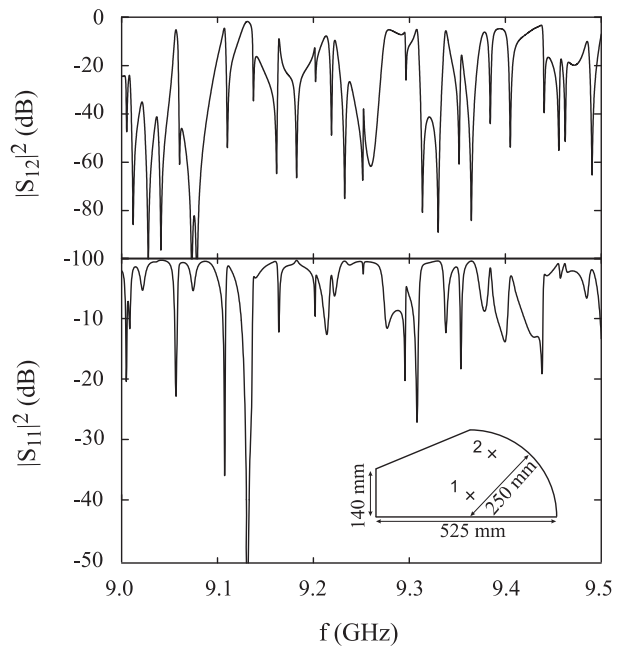}
	\caption{Transmitted (upper part) and reflected (lower part)
	intensity versus frequency between $9.0$ and $9.5~{\rm
	GHz}$. The resonances overlap and create a fluctuation
	pattern. The shape of the two--dimensional quantum billiard
	used in the experiment is shown in the inset. Points 1 and 2
	indicate the positions of the antennas. From~\cite{Die08}.}
	\label{fig:AR6}
\end{figure}

A thorough experimental investigation of chaotic scattering in
microwave billiards has recently also been performed in the regime of
weakly overlapping resonances ($\Gamma \lesssim d$), and the results
have been analyzed with the help of the formulas of
Section~\ref{general}. A normally conducting flat microwave resonator
made of copper was used as a quantum billiard~\cite{Die08}. The
resonator shown in the inset of Fig.~\ref{fig:AR6} has the shape of a
tilted stadium~\cite{Pri94}. The stadium has fully chaotic classical
dynamics. The resonator carried two antennas, and the complex elements
of the symmetric scattering matrix $S_{ab}(f)$ were measured versus
frequency. Figure~\ref{fig:AR6} shows examples of the measured
transmission ($|S_{12}|^2$) and reflection ($|S_{11}|^2$) intensities
as functions of resonance frequency. We note the strong fluctuations
of both quantities with frequency.

\begin{figure}[ht]
	\centering
	\includegraphics[width=0.45\textwidth]{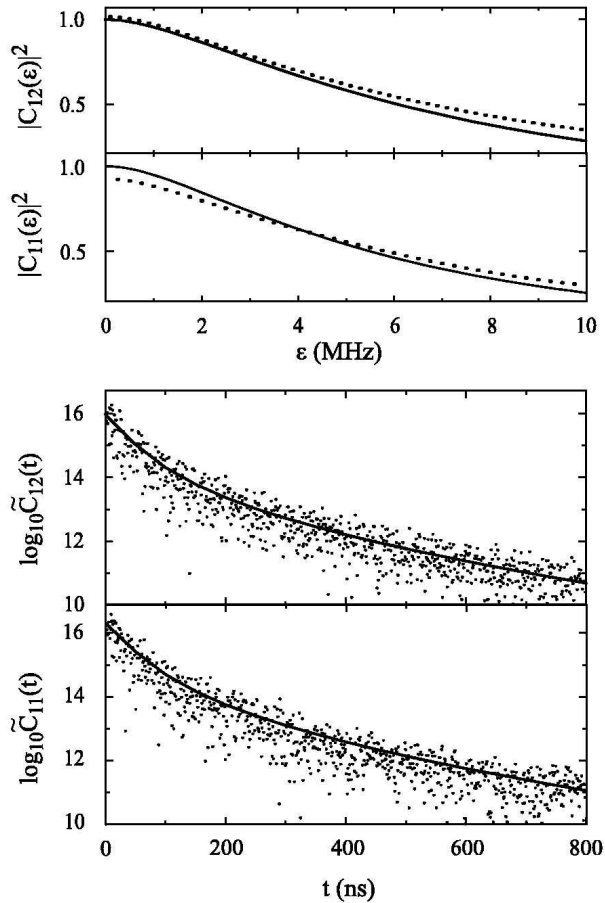}
	\caption{Upper part: The autocorrelation function $|C_{a
	b}(\varepsilon)|^2$ for values of $a$ and $b$ as
	indicated. The values of $|C_{a b}(\varepsilon)|^2$ were
	calculated from the measured $S$--matrix elements (points) and
	from the fit of Eq.~(\ref{60}) of the statistical theory to the
	data (solid line). Lower part: Fourier coefficients
	$\tilde{C}_{a b}(t)$ of the autocorrelation function (points)
	and the Fourier transform of the fit of $C_{a b}(\varepsilon)$
	(Eq.~(\ref{60})) to the data (solid line). The non--exponential
	decay in time in the region of weakly overlapping resonances
	($\Gamma/d \approx 0.2$) is striking. From~\cite{Die08}.}
	\label{fig:AR7}
\end{figure}

As in the case of isolated resonances the $S$--matrix autocorrelation
function $C_{a b} = \langle S_{ab}(f)\, S_{ab}^\ast(f+\varepsilon)
\rangle - | \langle S_{ab} \rangle |^2$ for $a, b = 1$ or $2$ was
computed from the data. The upper part of Fig.~\ref{fig:AR7} shows
that the values of the scattering matrix $S_{a b}(f)$ are correlated
with a correlation width $\Gamma$ of order several MHz. The values of
$S_{a b}(f)$ measured at $M$ equidistant frequencies with step width
$\Delta$ have been Fourier--transformed. The complex Fourier
coefficients are denoted by $\tilde{S}_{a b}(t)$ with $t \ge 0$. We
use the discrete time interval $t = d / M \Delta$ elapsed after
excitation of the billiard resonator instead of the Fourier index
$k$. The Fourier transform $\tilde{C}_{a b}(t)$ of
$C_{ab}(\varepsilon)$ has Fourier coefficients $| \tilde{S}_{a b}(t)
|^2$ and any two coefficients are uncorrelated random
variables~\cite{Eri65}. The lower part of Fig.~\ref{fig:AR7} shows
data for $\log_{10}\, \tilde{C}_{a b}(t)$. We note that the Fourier
coefficients scatter over more than five orders of magnitude. The
cutoff at $t = 800~{\rm ns}$ is due to noise.  The decay in time is
strikingly nonexponential, i.e., powerlike, as for isolated resonances
(Fig.~\ref{fig:AR5}).

The solid lines in Fig.~\ref{fig:AR7} result from fitting
Eqs.~(\ref{60}) to (\ref{62}) of the statistical theory to the data
points. The parameters in Eqs.~(\ref{60}) to (\ref{62}) for the
$S$--matrix autocorrelation function $C_{a b}(\varepsilon)$ are the
transmission coefficients $T_a$ with $a = 1, 2$ for the open antenna
channels, $T_c$ with $c=3, 4, \ldots$ for additional fictitious
channels modeling Ohmic absorption~\cite{Sch03} in the walls of the
normally conducting resonator, and the average level spacing $d$. For
$a = 1, 2$ the transmission coefficients were calculated from $T_a = 1
- | \langle S_{a a} \rangle |^2$ with data on $S_{a a}$ as input. The
average level spacing $d$ was calculated from the Weyl
formula~\cite{Bal76}. The product in Eq.~(\ref{60}) over the
ficticious channels was replaced by an exponential function of the sum
$\tau_{\rm abs}$ of the transmission coefficients of these channels.
This is a good approximation when all $T_c \ll 1$ and left $\tau_{\rm
abs}$ as the only free parameter in the fit. In order to allow for
secular variations of $\tau_{\rm abs}$ the experimental data were
analyzed in $1~{\rm GHz}$ intervals. It was found that the sum $T_1 +
T_2 + \tau_{\rm abs}$ increases from the value $0.11$ in the interval
$3$--$4~{\rm GHz}$ to the value $1.15$ in the interval $9$--$10~{\rm
GHz}$. (The resulting increase of $\tau_{\rm abs}$ is consistent with
conductance properties of copper.)  Using these numbers and the
Weisskopf estimate $\Gamma \approx [d/2\pi] \sum_c T_c$ of
Eq.~(\ref{9}) for the correlation width $\Gamma$ one finds that
$\Gamma/d$ increases from $0.02$ to $0.2$ over the same frequency
range. Thus, the statistical theory of chaotic scattering was indeed
experimentally tested in the regime of weakly overlapping resonances.
The Fourier coefficients turn out to be uncorrelated, Gaussian
distributed random variables. That fact and the large number of such
coefficients ($2400$ per frequency interval) made it possible to
assess the quality of the agreement between data and the fits in terms
of a Goodness--of--Fit (GOF) test, with excellent results~\cite{Die08}.

\begin{figure}[ht]
	\centering
	\includegraphics[width=0.45\textwidth]{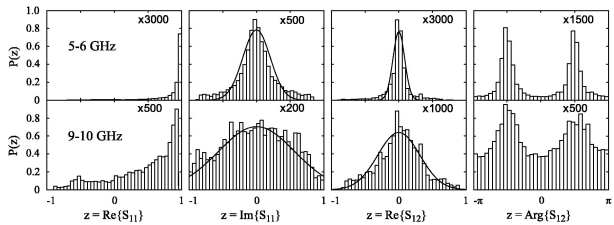}
	\caption{From left to right: Histograms for the scaled
	distributions of the real and imaginary parts of the
	reflection amplitude $S_{11}$ and the real part and the phase
	of the transmission amplitude $S_{12}$, respectively. The data
	were taken in the two frequency intervals $5$--$6~{\rm GHz}$
	(upper panels) and $9$--$10~{\rm GHz}$ (lower panels). The
	scaling factors are given in each panel. The solid lines are
	best fits to Gaussian distributions. From ~\cite{Die08}.}
	\label{fig:AR8}
\end{figure}

The experiment of \textcite{Die08} has produced interesting results
also on the distribution of moduli and phases of $S$--matrix elements
(see Section~\ref{sme}), and on the elastic enhancement factor, see
Sections~\ref{iso} and \ref{diag}. Concerning the first point, we
expect theoretically that for $\Gamma \ll d$ the distribution of
$S$--matrix elements is non--Gaussian. This is because unitarity
constrains the distribution of $S$--matrix elements. The constraints
are strongest for $\Gamma \ll d$, see Section~\ref{csf}.
Fig.~\ref{fig:AR8} shows that the distribution of the real part of
$S_{11}$ is strongly peaked near unity, especially for the lower
frequency interval from $5$ to $6~{\rm GHz}$ in which mostly isolated
resonances are found. But even in the regime of weakly overlapping
resonances, i.e., from $9$ to $10~{\rm GHz}$ the distributions of the
real and imaginary parts of $S_{11}$ deviate from Gaussians (solid
lines). The distributions of the phases (rightmost panels in
Fig.~\ref{fig:AR8}) are peaked. However, the valley between the two
peaks fills up as $\Gamma/d$ increases. For $\Gamma/d \gg 1$, i.e., in
the Ericson regime, the phases are expected to be uniformly
distributed and the $S$--matrix elements to be Gaussian
distributed. The elastic enhancement factor, defined as $W = \left(
\langle |S_{11}^{\rm fl}|^2 \rangle\, \langle |S_{22}^{\rm fl}|^2
\rangle \right)^{1/2} / \langle |S_{12}^{\rm fl}|^2 \rangle$, is
determined from the data as a function of $f$ in two ways. One may
either use the autocorrelation functions (Fig.~\ref{fig:AR7}) or the
widths of the distributions of the imaginary parts of the scattering
matrix. Both results agree and yield a smooth decrease of $W$ with $f$
from $W \approx 3.5 \pm 0.7$ for $4 \le f \le 5~{\rm GHz}$ to $W
\approx 2.0 \pm 0.7$ for $9 \le f \le 10~{\rm GHz}$, the errors being
finite--range--of--data errors.  The calculation of the enhancement
factor using Eqs.~(\ref{60}) to (\ref{62}) gives the values $W = 2.8$
and $2.2$ for the respective frequency intervals. We return to the
determination of $W$ once more in Section~\ref{time} below.

In summary, we have reviewed in this Section how a wealth of very
precise experimental data on chaotic quantum billiards that mimic the
CN and its resonances, can be used for a stringent test of the theory
of quantum chaotic scattering. Measurements of the phases of the
scattering matrix elements provide valuable additional information
that is usually not accessible in CN reactions. In the work of
\textcite{Alt95} on the statistics of partial widths of isolated
resonances ($\Gamma \ll d)$, all tests applied---Porter--Thomas
distribution, lack of correlations between partial widths in different
scattering channels, lack of correlations between partial widths and
resonance frequencies, non--Lorentzian decay of the $S$--matrix
autocorrelation function, and the related non--exponential time decay
of the chaotic quantum system---are in perfect agreement with GOE
predictions for the statistics of eigenfunctions and eigenvalues. In a
continuation of this work by \textcite{Die08} into the regime of
weakly overlapping resonances ($\Gamma \approx d$), the distributions
of $S$--matrix elements are found to be non--Gaussian while the
Fourier coefficients of these $S$--matrix elements do have an
approximately Gaussian distribution. These data were used for a highly
sensitive test of the statistical theory reviewed in
Sections~\ref{aver} and \ref{resu}. In particular, the predicted
non--exponential decay in time of isolated and weakly overlapping
resonances and the values of the elastic enhancement factors are
confirmed. The evidence for non--exponential decay in time, obtained
by Fourier--transforming measured $S$--matrix elements into time
space, is still indirect. A direct measurement of the decay time of an
excited nucleus might become possible at high--power laser facilities
such as the National Ignition Facility (NIF) where all the nuclear
resonances (or subsets of them) might be excited simultaneously by a
short laser pulse~\cite{Mos09}.

\subsection{Strongly Overlapping CN resonances: Ericson \\ Fluctuations}
\label{sec:EricFluc}

In the 1960s and 1970s, the newly discovered phenomenon of Ericson
fluctuations formed a central part of research in nuclear reactions,
see Section~\ref{eri}. Protons, deuterons and light ions up to oxygen
or so, but also fast neutrons have been used as projectiles in various
CN reactions. The field has been reviewed early~\cite{Eri66} and again
later~\cite{Ric74}. Both articles show that all of Ericson's
predictions were confirmed experimentally. We do not reiterate here
what has been known for many years about cross--section
autocorrelation functions, their Lorentzian shape, the mean coherence
width $\Gamma$, the vanishing of cross--correlation functions between
cross sections in different reaction channels in the absence of direct
reaction contributions, Hanbury Brown--Twiss behavior of
cross--correlation functions of cross sections measured at different
scattering angles, probability distributions of randomly fluctuating
cross sections, and about the interplay between direct and CN reaction
processes in general. From the few experiments performed lately in
nuclear physics, two are chosen to exemplify over and above what has
been stated in Section~II.D why Ericson fluctuations are now commonly
viewed as a paradigm for chaotic behavior of a quantum system.

\begin{figure}[ht]
	\centering \includegraphics[width=0.45\textwidth]{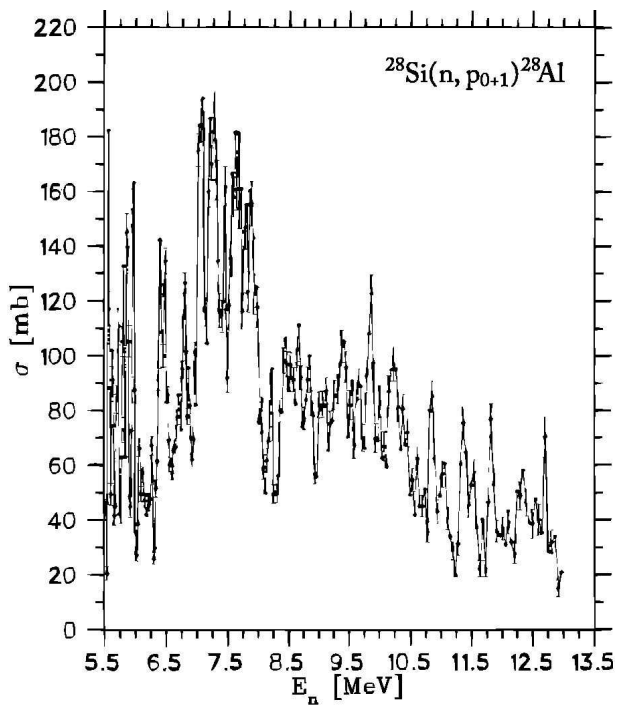}
	\caption{Fluctuating excitation function of the reaction
	$^{28}{\rm Si}(n, p_{0+1})^{28}{\rm Al}$ in the regime of
	strongly overlapping resonances ($\Gamma \gg d$) for the CN
	$^{29}{\rm Si}$. The cross sections of proton exit channels
	$p_0$ and $p_1$ leading to the ground and first excited states
	in $^{28}{\rm Al}$, respectively, were not resolved.
	From~\cite{Bat97}.}  \label{fig:AR9}
\end{figure}

Recently, a number of mainly neutron--induced CN reactions on
medium--heavy nuclei has been studied in the Ericson regime, primarily
in order to deduce nuclear level densities from the
data~(\textcite{Gri02} and Refs. therein). A striking example, the
excitation function for the reaction $^{28}{\rm Si}(n,
p_{0+1})^{28}{\rm Al}$, is shown in
Fig.~\ref{fig:AR9}~\cite{Bat97}. Decay of the CN $^{29}{\rm Si}$
populates the ground state and the first excited state of the final
odd--odd nucleus $^{28}{\rm Al}$. The two proton channels $p_{0+1}$
are not resolved. The high--resolution measurement of the CN cross
section reveals significant fluctuations with energy. The peaks and
minima of the excitation function are not caused by individual, more
or less isolated resonances, but instead result from the constructive
(or destructive) superposition of many overlapping CN resonances. The
amplitudes of the resonances are random variables. Therefore, the
curve connecting the measured points in Fig.~\ref{fig:AR9} has the
curious feature of being both reproducible and random~\cite{Wei90}. A
measurement of the same reaction in the same energy interval with the
same energy resolution will reproduce Fig.~\ref{fig:AR9}. Nonetheless,
the energy dependence of the cross section in the figure displays the
features of a random process.

The second example for the role of Ericson fluctuations in nuclei is
from the field of giant resonances where the question of direct versus
statistical decay plays a central role~\cite{Bor98,Har01}. The giant
resonance is a ``distinct'' and ``simple'' mode of excitation of the
nuclear ground state whose amplitude is usually spread over many
``complicated'' states for which the distinct mode acts as a
``doorway''. The strength function is typically of Breit--Wigner
shape~\cite{Boh69}. Doorway states were reviewed in Section~I.II.G. To
describe the approach, we have to distinguish several contributions to
the total width $\Gamma_0$ of the giant resonance. In good
approximation~\cite{Goe82} $\Gamma_0$ can be written as $\Gamma_0 =
\Delta \Gamma + \Gamma^\uparrow + \Gamma^\downarrow$. Here $\Delta
\Gamma$ stands for the Landau damping of the giant
resonance. (Electric dipole excitation of the ground state in the
first step produces a coherent superposition of one particle--one hole
(1p-1h) states that have different single--particle energies.) The
1p-1h states in turn couple to the continuum and acquire an escape
width $\Gamma^\uparrow$ which gives rise to a direct decay
contribution into dominant hole states of the daughter nucleus.
Finally, $\Gamma^\downarrow$ describes the spreading width resulting
from mixing the 1p-1h states with more complex 2p-2h and further
$n$p-$n$h configurations until an equilibrated compound nucleus is
reached. A primary goal of all giant--resonance high--resolution decay
experiments is, thus, to determine the relative contributions of the
widths $\Gamma^\uparrow$ and $\Gamma^\downarrow$ to the total width
$\Gamma_0$. The experimental signature of direct nuclear decay of a
giant resonance in a nucleus with mass number $A$ is the enhanced
population of hole states in the daughter nucleus $A - 1$. Statistical
decay can be identified either by a comparison of the measured decay
spectrum with predictions of the Hauser--Feshbach formula~(\ref{8}) or
by an Ericson--fluctuation analysis of the fine structure in the decay
spectrum measured with high resolution. Here we address the second
possibility.

\begin{figure}[ht]
	\centering
	\includegraphics[width=0.45\textwidth]{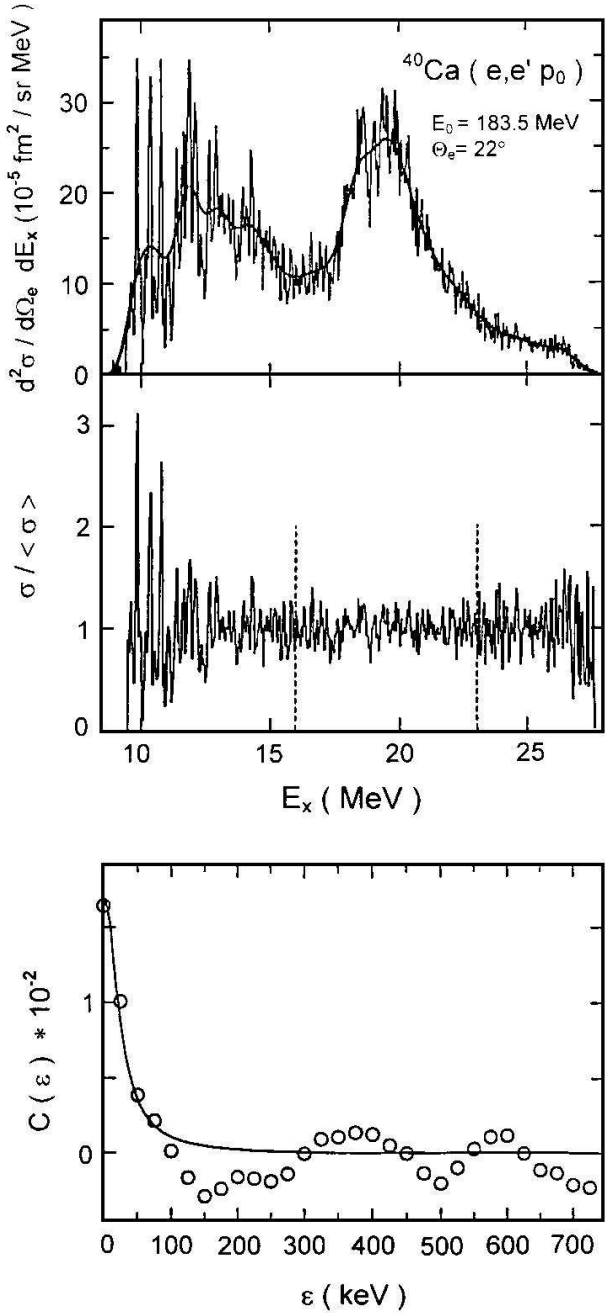}
	\caption{Upper part: Double differential cross section of the
	$^{40}{\rm Ca}(e, e' p_0)$ reaction at an electron energy $E_0
	= 183.5~{\rm MeV}$. Middle part: Ratio of the measured cross
	section and the smoothed cross section. Lower part:
	Autocorrelation function. From~\cite{Car01}.}
	\label{fig:AR10}
\end{figure}

Giant--resonance spectroscopy of $^{40}{\rm Ca}$ has recently been
performed through exclusive electroexcitation experiments of the type
$^{40}{\rm Ca}(e, e'x)$ where $x$ stands for either protons or alpha
particles detected in coincidence with the scattered
electron~\cite{Die01a,Die01b,Car01}. The description of the $(e,
e'x)$ reaction is based on the one--photon exchange mechanism. The
incident electron is inelastically scattered on $^{40}{\rm Ca}$, and a
virtual photon with energy $E_x$ and momentum $\vec{q}$ is transferred
to the $^{40}{\rm Ca}$ nucleus exciting it into the giant--resonance
region. The excited nucleus propagates in time and finally decays into
a residual nucleus ($^{39}{\rm K}$, $^{36}{\rm Ar}$) by emitting
particle $x$. It is assumed that excitation and decay can be treated
independently.  The upper part of Fig.~\ref{fig:AR10} shows the double
differential cross section for electrons that are scattered
inelastically on $^{40}{\rm Ca}$ and measured in coincidence with
protons that leave the residual nucleus $^{39}{\rm K}$ in its ground
state. The electrons were detected at an angle $\Theta_e =
22^\circ$. The protons were measured under various angles. The total
yield (integrated over all angles) was determined. The range of
excitation energies $E_x$ in $^{40}{\rm Ca}$ is $E_x \approx
10$--$27~{\rm MeV}$. For $E_x$ between $10$ and $15~{\rm MeV}$, a
number of isolated states is observed but the most prominent
excitation is clearly the MeV--wide peak at $E_x \approx 19~{\rm
MeV}$. It is the electric giant dipole resonance in $^{40}{\rm
Ca}$. Superposed upon the peak is considerable fine structure due to
Ericson fluctuations of the underlying overlapping CN resonances. Such
fluctuations have been seen already some time ago~\cite{Die73} in the
reaction $^{39}{\rm K}(p, \gamma_0)^{40}{\rm Ca}$. The original
spectrum has been smoothed with a Gaussian of ${\rm FWHM} = 800~{\rm
keV}$ (continuous solid line). The middle panel shows the ratio of the
actual cross section and the smoothed one. The data fluctuate around
the value unity. The autocorrelation function shown in the lower part
was computed from the fluctuating cross section within $16~{\rm MeV}
\leq E_x \leq 23~{\rm MeV}$ (dashed lines in the middle part of the
figure). The Lorentzian predicted by Ericson, $C(\varepsilon) = C(0)\,
(\Gamma^\downarrow)^2 / ((\Gamma^\downarrow)^2 + \varepsilon^2)$, was
fitted to the experimental points (open circles). This determines the
spreading width with a value between about $15$ to $30$ keV, depending
on the method of averaging~\cite{Die73,Car01}. The scatter of the
points results from the finite range of the data~\cite{Ric74}.

The value of $C(\varepsilon)$ at $\varepsilon=0$, i.e.\ the normalized
variance $C(0) = \langle \sigma^2 \rangle/\langle \sigma \rangle^2 -
1$ is related to the direct part $y_d = | \langle S \rangle |^2 /
\langle | \langle S \rangle + S^{\rm fl} |^2 \rangle$ for the reaction
feeding the proton decay channel (with $S = \langle S \rangle + S^{\rm
fl}$) through the equation $C(0) = (1/n\,N)\, (1-y_d^2)$. Here, $N$
corresponds to the effective number of spin channels contributing to
the reaction, and $n$ describes a damping factor due to the finite
experimental energy resolution. The detailed analysis of the
fluctuating cross sections in the $^{40}{\rm Ca}(e, e' p)^{39}{\rm K}$
reaction yields as fraction $y_d$ of the direct cross section a value
between about $85$--$95~\%$ for the $p_0$ and $p_1$ decay into the
ground and first excited states in $^{39}{\rm K}$, respectively. From
the point of view of the shell model, these states are dominated by
the $1d_{3/2}^{-1}$ and $2s_{1/2}^{-1}$ single--particle
configurations, respectively. Furthermore, the ratio of the cross
sections for these two decay channels is close to the ratio of the
single--nucleon transfer spectroscopic factors from the $^{40}{\rm
Ca}(d, ^3{\rm He})^{39}{\rm K}$ reaction~\cite{Dol76}. Thus, the
analysis of Ericson fluctuations~\cite{Car01} has shown that in
$^{40}{\rm Ca}$, the escape width $\Gamma^\uparrow$ for direct proton
emission of the electric giant dipole resonance feeding low--lying
states of $^{39}{\rm K}$, is considerably larger than the spreading
width $\Gamma^\downarrow$. We note that for the nuclear giant dipole
resonance, the ratio $\Gamma^\downarrow / \Gamma^\uparrow$ strongly
increases with mass number so that $\Gamma^\downarrow$ dominates in
heavy nuclei.

\begin{figure}[ht]
	\centering
	\includegraphics[width=0.45\textwidth]{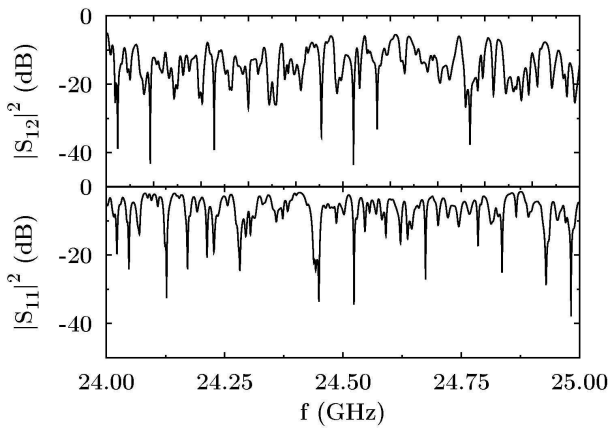}
	\caption{Same as in Fig.~\ref{fig:AR6} but in the Ericson
	regime of overlapping resonances. From~\cite{Sch09}.}
	\label{fig:AR11}
\end{figure}

\begin{figure}[ht]
	\centering
	\includegraphics[width=0.45\textwidth]{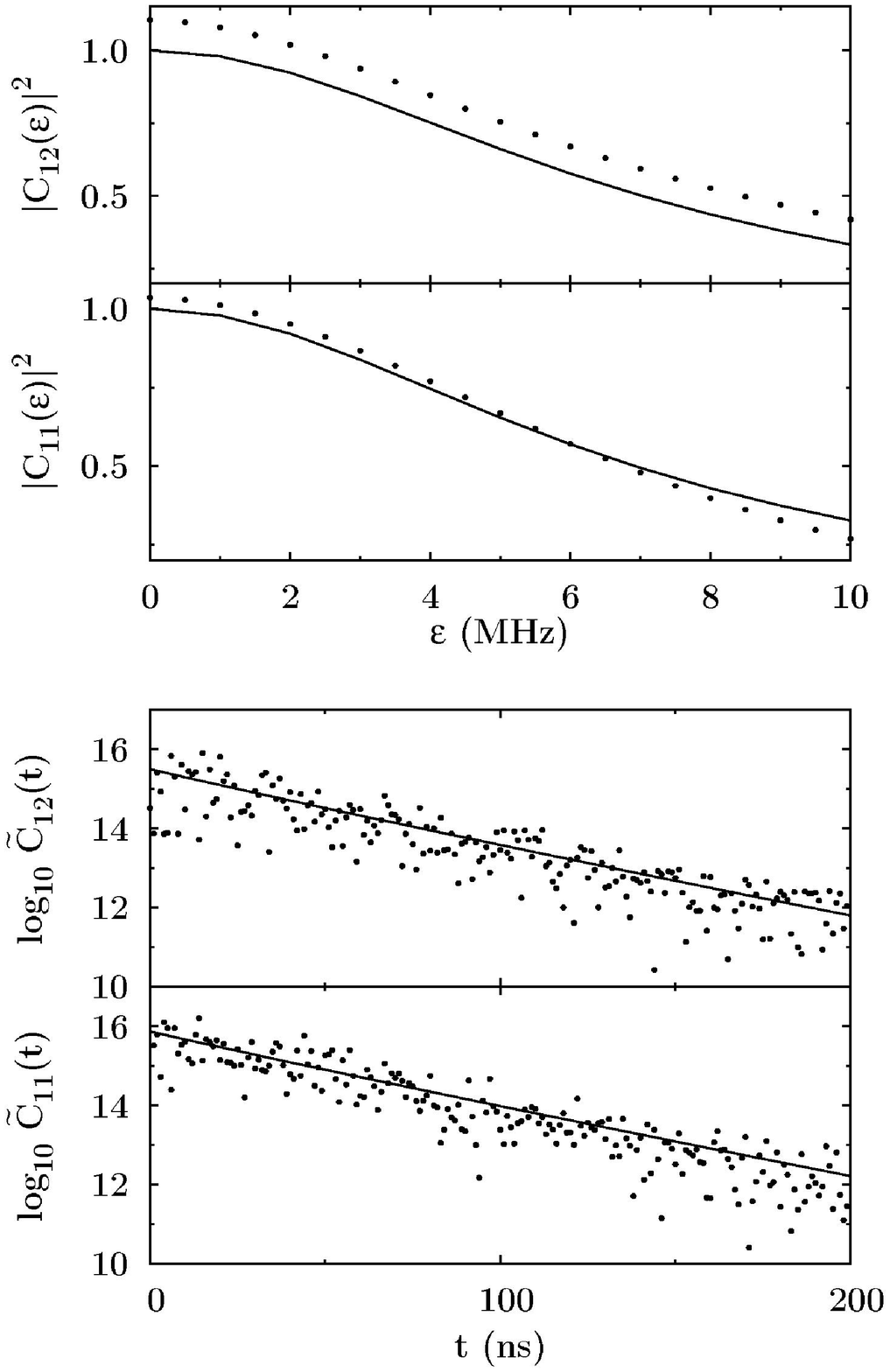}
	\caption{Autocorrelation functions (upper part) and Fourier
	coefficients (lower part) with a fit of Eqs.~(\ref{60}) to
	(\ref{62}) of the statistical theory. Same as in
	Fig.~\ref{fig:AR7} but for the Ericson regime. The exponential
	decay in time in the region of overlapping resonances
	($\Gamma/D \approx 1.06$) is striking. From~\cite{Sch09}.}
	\label{fig:AR12}
\end{figure}

\begin{figure}[ht]
	\centering
	\includegraphics[width=0.45\textwidth]{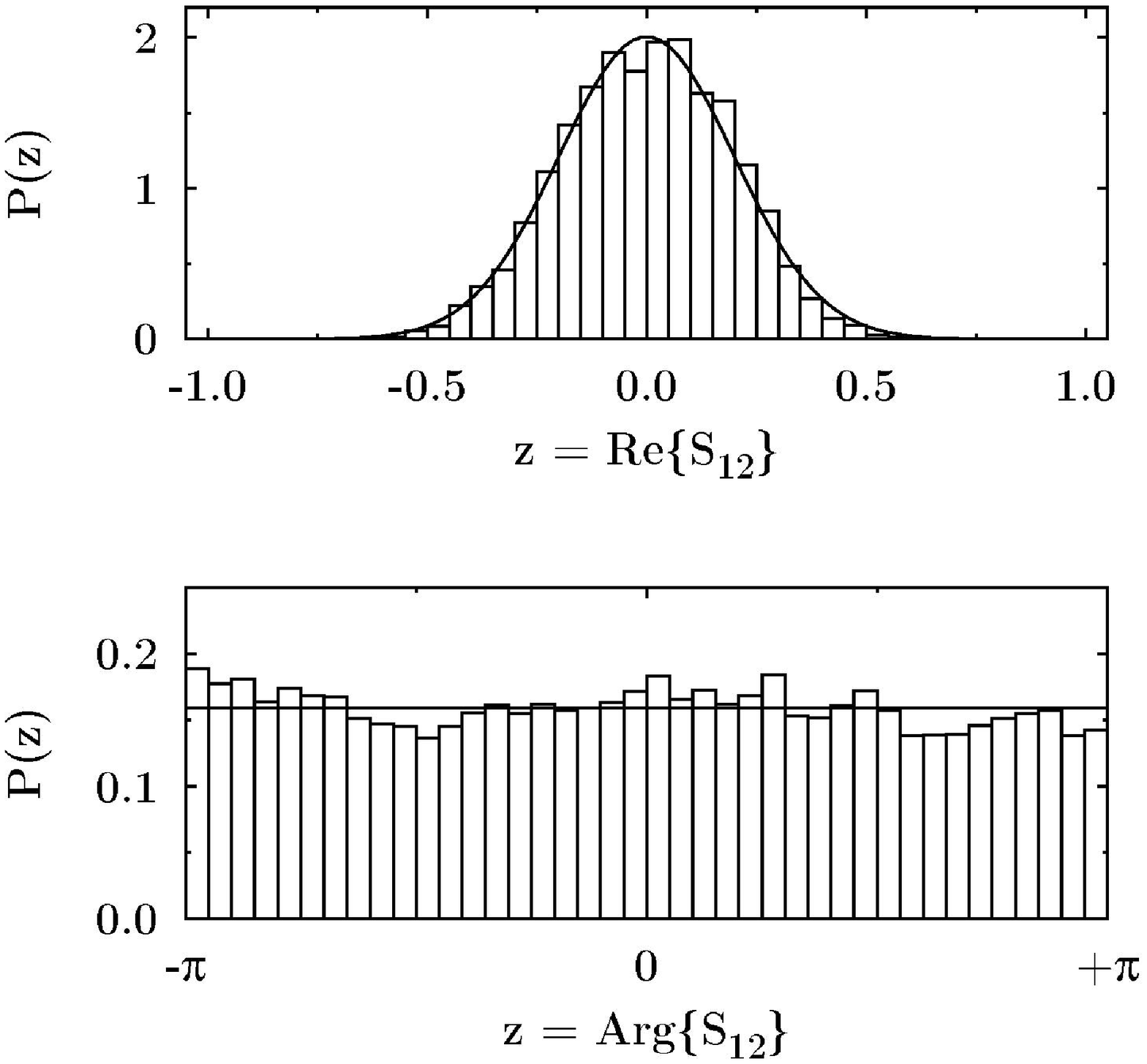}
	\caption{Upper part: Histogram for the distribution of the
	real part of the matrix element $S_{12}$. Lower part:
	Histogram for the distribution of the phase of $S_{1 2}$. The
	data are taken in the Ericson regime ($\Gamma/D \approx
	1.06$). From~\cite{Sch09}.}
	\label{fig:AR13}
\end{figure}

We return once more to the analogy between a flat chaotic microwave
resonator---a quantum billiard---and a CN. The data on billiards
described so far relate to the cases of isolated or weakly overlapping
resonances. An extension of the measurements with the chaotic tilted
stadium billiard (inset of Fig.~\ref{fig:AR6}) into the Ericson regime
($\Gamma > d$) was reported by~\textcite{Die09b}. The absolute squares
of the strongly fluctuating matrix elements $S_{12}$ and $S_{11}$,
respectively, taken at high excitation frequencies, are plotted versus
frequency in Fig.~\ref{fig:AR11}. The autocorrelation functions and
their Fourier transforms are shown in Fig.~\ref{fig:AR12}. The data
show that already for a value of $\Gamma/D \approx 1.06$ the system
decays exponentially in time. Again, the decay pattern and the
autocorrelation functions of the $S$--matrix elements in
Fig.~\ref{fig:AR12} are very well described by the statistical theory.
Furthermore, as shown in Fig.~\ref{fig:AR13} for $S_{12}$, the
$S$--matrix elements have a Gaussian distribution, and the phases are
uniformly distributed in the interval $\{0, 2 \pi \}$~\cite{Die09b}.
We note that the doorway--state phenomenon can also be modelled in
terms of a microwave billiard~\cite{Abe08}.

\textcite{Kuh08} have taken a different approach to chaotic scattering
in microwave cavities in the Ericson regime. They used the method of
harmonic analysis to determine the locations of the poles of $S$ in
the complex energy plane. The resulting width distribution was
compared with theoretical results by~\textcite{Som99}.

These examples from recent experiments together with the many others
summarized in the earlier reviews~\cite{Eri66,Ric74} should suffice to
demonstrate the importance of Ericson fluctuations in nuclei. The
phenomenon occurs, however, also in other quantum systems.
\textcite{Blu88} analyzed numerically the effect of irregular
classical scattering on the corresponding quantum--mechanical
scattering matrix. Using semiclassical arguments they showed that the
fluctuations of the $S$--matrix and the cross section are consistent
with Ericson fluctuations.~\textcite{Wei90} and~\textcite{Sor09}
compared universal conductance fluctuations of mesoscopic systems in
the metallic regime with Ericson fluctuations of CN cross sections and
pointed out the common stochastic features of the resonances in both
phenomena but also the differences. These arise because the
conductance is a sum over many channels, and because the length of the
mesoscopic system can be varied continuously. \textcite{Main1992}
reported exact quantum calculations for the photoionization cross
sections of the hydrogen atom in crossed magnetic and electric fields
and found strong Ericson fluctuations as a characteristic feature of
chaotic scattering. A feature article in physical chemistry by
\textcite{Rei96} emphasized the manifestation of interfering
overlapping resonances and Ericson fluctuations in the unimolecular
dissociation reaction of ${\rm NO}_2$ molecules into ${\rm NO} + {\rm
O}$ final states. They treat this process as resonance scattering
within the formalism of random--matrix theory. In a pioneering work,
\textcite{Sta05} reported the first experimental observation of
Ericson fluctuations in atomic and molecular systems. Quantum chaotic
scattering was studied in $^{85}{\rm Rb}$ in strong crossed magnetic
and electric fields in an energy regime beyond the ionization
threshold. The impressive experimental results of photoexcitation
cross sections were later supported by exact numerical
calculations~\cite{Mad05}. Very recently, Ericson fluctuations have
also been predicted for the inelastic electron scattering cross
section on a helium atom near the double ionization
threshold~\cite{Xu08}. The universality of Ericson fluctuations in
quantum chaotic scattering is, thus, very well established.

\section{Violation of Symmetry or Invariance}

As mentioned towards the end of Section~\ref{stoc}, the theoretical
framework of Eqs.~(\ref{C1},\ref{C2},\ref{C3}) and (\ref{C4}) is quite
flexible and allows for extensions of the theory that describe
violation of isospin symmetry, of parity, or of time--reversal
invariance in CN reactions. These are reviewed in turn. In the case of
both isospin and parity violation, the statistical theory is taken for
granted and used as a means to analyze the data. The same is true for
tests of time--reversal invariance in nuclear reactions. But in
experiments on induced violation of time--reversal symmetry in
microwave billiards, the underlying theoretical framework has been
thoroughly tested.

\subsection{Isospin}
\label{isos}

Violation of isospin symmetry is treated by replacing $H^{\rm GOE}$ in
Eq.~(\ref{C4}) by a Hamiltonian $H$ with block structure~\cite{Ros60}
as done in Eq.~(I.30). We simplify the presentation and consider only
two classes of states with different isospins $T_1$ and $T_2$ (where
typically we have $T_2 = T_1 + 1$) containing $N_1$ ($N_2$) elements,
respectively. The generalization to more classes with different
isospins is straightforward but is not needed in practice. The matrix
representation of $H$ has the form
\be
\left( \matrix{ H^{(1)}_{\mu \nu} & V_{\mu \sigma} \cr
       V_{\rho \nu}    & H^{(2)}_{\rho \sigma} \cr} \right) \ .
\label{isos1}
\ee
Here $H^{(1)}$ ($H^{(2)}$) are the Hamiltonian matrices acting on
states that carry the isospin quantum numbers $T_1$ and $T_2$,
respectively; each is taken from the GOE. The isospin--breaking
interaction is represented by the rectangular matrix $V$ residing
in the non--diagonal blocks. The elements of $V$ are due to the
Coulomb interaction between protons and to other isospin--violating
effects, see Section~I.III.D.1, and are typically small compared to
those of $H^{(1)}$ and $H^{(2)}$. A similar model is also used to
describe parity mixing, with $H^{(1)}$ and $H^{(2)}$ now denoting the
Hamiltonian matrices of states of positive and negative parity,
respectively, and $V$ the induced parity--violating nucleon--nucleon
interaction.

The density of states with isospin $T_1$ is usually considerably
larger than that of states with isospin $T_2 = T_1 + 1$. Isospin
mixing has been extensively investigated for two cases: (i) In the
Ericson regime. For both isospin classes the resonances overlap
strongly. (ii) For isobaric analogue resonances. Resonances with
isospin $T_2$ are well separated. Because of isospin mixing, each acts
as a doorway for the weakly overlapping resonances with isospin
$T_1$. We deal with both cases in turn.

\subsubsection{Ericson Regime}

Isospin violation in the Ericson regime (strongly overlapping
resonances for both values $T_1$ and $T_2$ of isospin) was extensively
reviewed in~\cite{Har86a}. Since then, essential new developments
have not occurred either in theory or in experiment. We confine
ourselves here to a brief summary.  The only recent experimental
information on isospin mixing in very highly excited CN is from
measurements of $\gamma$--ray spectra in heavy--ion fusion reactions
which we address below. We first sketch the modifications of the
general framework of Sections~\ref{stoc}, \ref{aver} and \ref{resu}
that are required if two values of isospin contribute to the
reaction. We then turn to a summary of the main experimental results.

In order to model isospin violation theoretically in the Ericson
regime one must, in addition to Eq.~(\ref{isos1}) for the Hamiltonian,
also specify the isospin properties of the channels. The physical
channels labeled $(a t)$ carry the quantum number $t$. It is given by
the projection of total isospin onto some axis and equals half the
difference of neutron and proton numbers in both fragments (projectile
plus target). The background matrix $S^{(0)}$ that describes
scattering in the absence of resonances and is given by Eq.~(\ref{C1})
as $S^{(0)} = U U^T$, is assumed to be diagonal with respect to the
physical channels, $S^{(0)}_{a t, b t'} = \delta_{a b}
\delta_{t t'} \exp \{ 2 i \delta^{(0)}_{a t} \}$. In other words, in
Eq.~(\ref{30}) we put both ${\cal O}^{(0)}$ and ${\cal O}^{\rm CN}$
equal to unit matrices. Then the physical $S$--matrix differs from the
matrix $S^{\rm CN}$ in Eq.~(\ref{C2}) only by phase factors; these are
suppressed in what follows. Our assumption neglects direct transitions
between ``mirror'' channels that are related by neutron--proton
symmetry; in~\cite{Har86a} it is shown that in the Ericson regime
that neglect is irrelevant.

Although the theory~\cite{Har86a} is more general, we focus attention
here on the main mechanism of isospin mixing. It is due to ``internal
mixing'' induced by the matrix elements of $V$ in Eq.~(\ref{isos1}).
Because of charge effects, isospin mixing also occurs in the channels
but is negligible. The coupling matrix elements $W^T_{a \mu}$ in
Eq.~(\ref{C2}) carry the additional isospin quantum number $T$ and so
do the transmission coefficients labeled $\tau_{a T}$ (we deviate here
from our standard notation to distinguish the transmission
coefficients from the isospin quantum number). In the physical
channels the transmission coefficients $\tau^{t t'}_{a T}$ are not
diagonal and are given by projecting (with the help of
angular--momentum coupling coefficients $\langle a T | a t \rangle$)
the transmission coefficient $\tau_{a T}$ onto $t, t'$ by $\tau^{t
t'}_{a T} = \langle a T | a t \rangle \langle a T | a t' \rangle
\tau_{a T}$. The autocorrelation function of $S^{\rm fl}$ is then
given by \ba && \langle S^{\rm fl}_{a t, b t'}(E_1) (S^{\rm fl}_{c
t'', d t'''}(E_2))^* \rangle = \delta_{a c} \delta_{b d} \sum_{m n}
\tau^{t t''}_{a m} \Pi_{m n} \tau^{t' t'''}_{b n} \nonumber \\ &&
\qquad \qquad + \delta_{a d} \delta_{b c} \sum_{m n} \tau^{t t'''}_{a
m} \Pi_{m n} \tau^{t' t''}_{b n} \ .
\label{isos8}
\ea
Here $\Pi_{m n}$ is a $2 \times 2$ matrix in isospin space. With $\ve
= E_2 - E_1$, the inverse of $\Pi$ is given by
\ba
\Pi^{-1} = \left( \matrix{ {\cal N}_1 + z + 2 i \pi \ve / d_1 & - z
\cr - z & {\cal N}_2 + z + 2 i \pi \ve / d_1 \cr} \right) \ .
\label{isos9}
\ea
The average level spacing for the states with isospin $T_m$, $m = 1,
2$ is denoted by $d_m$, and ${\cal N}_m = \sum_a \tau_{a T_m}$.
Eqs.~(\ref{isos8}) and (\ref{isos9}) generalize Eq.~(\ref{52}) for the
Ericson regime to the case of isospin mixing. The strength of isospin
mixing is characterized by a single dimensionless parameter $z$ given
by
\be
z = 4 \pi^2 \frac{\langle V^2 \rangle}{d_1 d_2} \ .
\label{isos10}
\ee
The parameter $z$ bears an obvious close analogy to the spreading
width $\Gamma^\downarrow$ introduced in Section~I.II.G. The
matrix~(\ref{isos9}) becomes diagonal if $z=0$. In the full theory not
reviewed here, the parameter $z$ comprises both internal isospin
mixing (via the Coulomb interaction) and external mixing (via the
reaction channels). The theory contains also both, the predictions
from the static criterion and the dynamic criterion for isospin
symmetry breaking: The effect is large when either the Coulomb mixing
matrix elements are of the order of the mean level spacing (so that $z
\approx 1$) or when the spreading widths $z d_1$ ($z d_2$) are
comparable to the decay widths $2 \pi {\cal N}_1 / d_1$ ($2 \pi {\cal
N}_2 / d_2$, respectively).

In~\cite{Har86a}, the theory was applied to a number of experimental
examples of isospin mixing from which $z$ was determined. The examples
can be divided into four classes. In the first class, the average
cross section of a reaction forbidden by an isospin selection rule is
compared to the average cross section of an isospin--allowed
reaction. The resulting suppression factor $f$ is related to the
mixing parameter $z$ and is a measure of the average mixing
probability of the CN levels with isospin $T_2$ with those that have
isospin $T_1$. Nuclear reactions in this first category use
self--conjugate target nuclei (i.e., nuclei with equal neutron and
proton numbers) and are of the type $(d,\alpha)$, $(d,d')$,
$(\alpha,\alpha')$, $(^6{\rm Li},\alpha)$, \ldots. For CN excitation
energies close to neutron threshold one finds suppression factors $f$
around $0.3$--$0.5$. These decrease to values of only a few percent at
high excitation energies suggesting that at those energies, isospin
symmetry is restored. The detection of isospin--forbidden dipole
radiation from alpha and heavy--ion capture reactions supports that
observation. While the former yields suppression factors $f \approx
0.15$ for compound nuclei in the $sd$--shell nuclei at about $14~{\rm
MeV}$ excitation energy, in the fusion reaction $^{12}{\rm C}$ $+$
$^{16}{\rm O}$ populating the CN $^{28}{\rm Si}$ at $E_x = 34~{\rm
MeV}$, $f$ was found to be $\lesssim 0.05$. Since the early
experiments~\cite{Sno84,Har86} several more heavy--ion fusion
reactions were studied. \textcite{Kic94} e.g.\ investigated isospin
mixing in the CN $^{26}{\rm Al}$ and $^{28}{\rm Si}$ up to excitation
energies $E_x \approx 63~{\rm MeV}$, \textcite{DiP01} in $^{24}{\rm
Mg}$ up to $E_x \approx 47~{\rm MeV}$, \textcite{Kic05} in $^{32}{\rm
S}$ and $^{36}{\rm Ar}$, and lately \textcite{Woj07} also in
$^{44}{\rm Ti}$ and $^{60}{\rm Zn}$ at $E_x \approx 50~{\rm
MeV}$. (The isospin mixing parameter $\alpha^2$ determined in these
heavy--ion capture $\gamma$--ray reactions is directly related to the
parameter $f$ introduced in~\cite{Har86a}). The emerging systematics
on the very small suppression factors $f$ supports the statement that
isospin is quite pure at high excitation energies (several 10 MeV) of
the CN.  Other experiments on isospin mixing in the Ericson regime
that belong to the first category are measurements of the
isospin--forbidden neutron decay of the giant dipole resonance in
medium--heavy ($^{60}{\rm Ni}$) to heavy nuclei ($^{88}{\rm Sr}$,
$^{89}{\rm Y}$, $^{90}{\rm Zr}$). One finds suppression factors
ranging from $f=0.48$ to $0.84$ which indicate a fairly sizable
isospin mixing at nuclear excitation energies of $E_x \approx 20~{\rm
MeV}$. Information on isospin mixing in highly excited compound nuclei
is also deduced from evaporation spectra in $(\alpha, \alpha')$,
$(p,p')$, $(p,\alpha')$ and $(\alpha, p')$ reactions. In short, if the
cross--section ratio $R = \sigma_{\alpha \alpha'}\, \sigma_{p p'} /
\sigma_{\alpha p'} \sigma_{p \alpha'}$ is approximately equal to unity
then the isospin selection rule does not play any role. This follows
from the Bohr hypothesis (independence of formation and decay of the
CN). More generally, an expression for $R$ can be obtained and
compared with the experimental results for any degree of isospin
mixing from the generalized Hauser--Feshbach
expression~(\ref{isos8})~\cite{Har86a}.

Another highly sensitive test of isospin violation is provided by a
comparison of cross--section fluctuations in the Ericson regime for
pairs of isobaric mirror channels (channels that are linked by the
neutron $\leftrightarrow$ proton transformation). Such reactions were
studied by \textcite{Sim78} and form the second class of nuclear
reactions that test isospin violation. An example is the reaction
$^{14}{\rm N}$ $+$ $^{12}{\rm C}$ leading to the highly excited CN
$^{26}{\rm Al}^\ast$ which subsequently decays into $^{23}{\rm Mg}$
$+$ $^3{\rm H}$ or into the mirror channel $^{23}{\rm Na}$ $+$ $^3{\rm
He}$. The most sensitive test for isospin violation is provided by the
value of the cross--correlation function $C_{tt'}(\varepsilon)$ at
$\varepsilon = 0$ for the two mirror channels $t$ and $t'$. If isospin
is a good quantum number (so that the mixing parameter $z = 0)$, the
cross sections in the mirror channels are strongly correlated. If
isospin mixing is complete (so that $z/{\cal N}_2 \rightarrow
\infty$), the two cross sections should fluctuate in an uncorrelated
way, as does any pair of cross sections pertaining to different final
states~\cite{Eri66,Ric74}. In several pairs of mirror channels the
measured cross--correlations are significantly larger than for
arbitrary pairs of cross sections.

In the third class of CN reactions sensitive to isospin mixing, the
shape of the cross--section autocorrelation function in the Ericson
regime is used as a test. The relevant observable is the correlation
width $\Gamma$. The Lorentzian form of the autocorrelation function is
obtained only for very strong isospin mixing. In the case of strict
isospin conservation, a superposition of two Lorentzians is expected.
The theoretical expressions for the case of partial isospin symmetry
breaking obtained from Eq.~(\ref{isos8})~\cite{Har86a} have been
compared with very precise cross--section fluctuation data mainly from
the $^{32}{\rm Si} (d, \alpha) ^{30}{\rm P}$ reaction~\cite{Spi78}
leading to final states with isospins $T = 0$ and $T = 1$.

Charge--exchange reactions like $(p,n)$ or $(n,p)$ that populate the
isobaric analogue state of the target (see Section~\ref{fine}) form
the fourth class of CN reactions sensitive to isospin effects. If
isospin were totally conserved, such reactions could be viewed as
elastic scattering processes in isospin space. In the Ericson regime,
the elastic enhancement factor has the value $W = 2$, see
Eq.~(\ref{52}).  That same value for $W$ is expected for a
charge--exchange reaction if isospin is conserved while $W = 1$ for
strong isospin mixing. It seems that this interesting test has not
been used so far.

We have briefly summarized what is known about isospin--symmetry
breaking in CN reactions~\cite{Har86a}. The data show that in the
Ericson regime, isospin--symmetry breaking is neither so weak as to be
altogether negligible nor so strong as to reduce CN scattering to a
Hauser--Feshbach situation without any reference to the isospin
quantum number. Isospin mixing is expected to become weaker with
increasing excitation energy, see~\cite{Sok97} and references
therein. As shown in~\cite{Har86a}, data in the Ericson regime can be
used to determine the average Coulomb matrix elements $\langle V^2
\rangle^{1/2}$ and spreading widths $\Gamma^\downarrow_1 = 2 \pi z^2
d_2$ or $\Gamma^\downarrow_2 = 2 \pi z^2 d_1$ for isospin violation in
nuclei. While the values of the average Coulomb matrix elements vary
over many orders of magnitude, the values of the spreading widths are
nearly constant versus excitation energy and mass number, see Fig.~15
in~\cite{Har86a}. This fact provides a meaningful consistency check on
both theory and data analysis. Similarly to the experiments reviewed
in Section~\ref{wocr} and to the study of symmetry breaking in the
regime of isolated resonances~\cite{Alt98,Die06}, further subtle
aspects of the theory in the regime $\Gamma \gg d$ might be tested
with the help of experiments on two coupled microwave billiards.

\subsubsection{Fine Structure of Isobaric Analogue Resonances}
\label{fine}

\begin{figure}[ht]
	\centering
	\includegraphics[width=0.45\textwidth]{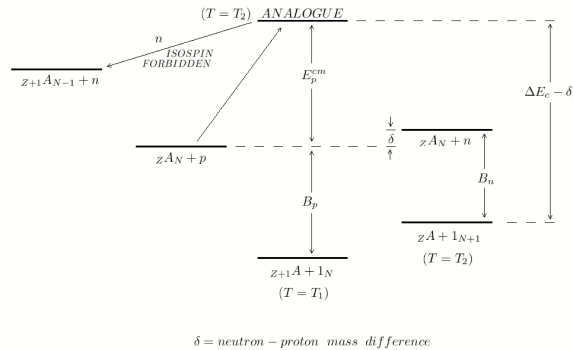}
	\caption{Level scheme showing the parent nucleus, the isobaric
	analogue state, and a state of lower isospin (schematic).}
	\label{fig:GM1}
\end{figure}

If isospin $T$ were a good quantum number, isospin multiplets
consisting of degenerate states with fixed $T$ but with different
$z$--quantum number $T_z$ (``isobaric analogue states'') would exist
in nuclei with the same mass number $A$ but different neutron and
proton numbers. The degeneracy is lifted by the isospin--breaking
interaction (mainly the Coulomb interaction between protons), and the
energies of the members of a multiplet increase with increasing proton
number. For proton numbers $Z \approx 20$, the energy difference
between neighboring members of a multiplet is of the order of $10$
MeV. In a medium--weight nucleus with ground--state isospin $T_1$, the
lowest state with next--higher isospin $T_2 = T_1 + 1$ (an isobaric
analogue state of the ``parent state'', here: the ground state of a
nucleus with the same mass number but one proton replaced by a
neutron) typically has an excitation energy of several
MeV. Higher--lying states with isospin $T_2$ follow with typical
spacings of several $100$ keV. These states may be unstable against
proton decay. The resulting resonances (``isobaric analogue
resonances'', IARs) are then observed in elastic proton
scattering. The proton channel does not have good isospin and couples
to both the IARs and the numerous background states with isospin
$T_1$. The situation is schematically illustrated in
Fig.~\ref{fig:GM1}. The parent state is the ground state of the
nucleus $_Z$(A+1)$_{N+1}$ with isospin $T_2$. In its ground state the
CN nucleus $_{Z+1}$(A+1)$_N$ has isospin $T_1$. The isobaric analogue
state in that nucleus occurs at an excitation energy of several MeV.

Low--lying IARs correspond to simple states of the shell
model. Therefore, their elastic widths (typically several keV) are
much larger than those of the complicated CN background states. On the
other hand, the elastic width of the IAR is typically small compared
to the average spacing between IARs which will thus be considered as
isolated. The isospin--breaking interaction mixes each IAR with CN
background resonances. The mixing strongly enhances the elastic widths
of all CN resonances that occur in the vicinity of the IAR. The
mechanism is similar to that of a doorway state, see Section~I.II.G,
except that now all states are resonances, and that the mixing
mechanism is rather special. The resulting ``fine structure'' of an
IAR is a topic of special interest. While it is sometimes possible to
apply nuclear--structure theory to individual IARs, the background
states are too numerous and too highly excited to allow for anything
but a random--matrix approach. Thus, the theoretical description uses
for the Hamiltonian the matrix~(\ref{isos1}) with the proviso that the
submatrix $H^{(1)}$ is taken from the GOE while the submatrix
$H^{(2)}$ has dimension one.

IARs as resonances in the CN were discovered by the Florida State
group which partially resolved a proton $s$--wave IAR in
$^{92}$Mo(p,p)~\cite{Ric64}. If the CN background resonances overlap
only weakly, the excitation curves fluctuate strongly, and the fine
structure of an IAR can be completely resolved experimentally. That is
possible mainly in the nuclear $1f$--$2p$ shell, see Section~I.IV.A.
The fine structure is investigated in elastic and inelastic proton
scattering and sometimes in the $(p, n)$ reaction.  The latter is
isospin forbidden and gives direct evidence for symmetry breaking.
Depending on the mean level spacing $d$ of the background states, we
distinguish three cases: almost all of the original proton strength is
retained by the analogue state (weak mixing), the strength is spread
among many states with no state being dominant (strong mixing), and
cases between these extremes (intermediate mixing). The background
states have their own (small) proton widths. Thus, the observed fine
structure is the combination of two amplitudes which may display
interference effects. Special interest was shown in the resulting
asymmetry~\cite{Rob65} that is found in many fine--structure
distributions. Another phenomenon in nuclear physics with similar
mixing patterns is that of fission doorways~\cite{Lyn69}.

The observation of fine structure of IARs requires excellent energy
resolution for the incident proton beam. Since most of the
fine--structure data were obtained by the Triangle Universities'
Nuclear Laboratory (TUNL) group, we limit discussion of the
experimental techniques to a brief description of their method. The
requirements of very good beam--energy resolution (needed to resolve
the fine structure) and of high beam intensity (needed to have good
statistics) seem contradictory, particularly because of the
time--dependent fluctuations in beam energy. The method adopted by the
TUNL group to resolve the resolution--intensity problem uses two
beams. One high--intensity (H$^{+}$) beam is used to perform the
experiment, while the other (HH$^{+}$) beam is used to generate a
feedback signal that follows the beam--energy fluctuations. This
signal generates a voltage difference which is applied to the target,
thus canceling the time--dependent energy fluctuations. The
experimental details are covered in the review by~\textcite{Bil76}.

\begin{figure}[ht]
	\centering
	\includegraphics[width=0.45\textwidth]{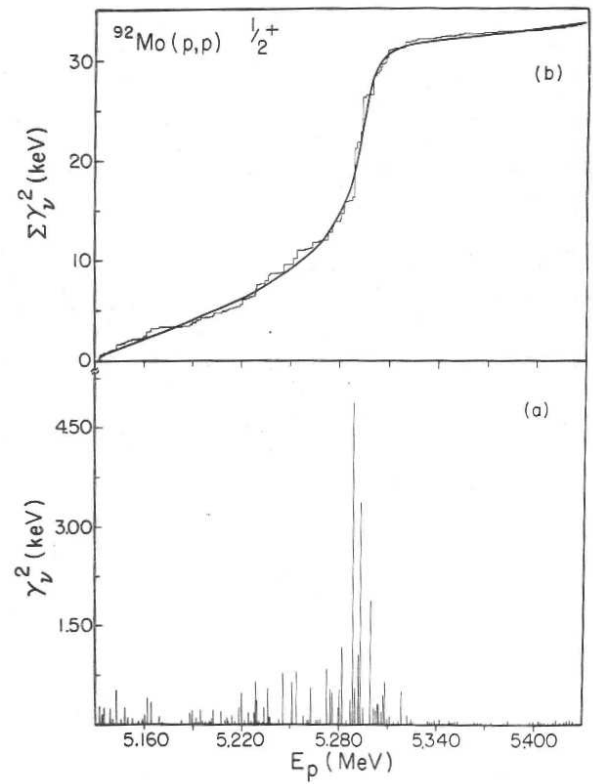}
	\caption{Partial widths of individual resonances (lower panel)
	and their integral (upper panel) versus proton bombarding
	energy $E_p$ in the elastic proton scattering on $^{92}$Mo.
        From~\cite{Bil74}.}
	\label{fig:GM2}
\end{figure}

\begin{figure}[ht]
	\centering
	\includegraphics[width=0.45\textwidth]{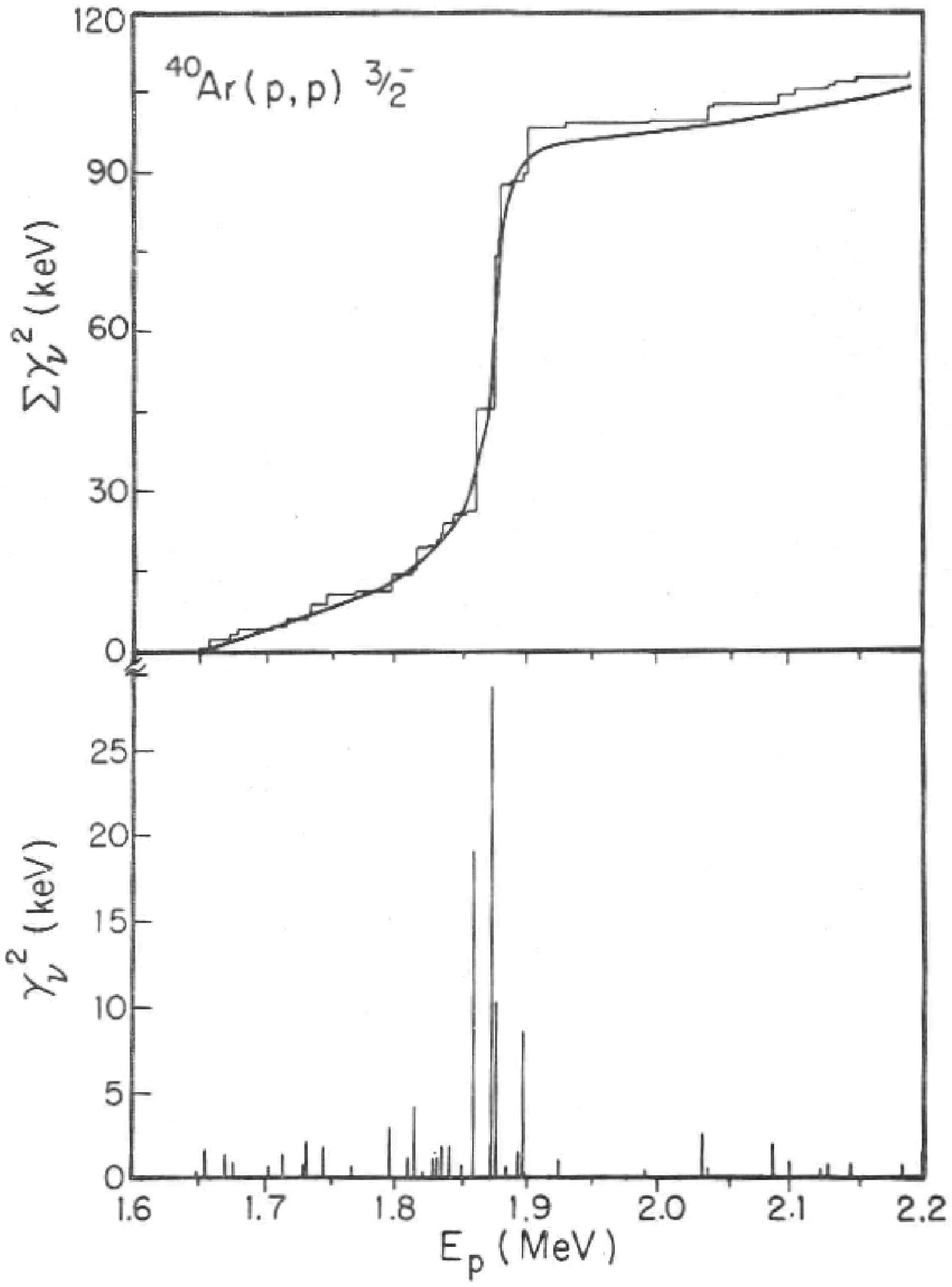}
	\caption{Same as in FIG.~\ref{fig:GM2} but for the target
	nucleus $^{40}$Ar. From~\cite{Key68}.}
	\label{fig:GM3}
\end{figure}

After early work by the Florida State group on an $s$-wave IAR in
$^{92}$Mo(p,p)~\cite{Ric64}, later measurements on this analogue with
better resolution~\cite{Bil74} provided the fine--structure pattern
with the largest number of individual states ever, see
Fig.~\ref{fig:GM2}. Prior to these data the existence of fine
structure was most clearly shown by~\textcite{Key66,Key68}, see
Fig.~\ref{fig:GM3}, who used a windowless gas target. Their data
definitively established the essential correctness of the view that in
the A ${\approx}$ 40 mass region, the analogue (doorway) state is
mixed into the CN background states with a spreading width (see
Section~I.II.G) of the order of 10 keV. The bulk of the
fine--structure data consists of 15 elastic proton scattering
excitation functions on thin solid targets (of order 1 ${\mu}$g) in
the mass region 40 ${\ <}$ A ${\ <}$ 64. There are also some data on
other channels, including the (p,p'), (p,${\gamma}$), and (p,n)
reactions. Almost all of these results are included in the review
by~\textcite{Bil76}.

Many data on strength functions of IAR display the ``Robson
asymmetry''~\cite{Rob65}, characterized by a dip in the strength
function located above the energy of the resonance maximum. For
simplicity we illustrate the origin of that asymmetry for the case of
a single open channel (the proton channel). That case suffices to
display the very peculiar nature of the doorway mechanism in IARs. The
IAR is mixed with the background states by two mechanisms: The matrix
elements of the Coulomb interaction $V$ appearing in the non--diagonal
blocks of expression~(\ref{isos1}), and the elements $F_{\mu \nu}$ of
the matrix~(\ref{28}) connecting IAR and background states. The latter
do not vanish automatically because isospin is not a good quantum
number in the proton channel. Moreover, in contrast to most other
applications of the statistical theory reviewed in this paper, the
elements of the shift matrix are not small, and the characteristic
features of the IARs are in fact due to the real part of $F_{\mu
\nu}$. This is because IARs occur below or at the Coulomb barrier
where the energy dependence of the matrix elements $W_{\mu c}(E)$
cannot be neglected. To display the effect most clearly we follow
Robson, neglect $V$ (that approximation is often referred to as ``no
internal mixing''), and consider only mixing due to $F_{\mu \nu}$
(``purely external mixing''). In other words, isospin mixing is solely
due to the proton channel to which both the IAR and the background
states are coupled.

We omit the channel index and express the scattering function $S(E)$
in terms of the $K$--function, see Eqs.~(\ref{22a}) and (\ref{22b}).
We replace $\sqrt{2 \pi} W_\mu$ by $\gamma_\mu$ and have
\be
K(E) = \frac{1}{2} \sum_\mu \frac{\gamma^2_\mu}{E - E_\mu} \ .
\label{isos2}
\ee
The parameters $\gamma_\mu$ and $E_\mu$ of the $K$--function are
experimentally obtained by a multi--level $R$--matrix fit to
fine--structure data, see Section~\ref{earl}. The object of interest
is the strength function $\langle \gamma^2_\mu \rangle / d$ obtained
as an average over a number of neighboring resonances, see
Figs.~\ref{fig:GM2} and \ref{fig:GM3}. We replace in Eq.~(\ref{isos2})
$E$ by $E + i I$ and obtain $- (1 / \pi) \Im K(E + i I) \approx
\langle \gamma^2_\mu \rangle / d$, see Section~\ref{calc}.  The
averaging interval $I$ should contain many CN resonances (to reduce
the statistical error), but be small compared to the spreading width
of the IAR defined below (to display the resonance enhancement and
asymmetry). In the analysis of actual data it may be hard to meet both
requirements. To calculate $K(E)$ we drop $V$ in Eq.~(\ref{isos1}) and
use Eqs.~(\ref{C2}) and (\ref{C3}) keeping the shift function
$F$. After a little algebra we obtain
\be
K(E) = \pi \sum_{i j} W_i ({\cal B}^{-1})_{i j} W_j \ .
\label{isos3}
\ee The matrix ${\cal B}$ has the same form as
expression~(\ref{isos1}). The indices $(i, j)$ take the values $1$ to
$N$ (for the background states) and zero (for the analogue state)
while $\mu$ and $\nu$ run from $1$ to $N$ as before. Explicitly we
have
\be
{\cal B} = \left( \matrix{ (E - \ve_\mu) \delta_{\mu \nu} & -
\Re F_{\mu 0} \cr - \Re F_{0 \nu} & (E - E_0 + F_{0 0}) \cr} \right) \ .
\label{isos4}
\ee
We have assumed that the matrix $H^{(1)}_{\mu \nu} + \Re F_{\mu \nu}$
has been diagonalized. The resulting eigenvalues are denoted by
$\ve_\mu$ but the notation on the transformed matrix elements $W_\mu$
and $\Re F_{\mu 0}$ has not been changed. The $W_\mu$ are random
Gaussian variables which also appear as arguments of the integrals
defining the matrix elements $\Re F_{\mu 0}$. Thus, $W_\mu$ and $\Re
F_{\nu 0}$ are correlated for $\mu = \nu$. The energy of the
unperturbed IAR is denoted by $E_0$. The matrix element $W_0$ is not
random.

For a common doorway state, the spreading width $\Gamma^{\downarrow}$
is defined as $\Gamma^{\downarrow} = 2 \pi v^2 / d$, see Eq.~(I.26).
Here $d$ is the mean spacing of the background states and $v^2$ is the
average squared coupling matrix element. Using the analogy between
Eq.~(\ref{isos4}) and the matrix description Eq.~(I.24) for a doorway
state, we define the spreading width of an IAR as
\be
\Gamma^{\downarrow} = 2 \pi \langle (\Re F_{0 \mu})^2 \rangle / d \ .
\label{isos5}
\ee
This equation shows once again that isospin mixing is due to the
proton channel. 

The strength function is obtained from Eq.~(\ref{isos3}) by replacing
$E$ by $E + i I$ and assuming $d \ll I \ll \Gamma^{\downarrow}$. The
explicit calculation uses the statistical assumptions mentioned above
and may, for instance, be found in Chapter~$13$ of~\cite{Mah69}. The
result is
\be
\frac{\langle \gamma^2_\mu \rangle}{d} = s^{bg} \ \frac{(E - E_0)^2}
{(E - E_0 + \Re F_{0 0})^2 + (1/4) (\Gamma^{\downarrow})^2} \ . 
\label{isos6}
\ee
Here $s^{bg}$ is the strength function of the background states in the
absence of the IAR. The IAR enhances the strength function in the
vicinity of the analogue state. Since $\Gamma^{\downarrow} \gg d$ the
resonance is completely mixed with the background states and is seen
only through the enhancement factor in Eq.~(\ref{isos6}). That factor
has the shape of an asymmetric Lorentzian and approaches the value
unity far from the resonance. The width of the Lorentzian is given by
the spreading width. The strength function vanishes at $E = E_0$ where
there is no isospin mixing. The zero occurs above the resonance energy
$E_0 - \Re F_{0 0 }$: Because of the Coulomb barrier, the main
contribution to the integral defining $\Re F_{0 0 }$ stems from states
with energies larger than $E_0$, and $\Re F_{0 0 }$ is, therefore,
positive.

In general (several open channels, both external and internal mixing)
the asymmetry of the strength function in channel $c$ in the vicinity
of an analogue state is reduced. For purposes of fitting data the
expression given, for instance, by~\textcite{Lan69} can be written in
the form~\cite{Bil76}
\ba
\frac{\langle \gamma^2_{c \mu} \rangle}{d} &=& s^{bg}_c + \frac{2
s^{bg}_c \Delta_c (\ve_0 - E)}{(\ve_0 - E)^2 + (1/4)
(\Gamma^{\downarrow})^2} \nonumber \\
&& + \frac{\gamma^2_c \Gamma^{\downarrow} / (2 \pi)}{(\ve_0 - E)^2 +
(1/4) (\Gamma^{\downarrow})^2} \ .
\label{isos7}
\ea
Here $s^{bg}_c$ is the background strength function in channel $c$,
$\ve_0$ is the resonance energy, $\Delta_c$ is the asymmetry parameter
in channel $c$, and $\gamma^2_c$ is the total reduced width of the
analogue state in channel $c$, i.e., the sum over the fine--structure
contributions.

Eq.~(\ref{isos7}) was originally derived under the assumption of
strong mixing, $\Gamma^{\downarrow} \gg d$. However, the experimental
data indicate intermediate or weak mixing in essentially all
cases. That case was theoretically considered in~\cite{Lan74}. It was
shown that the form of Eq.~(\ref{isos7}) remains unchanged and that
$\langle \gamma^2_{c \mu} \rangle / d$ as given by Eq.~(\ref{isos7})
must be considered an ensemble average of the strength function, all
members of the ensemble having the same physical parameters. To
analyze the data one avoids the need of using energy--averaging
intervals and fits the accumulated strength to its ensemble average
$\int^E \langle \gamma^2_{c \mu} \rangle / d$.

Due to the limited number of fine--structure states and to the width
fluctuations (we recall that the latter follow the Porter--Thomas
distribution, see Section~I.II.D.1), the extracted parameters have
rather large uncertainties. The general status of these parameters is
as follows: The best--fit values for the background strength function
$s^{bg}_{c}(E)$ are generally in agreement with those expected from
systematics. The values of the energy $\ve_0$ of the analogue
resonance agree with systematics on shifts between analogue state and
parent state, see~\cite{Jae69}. The values of the reduced widths
$\gamma^2_c$ can be used to calculate the proton spectroscopic factors
S$_p$. The latter essentially measure the probability to find a proton
of fixed angular momentum in the projection of the resonance wave
function onto the target nucleus. When compared with the neutron
spectroscopic factors S$_n$ of the parent states, the analogue
spectroscopic factors are significantly lower (30-50 ${\%}$) than
expected, even after Coulomb corrections. To the best of our
knowledge, that discrepancy has never been satisfactorily resolved.
The statistically significant best--fit values for the asymmetry
parameter $\Delta_c$ are all negative and usually agree (within a
factor of 2) with the values predicted by~\textcite{Rob65}.  More
detailed analysis indicates that the effects of the inelastic channels
do not dominate the elastic channel, but are not negligible either (as
assumed by the Robson model). Instead these effects are comparable to
that of the elastic channel for most analogues. The best--fit values
of the spreading width $\Gamma^{\downarrow}$ agree rather well with
the Robson model (only external mixing that occurs only in the elastic
channel). There is evidence that the inelastic channels contribute
significantly to $\Gamma^\downarrow$ in some cases.

The analogue state can decay via several inelastic and/or
capture--gamma channels and is then a doorway common to these
channels. Such decay processes provide unique information on the
analogue state. For an isolated doorway common to two channels $c$ and
$c'$~\textcite{Lan71} predicted that the reduced width amplitudes
$\gamma_{c \mu}$ and $\gamma_{c' \mu}$ of the fine--structure
resonances labeled $\mu$ should be maximally correlated. More
precisely, the normalized linear correlation coefficient
$\rho(\gamma_{c}, \gamma_{c'})$ defined in Eq.~(I.29) should be equal
to unity and the product $\gamma_{c \mu} \gamma_{c' \mu}$ should have
the same sign for all fine--structure resonances
$\mu$.~\textcite{Gra74} and~\textcite{Dav75} indirectly showed the
expected constancy of the relative phase. The two predictions were
proved directly by~\textcite{Mit85}. We consider the fragmented
3/2$^{-}$ analogue state at proton bombarding energy E$_{p}$ = 2.62
MeV in $^{45}$Sc as an example.

\begin{figure}[ht]
	\centering
	\includegraphics[width=0.45\textwidth]{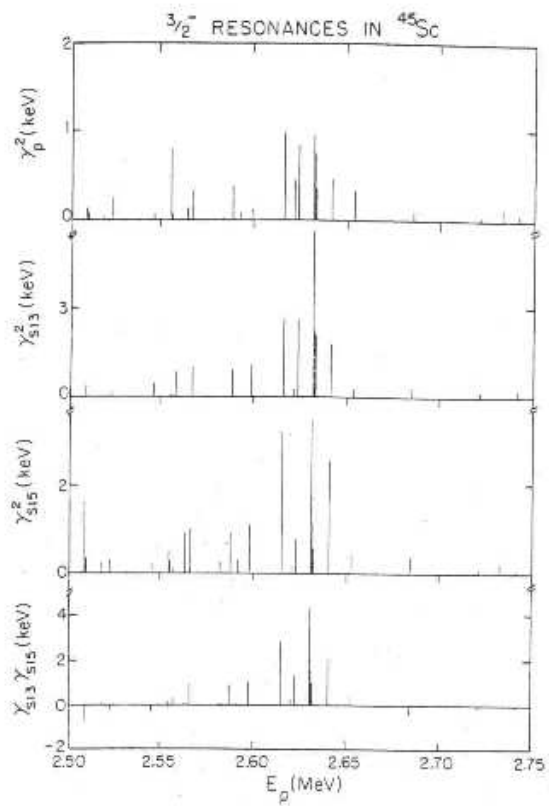}
	\caption{Fine structure of the analogue resonance with spin
	3/2 in the CN $^{45}$Sc. The partial width amplitudes $\gamma$
	are indexed by the channel angular momentum and spin. The
	Robson asymmetry is clearly displayed. From~\cite{Mit85}.}
	\label{fig:GM4}
\end{figure}

That state has a strong decay to the 2$^{+}$ first excited state in
$^{44}$Ca. In the channel--spin representation, the two p--wave proton
inelastic decay channels have spins $s$ = 3/2 and $s$ = 5/2. These two
channels also display a well--developed fine structure pattern (see
Fig.~\ref{fig:GM4}); both show very clearly the Robson asymmetry. With
the method described by~\textcite{Mit85}, one can determine the
correlation between the decay amplitudes. The measured value
$\rho(\gamma_{s=3/2}, \gamma_{s=5/2}) = 0.93$ is in agreement with the
predicted value of unity. The relative sign of the decay amplitudes is
the same for the 15 consecutive resonances that make up the
analogue. To quote~\textcite{Mah79}, ``isobaric analogue resonances
provide the best--understood example of isolated doorway states.``

\subsection{Parity}
\label{pari}

In this Section we discuss data on parity violation obtained by
low--energy neutron scattering, and their statistical analysis. We
keep the discussion short since a general survey of experiments on
parity violation with neutrons and their analysis was given
by~\textcite{Mit99}. A comprehensive review of the experiments and
analysis of the TRIPLE (Time--Reversal Invariance and Parity violation
at Low Energies) collaboration was presented by~\textcite{Mit01}.

The study of parity violation in nuclei has a long history. Following
the initial discovery of parity violation in beta decay, efforts
towards detecting and understanding the induced parity--violating
nucleon--nucleon interaction focused on precise measurements of
observables indicating parity violation. Most of these involved
electromagnetic transitions between nuclear states at excitation
energies of a few MeV. In spite of many efforts and elegant
experimental results, the work has been only partially successful. The
difficulty is in the theoretical analysis -- calculating the induced
effective parity--violating interaction required a more precise
knowledge of the wave functions of the nuclear states involved than
can be attained with present--day nuclear theory~\cite{Ade85}. The
situation changed when~\textcite{Sus80} predicted two enhancement
factors which together would lead to large parity--violating effects
in the CN scattering of low--energy neutrons, and when statistical
concepts were used to analyze the data.

The two enhancement factors are usually referred to as dynamical
enhancement and kinematical enhancement. Dynamical
enhancement~\cite{Bli60,Sus80} arises because in heavy nuclei the
average spacing of neutron resonances of opposite parity is small and
typically $10$ eV or so. With $V$ the parity--violating part of the
nucleon--nucleon interaction and $E_1$ and $E_2$ the energies of two
levels $| 1 \rangle$ and $| 2 \rangle$ of opposite parity, the mixing
of the two states is given by $\langle 1 | V | 2 \rangle / (E_1 -
E_2)$. The mixing obviously increases with decreasing spacing $(E_1 -
E_2)$. The increase is not inversely proportional to the spacing,
however, because the complexity of the wave functions of the states $|
1 \rangle$ and $| 2 \rangle$ also increases with increasing level
density, reducing the overlap in $\langle 1 | V | 2 \rangle$. The
result is an increase of the mixing that is inversely proportional to
the square root of the spacing~\cite{Sus82,Fre88a,Fre88b}. With
typical spacings of states of opposite parity in the ground--state
domain around $100$ keV, the resulting enhancement factor is $\approx
10^2$ and does not change rapidly with excitation energy. Kinematical
enhancement arises because of the unequal resonance strength of the
states mixed by the parity--violating interaction.  At low neutron
energies only resonances with orbital angular momentum 0 or 1 are
populated. Because of the angular momentum barrier, the $s$--wave
resonances have much larger widths than the $p$-wave resonances.
Thus, $s$--wave decay of a $p$--wave resonance with an $s$--wave
resonance admixture is enhanced over the regular $p$--wave decay of
that resonance by a factor given by the ratio of the two barrier
penetrabilities -- approximately $(kR)^{-1}$. Here $R$ is the nuclear
radius and $k$ the wave number. The factor $(kR)^{-1}$ is strongly
energy dependent, significant enhancement occurs only near neutron
threshold. The product of the two enhancement factors is about
10$^{5}$. Since the weak interaction is approximately a factor of
10$^{7}$ smaller than the strong interaction, the total enhancement
leads to expected parity--violating effects of the order of percent.
That prediction of~\textcite{Sus80} was confirmed shortly afterwards
at Dubna (Alfimenkov {\it et al.}, 1982, 1983). The results at Dubna
were extremely interesting. However, there were both experimental and
theoretical limitations.

\begin{figure}[h]
       \centering
       \includegraphics[width=0.45\textwidth]{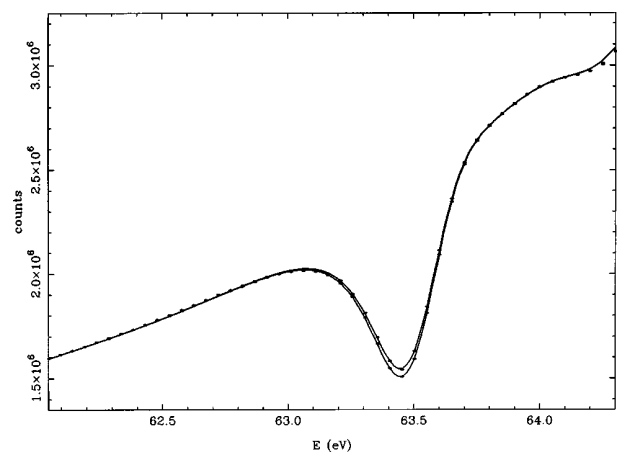}
       \caption{Neutron transmission spectra for two helicity states
       near the $63$ eV resonance in $^{238}$U. The resonance appears
       as a dip in the transmission curve. The transmission at the
       resonance differs significantly for the two helicity states and
       parity violation is apparent by inspection. From~\cite{Mit99}.}
\label{figpar1}
\end{figure}

The experimental limitations at the Dubna facility were severe: the
neutron flux dropped dramatically above a few tens of eV, limiting the
experiments on parity violation to neutron energies below 20 or 30
eV. Subsequently the TRIPLE collaboration was formed to extend the
experiments on parity violation to higher energies. These experiments
were performed at the Los Alamos Neutron Science Center (LANSCE). A
moderated and collimated beam of neutrons is produced by spallation. A
beam of longitudinally polarized protons is produced by scattering
from a longitudinally polarized proton target. The neutron spin
direction is reversed by a system of magnetic fields. The neutrons
pass through the target and are detected in a highly segmented system
of liquid detectors located at a distance of approximately 57
meters. CN resonance energies are determined by the time--of--flight
method. A sample result is shown in Fig.~\ref{figpar1}. The details of
the experimental system are given in the review by~\textcite{Mit01}.

Although the observation of large parity violation in neutron
resonances was certainly very impressive, the Dubna results were first
considered of only anecdotal interest, since the resonance wave
functions were too complicated to be theoretically accessible. This
problem was overcome by the statistical approach. Since $s$--wave
neutron resonances obey GOE statistics, it is safe to assume that
$p$--wave resonances do, too. Then the matrix elements of the
parity--violating interaction connecting $s$--wave and $p$--wave
resonances have a Gaussian distribution with mean value zero. The
variance of the distribution (or the mean--squared matrix element
$\overline{v^2}$ for parity violation) can be determined from a set of
$p$--wave resonances with parity violation in a given nuclide. This is
described below. However, $\overline{v^2}$ is not a good measure for
the strength of the parity--violating interaction, since it decreases
rapidly with increasing complexity of the wave functions and, thus,
with increasing mean level density. A measure that is roughly
independent of excitation energy and mass number is the spreading
width $\Gamma^{\downarrow} = 2 \pi \overline{v^2} / d$ (see
Sections~I.II.G and \ref{isos}). The convention is to adopt for $d$
the mean spacing of the $s$--wave resonances. The expected size of the
weak spreading width can be estimated from the spreading width for the
strong interaction (which is experimentally of order MeV) and by
adopting for the ratio of the square of the strength of the weak
interaction to the strength of the strong interaction the value
10$^{-13}$. Thus one expects ${\Gamma}^{\downarrow} \approx 10^{-6}$
eV.

The observable for parity violation is the longitudinal asymmetry
(often simply referred to as ``asymmetry'')
\begin{equation}
P = \frac{\sigma^p_+ - \sigma^p_-}{\sigma^p_+ + \sigma^p_-},
\label{reduced}
\end{equation}
where $\sigma^p_{\pm}$ is the total $p$--wave cross section for
neutrons with helicities $\pm$. Clearly we have $P = 0$ if parity is
conserved. Asymmetries for a set of resonances were determined
separately for each run of approximately 30 minutes and a histogram
created for each nuclide and each resonance measured. The mean of this
histogram was the value adopted for $P$.

We illustrate the analysis that determines $\overline{v^2}$ and the
weak spreading width $\Gamma^{\downarrow}$ by considering a target
nucleus with spin $I = 0$ and positive parity. This case illustrates
most of the principles involved. The $s$--wave resonances have spin
and parity $1/2^{+}$ and the $p$--wave resonances $1/2^{-}$ or
$3/2^{-}$. Only the $1/2^{-}$ resonances are considered as only these
are mixed with the $1/2^{+}$ resonances by the parity--violating
interaction. We use the formalism of Section~\ref{stoc}. We neglect
direct reactions, use the scattering matrix as given by
Eqs.~(\ref{C2}) and (\ref{C3}), and replace the Hamiltonian $H^{\rm
GOE}$ in Eq.~(\ref{C4}) by an expression of the
form~(\ref{isos1}). The upper indices $1$ and $2$ now stand for states
with positive and negative parity, respectively, and $V$ denotes the
induced parity--violating nucleon--nucleon interaction. We use the
diagonal representation of $H^{(1)}$ and $H^{(2)}$ and denote the
eigenvalues by $E^{(1)}_\mu$ and $E^{(2)}_\nu$ and the eigenfunctions
by $| 1 \mu \rangle$ and $| 2 \nu \rangle$, respectively. The
matrix~(\ref{isos1}) takes the form
\be
\left( \matrix{
E^{(1)}_\mu \delta_{\mu \nu} & \langle 1 \mu | V | 2 \rho \rangle \cr
\langle 2 \sigma | V | 1 \nu \rangle & E^{(2)}_\rho \delta_{\rho 
\sigma} \cr} \right) \ .
\label{}
\ee
In the diagonal representation of $H^{(1)}$ and $H^{(2)}$, the partial
width amplitudes of the states with positive and negative parity are
denoted by $g^{(1)}_\mu$ and $g^{(2)}_\rho$, respectively. We assume
that the $p$--wave resonances are isolated and focus attention on a
single one. We take the bombarding energy $E$ in the center of that
resonance, $E = E^{(2)}_\rho$. We neglect the total width of that
resonance and of the admixed $s$--wave resonances since in all cases
investigated so far, the spacings $|E^{(1)}_{\mu}-E^{(2)}_{\rho}|$ are
large compared to the total widths. This yields~\cite{Sus80,Bun81}
\begin{equation}
P_{\rho}=2 \sum_{\mu} \frac{\langle 1 \mu | V| 2 \rho \rangle}
{E^{(1)}_{\mu} - E^{(2)}_{\rho}} \ \frac{g^{(1)}_{\mu} g^{(2)}_{\rho}}
{\Gamma^{(n)}_{\rho}} \ ,
\label{Vb1}
\end{equation}
where $\Gamma^{(n)}_{\rho} = (g^{(2)}_{\rho})^2$ is the the partial
width for neutron decay of the $p$--wave resonance with label $\rho$.
The ratio $g^{(1)}_{\mu} g^{(2)}_{\rho} / \Gamma^{(n)}_{\rho} =
g^{(1)}_{\mu} / g^{(2)}_{\rho}$ contains the kinematical enhancement
factor, and the first term on the right--hand side, the dynamical
enhancement factor.

For spin--zero target nuclei, the resonance parameters are usually
known. For the $s$--wave resonances, the information is available from
previous work on $s$--wave neutron scattering. For the $p$--wave
resonances, most of the information was obtained in the framework of
the TRIPLE experiments. In practice one may assign the spin value
$1/2$ to a $p$--wave resonance by the presence of parity
violation. Unfortunately one cannot determine the individual matrix
elements $\langle 1 \mu | V | 2 \rho \rangle$ since there are too few
equations and too many unknowns.  But using the fact that the matrix
elements $\langle 1 \mu | V | 2 \rho \rangle$ have a Gaussian
distribution with mean value zero and a second moment given by
$\overline{v^2}$, we write Eq.~(\ref{Vb1}) in the form $P_{\rho} =
\sum_{\mu} A_{\mu \rho} \langle 1 \mu | V | 2 \rho \rangle$, with
coefficients $A_{\mu \rho} = (2/(E^{(1)}_{\mu} - E^{(2)}_{\rho}))
(g^{(1)}_{\mu}/g^{(2)}_{\rho})$. Then, $P_{\rho}$ is a linear
combination of equally distributed Gaussian random variables and,
therefore, is itself a Gaussian random variable with mean value
zero. The variance of $P_{\mu}$ with respect to both ${\mu}$ and the
ensemble generated by a sequence of runs is given by $A^2 \
\overline{v^2}$, where $A^2 = (1/N)\sum_{\mu \rho} A^2_{\mu \rho}$ and
where $N$ is the number of $p$--wave resonances. It follows that
\begin{equation} 
\label{Vb2}
\overline{v^2} = \frac{{\rm var}(P_{\mu})}{A^2} \ .
\end{equation}
Eq.~(\ref{Vb2}) is the central result of the statistical approach. It
yields $\overline{v^2}$ from the data in spite of the fact that the
signs of the partial width amplitudes in Eq.~(\ref{Vb1}) are usually
not known. The analysis is more difficult for target nuclei with
non--zero spin values. Moreover, usually some but not all
spectroscopic information is available. Suitable methods of analysis
were developed for all such cases~\cite{Mit01}, but the spirit of the
approach is the same. It yields the values of $\overline{v^2}$ and of
$\Gamma^{\downarrow}$ although the wave functions of the individual
nuclear states are not known.

\begin{figure}[ht]
        \centering
	\includegraphics[width=0.45\textwidth]{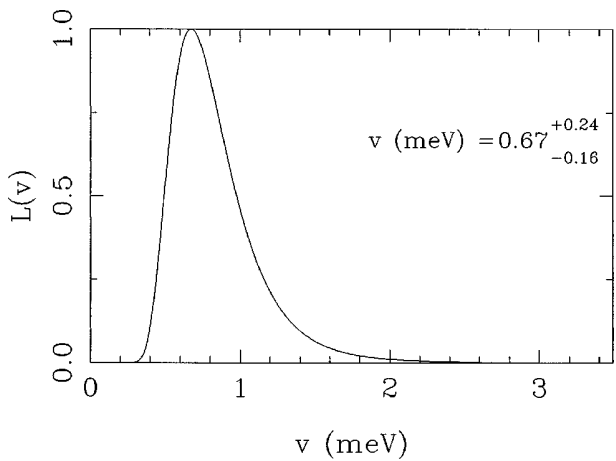}
        \caption{Plot of the maximum--likelihood function $L(v)$
        versus the root--mean--square matrix element $v$ for
        $^{238}$U. The spins of all the $p$--wave resonances are
        known. From~\cite{Mit99}.}
\label{figpar2}
\end{figure}

For the actual determination of $\overline{v^2}$, a maximum likelihood
approach was adopted. The value of $P_{\rho}$ is a realization of a
random variable. For a number of independent resonances the likelihood
function is the product of the functions for the individual
resonances. One inserts the values of the experimental asymmetries and
their errors, determines the spectroscopic term $A$ from the known
resonance parameters, and calculates the likelihood function. The
location of the maximum gives the most likely value $v_{L}$ of the
root mean square matrix element $v = \sqrt{\overline{v^2}}$. A
maximum--likelihood plot for $^{238}$U is shown in Fig.~\ref{figpar2}.
The likelihood function is found to have a well--defined maximum so
that $v$ is rather well determined. When that analysis was applied to
$^{232}$Th, the mean value of $\langle 1 \mu | V | 2 \rho \rangle$ was
found to differ from zero, and a Gaussian with non--zero mean value
was required to fit the data -- all parity violations had the same
sign. This result raised serious questions about the statistical
approach and led to much theoretical activity. However, the anomaly
only occurred in Thorium, and subsequent studies at higher energies in
$^{232}$Th revealed parity violations of opposite sign~\cite{Sha00}.
It was concluded that this ``sign effect`` was due to a local doorway
state (see Section~I.II.G).

Parity violation was studied in 20 nuclides -- especially in those
regions of mass number $A$ where the $p$--wave strength function has
maxima. (The strength function is defined and discussed in
Sections~\ref{opt} and \ref{wocr}). Maxima occur near $A = 238$, i.e.,
for $^{232}$Th and $^{238}$U (maximum of the $4p$ strength function)
and near $A = 100$ where most of the other nuclides were studied
(maximum of the $3p$ strength function). Parity violation was observed
in all but one of these targets ($^{93}$Nb). For 15 nuclei
sufficiently many resonances with parity violation were observed to
determine the weak spreading width; for the other nuclides only very
approximate values or limits could be determined. The values of
$\Gamma^{\downarrow}$ lie around 10$^{-6}$ eV as expected and are
approximately independent of $A$, with some indications of local
fluctuations. Such fluctuations have been observed in the spreading
width for isospin mixing (Section~\ref{isos} and~\cite{Har86a}).

In summary: Except for the Thorium anomaly, the data are consistent
with the statistical model. With the help of a statistical analysis,
it is possible to determine the root--mean--square matrix element $v$
for parity violation and the weak spreading width
$\Gamma^{\downarrow}$ without knowledge of the wave functions of
individual nuclear states. The values found for $\Gamma^{\downarrow}$
are consistent with expectations based on the strength of the weak
interaction, see~\cite{Tom00} for an analysis. For lack of space we
have not discussed experiments on parity violation in
fission~\cite{Koe00}.

\subsection{Time Reversal}
\label{time}

Time--reversal (${\cal T}$) invariance implies symmetry of the
$S$--matrix, $S_{a b} = S_{b a}$ and, hence, detailed balance, $|S_{a
b}|^2 = |S_{b a}|^2$. Tests of ${\cal T}$ invariance compare resonance
scattering cross sections for the two--fragment reactions $a \to b$
and $b \to a$ at the same center--of--mass energy and aim at
establishing an upper bound on the strength of the ${\cal
T}$--invariance violating interaction in nuclei.  Such experiments
have been performed both for isolated and for strongly overlapping
resonances more than twenty years ago and are well documented in the
literature. Thus, we can be brief.

To test detailed balance, one compares the reaction rates $A_1 + A_2
\rightleftharpoons B_1 + B_2$. The fragments $A_1, A_2, B_1, B_2$ are
in their ground states and unpolarized. A difference in the rates $a
\to b$ and $b \to a$ indicates a violation of ${\cal T}$ invariance.
Detailed balance was tested by \textcite{Dril79} in the reactions
$^{27}{\rm Al} + p$ $\rightleftharpoons$ $^{24}{\rm Mg} + \alpha$
populating an isolated $J^\pi = 2^+$ resonance at an excitation energy
$E_x = 12.901~{\rm MeV}$ in the CN $^{28}{\rm Si}$. For a reaction
through an isolated resonance, detailed balance would normally hold
automatically~\cite{Hen59}. The present case is different because for
the fragmentation $^{27}{\rm Al} + p$, the partial waves with angular
momenta $l=0$ and $l=2$ and channel spin $s = 2$
interfere~\cite{Pea75}. The cross sections for the reactions $a \to b$
and $b \to a$ were found to agree within $\delta = 0.0025 \pm
0.0192\%$. This result is consistent with $\delta = 0$ and, thus, with
${\cal T}$ invariance.

\textcite{Bun89} pointed out that in detailed--balance tests that use
two close--lying CN resonances in the regime $\Gamma \ll d$, large
enhancement factors amounting to several orders of magnitude for
${\cal T}$--invariance violation may arise. \textcite{Mit93}
investigated specific experimental possibilities and showed that the
difference $\delta$ of the two reaction cross sections depends
sensitively on energy, angle, and on the parameters of both resonances
and may vary by many orders of magnitude. These theoretical
predictions have been partially tested experimentally in
billiards~\cite{Die07}.

\begin{figure}[ht]
	\centering
	\includegraphics[width=0.45\textwidth]{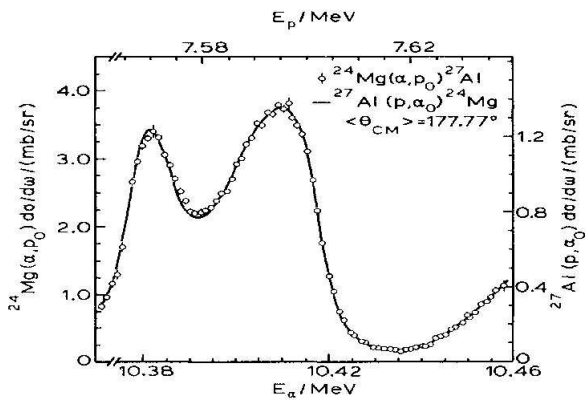}
	\caption{Normalized cross sections for the reaction $^{27}{\rm
	Al}(p, \alpha_0)^{24}{\rm Mg}$ (solid line) and for the
	inverse reaction $^{24}{\rm Mg}(\alpha, p_0)^{27}{\rm Al}$
	(open circles) near and at a deep minimum. The two cross
	sections are equal, and detailed balance
	holds. From~\cite{Blan83}.}
	\label{fig:AR14}
\end{figure}

In the most precise test of detailed balance in CN reactions so far,
the reactions $^{27}{\rm Al}(p, \alpha_0)^{24}{\rm Mg}$ and $^{24}{\rm
Mg}(\alpha, p_0)^{27}{\rm Al}$ were compared in the Ericson regime
$\Gamma \gg d$~\cite{Blan83}. As in a predecessor of this
experiment~\cite{Wit68}, both reaction rates were measured at a
scattering angle of $\Theta \approx 180^\circ$ and were normalized at
a suitable cross--section maximum. The results were then compared at a
cross--section minimum. This was done in order to maximize the
sensitivity for ${\cal T}$--invariance violation. The result is shown
in Fig.~\ref{fig:AR14}. The measured relative differential cross
sections agree within the experimental uncertainty $\delta = \pm
0.51~\%$ and are, thus, consistent with ${\cal T}$--invariance. From
this result an upper bound $\xi \leq 5\cdot10^{-4}$ ($80~\%$
confidence) for a possible ${\cal T}$--noninvariant amplitude in the
CN reaction was derived.

At the time of the experiment, there was no adequate theoretical
framework to interpret that upper bound. The development of the
statistical theory made such an interpretation possible~\cite{Boo86a,
Boo86b} and yielded upper bounds for both, the strength of and the
spreading width $\Gamma^\downarrow$ for the ${\cal T}$--invariance
violating part of the nuclear Hamiltonian. Both bounds compare
favorably with those derived from an analysis of spectral fluctuations
(see Section~I.III.D.4 and~\cite{Fre85,Fre88b}). As in the cases of
violation of isospin and of parity in Sections~\ref{isos}
and~\ref{pari}, the fundamental parameter is the spreading width, and
the upper bound on that quantity is $\Gamma^\downarrow \leq 9\cdot
10^{-2}$ eV. The upper bound from the detailed balance experiment
of~\cite{Blan83} has been improved by two orders of magnitude in an
experiment on polarized neutron transmission through
nuclear--spin--aligned Holmium~\cite{Huf97}. The measurements test
reciprocity and could possibly improved by another order of magnitude
with more intense neutron beams at new spallation sources and with
other targets~\cite{Bar05}. The expected upper bounds for the strength
of possible parity--conserving, time--reversal--violating interactions
will, however, still be several orders of magnitude larger than the
ones provided by the upper limit on the electric dipole moment of the
neutron for parity and time--reversal violating
interactions~\cite{Bak06,Har07}.

To include violation of ${\cal T}$--invariance in the statistical
theory, the $T$--invariant Hamiltonian ensemble $H^{\rm GOE}$ on the
right--hand side of Eq.~(\ref{C4}) is generalized, see
Section~I.III.D.4,
\be
H^{\rm GOE} \to H = \frac{1}{\sqrt{1 + (1/N) \alpha^2}} (H^{\rm GOE}
+ i \frac{\alpha}{\sqrt{N}} A ) \ .
\label{T1}
\ee
The independent elements of the real, antisymmetric, $N$--dimensional
matrix $A$ are uncorrelated Gaussian--distributed random variables
with zero mean values and second moments given by $\overline{A_{\mu
\nu} A_{\rho \sigma}} = (\lambda^2 / N) ( \delta_{\mu \rho}
\delta_{\nu \sigma} - \delta_{\mu \sigma} \delta_{\nu \rho})$. The
parameter $\alpha$ measures the strength of ${\cal T}$--invariance
violation. As explained in more detail in Section~I.III.D.4, the
normalization factor $N^{-1/2}$ is chosen so that significant
invariance violation on the scale of the mean level spacing occurs for
$\alpha \sim 1$. With the replacement~(\ref{T1}) in Eq.~(\ref{C4}),
the calculation of measures for ${\cal T}$--invariance violation in
the statistical theory is a formidable task. In the Ericson regime,
~\textcite{Boo86a,Boo86b} solved the problem by combining a
perturbative treatment of $A$ with an asymptotic expansion in powers
of $d / \Gamma$. A treatment valid for all values of $\Gamma / d$ was
given in~\cite{Plu95,Ger96} and further developed for the analysis of
scattering data on microwave resonators with induced violation of
${\cal T}$--invariance in~\cite{Die09a}. These papers were based on
progress in understanding the GOE $\to$ GUE crossover transition in
spectra~\cite{Alt92,Alt93}. In~\cite{Die09a}, the parameter $\alpha$
is replaced by $\pi \xi$. The resulting expressions for measures of
${\cal T}$--invariance violation are complex and not reproduced
here. We confine ourselves to a discussion of the results.

In Sections~\ref{wocr} and~\ref{sec:EricFluc} it was shown that many
properties of chaotic scattering can be studied with the help of
microwave resonators. That statement applies also to the violation of
${\cal T}$--invariance. In a flat chaotic microwave resonator---a
quantum billiard---shown schematically in Fig.~\ref{fig:AR10}, ${\cal
T}$--invariance violation can be induced by placing a ferrite
(magnetized by an external field) into the resonator. The spins within
the ferrite precess with their Larmor frequency about the magnetic
field. This induces a chirality into the system. The magnetic--field
component of the radio frequency (rf) field in the resonator can be
split into two circularly polarized fields rotating in opposite
directions with regard to that static magnetic field. The component
that has the same rotational direction and frequency as the rotating
spins, is attenuated by the ferrite.  This causes ${\cal
T}$--invariance violation. The strongest effect is expected to occur
when Larmor frequency and rf frequency coincide.  Connecting the
resonator to two antennas, one defines a scattering system. The
violation of ${\cal T}$--invariance then causes the scattering
amplitudes $S_{1 2}$ and $S_{2 1}$ to differ.

Experiments on induced ${\cal T}$--reversal invariance violation in
microwave billiards offer two advantages. First, by a measurement of
phases and amplitudes of the reflected and transmitted rf signals it
is possible to test the reciprocity relation $S_{ab} = S_{ba}$ while
experiments with nuclear reactions typically test only the weaker
detailed--balance relation $| S_{ab} |^2 = | S_{ba} |^2$. Second, such
experiments offer the unique chance of a stringent and detailed test
of the statistical theory that is otherwise often taken for granted
and used to analyse data on CN reactions.

Induced violation of ${\cal T}$--invariance in microwave billiards has
so far been studied in two scattering experiments. In the first
one~\cite{Die07}, a vector network analyzer was used to measure
magnitudes and phases of $S$--matrix elements in a fully chaotic
``annular'' billiard~\cite{Dem00, Hof05} in the regime $\Gamma \ll d$.
By interchanging input and output antennas, both $S_{1 2}$ and $S_{2
1}$ were measured. For all eight isolated resonances (singlets) that
were investigated, the complex element $S_{1 2}$ agrees with $S_{2
1}$, i.e., reciprocity holds, in agreement with~\cite{Hen59}. For the
three pairs of partially overlapping resonances (doublets) that were
studied reciprocity was found to be violated.  The dependence of the
${\cal }T$--violating matrix elements of the effective Hamiltonian for
the microwave billiard on the magnetization of the ferrite could be
determined with the help of Eqs.~(\ref{T1}) and (\ref{C2},\ref{C4}).

\begin{figure}[ht]
	\centering
	\includegraphics[width=0.45\textwidth]{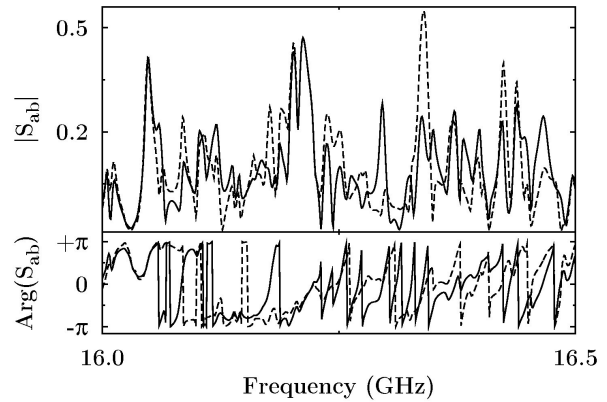}
	\caption{Magnitude and phase of $S_{1 2}$ (solid lines) and of
	$S_{2 1}$ (dashed lines) measured in the frequency interval
	from $16$ to $16.5$~GHz for a fixed magnetization of the
	ferrite. From~\cite{Die09a}.}
	\label{fig:AR15}
\end{figure}

\begin{figure}[ht]
	\centering
	\includegraphics[width=0.45\textwidth]{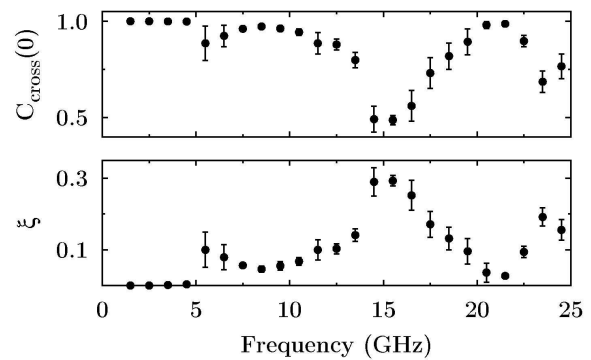}
	\caption{Upper panel: Values of $C_{\rm cross}(0)$ calculated
	from the fluctuating matrix elements $S_{1 2}$ and $S_{2 1}$
	of Fig.~\ref{fig:AR15}. Lower panel: The parameter $\xi$
	versus frequency as deduced from the
	data. From~\cite{Die09a}.}
	\label{fig:AR16}
\end{figure}

In the second experiment~\cite{Die09a}, induced violation of ${\cal
T}$--invariance was investigated in the regime of weakly overlapping
resonances.  A small cylindrical ferrite was placed within the fully
chaotic tilted--stadium billiard shown on the inset of
Fig.~\ref{fig:AR6}. The ferrite was magnetized by an external magnetic
field. Again the elements $S_{1 2}$ and $S_{2 1}$ of the
complex--valued $S$--matrix were measured versus resonance
frequency. Figure~\ref{fig:AR15} shows that magnitude and phase of the
$S$--matrix elements fluctuate strongly, and that reciprocity is
violated. The value of the normalized cross--correlation coefficient
\be 
C_{\rm cross}(0) = \Re \frac{\overline{S_{1 2}(f)
S^*_{2 1}(f)}}{[\overline{|S_{1 2}|^2} \ \overline{|S_{2 1}|^2}]^{1/2}}
\label{T2}
\ee
serves as a measure of the strength of ${\cal T}$--invariance
violation.  ${\cal T}$--invariance holds (is completely violated) for
$C_{\rm cross}(0) = 1$ ($C_{\rm cross}(0) = 0$, respectively). The
upper panel of Fig.~\ref{fig:AR16} shows that $C_{\rm cross}(0)$
depends strongly on frequency. Complete violation of ${\cal T}$
invariance is never attained. The lower panel shows the value of the
parameter $\xi = \alpha / \pi$ for ${\cal T}$--invariance violation
deduced from the data.

The data were used for a thorough test of the underlying
theory~\cite{Die09a}. The parameters of the theory (the transmission
coefficients $T_1$ and $T_2$ in the two antenna channels, a parameter
describing Ohmic absorption by the walls of resonator and ferrite, and
$\xi$) were fitted to the Fourier--transformed $S$--matrix elements.
As in Section~\ref{wocr}, a goodness--of--fit test was used to
establish the quality of the fit, with excellent results. Moreover,
values of the elastic enhancement factor versus frequency were
predicted correctly by the theory without further fit parameters. We
recall that for ${\cal T}$--invariant systems, that factor takes
values between $2$ and $3$, see the remarks below
Eq.~(\ref{47}). Experimentally, values well below $2$ were found in
some frequency intervals. These are possible only if ${\cal T}$
invariance is violated.

\section{Summary and Conclusions}

The Bohr assumption on independene of formation and decay of the CN
was the first of several insightful conjectures concerning
compound--nucleus scattering. It laid the ground for all later
developments and was followed by the Hauser--Feshbach formula, the
Weisskopf estimate, Ericson's idea of random fluctuations of nuclear
cross sections, and the generalization of the Hauser--Feshbach formula
for the case of direct reactions. By way of justification, all of these
developments referred to the fact that the resonances that dominate
compound--nucleus reactions, have statistical properties. The work
anticipated general features of quantum chaotic scattering discovered
only many years later.

The challenge to actually derive these conjectures from the
statistical properties of resonances was taken up early. But it took
several decades and the efforts of many people until a comprehensive
theory of compound--nucleus scattering was established. The theory is
based on a description of the statistics of resonances in terms of
Wigner's random matrices. That description applies generically to
resonances in chaotic quantum systems. As a result, the statistical
theory of nuclear reactions is, at the same time, a generic
random--matrix theory of quantum chaotic scattering. For systems with
few degrees of freedom, it competes with the semiclassical approach to
chaotic scattering. The latter incorporates system--specific features
in the form of short periodic orbits. In the statistical theory, such
features enter as input parameters. Unfortunately, the semiclassical
approach has so far not been extended to many--body systems.

The development of the theory posed essentially two major challenges:
(i) How to incorporate random--matrix theory into scattering theory
and (ii) How to calculate moments and correlation functions of
$S$--matrix elements from that input. The first problem required a
formulation of scattering theory in terms of the Hamiltonian governing
the quasibound states that turn into resonances as the coupling to the
channels is taken into account. The second problem was solved with the
help of methods adopted from quantum field theory. Such methods have
found wide application in condensed--matter theory. The connection to
this area of physics is not one of technicalities only. It actually
connects the statistical theory of nuclei with the statistical
mechanics of many--body systems. Conceptually, it shows how a
separation of scales is achieved in the limit of large matrix
dimension $N$: Universal features govern the system on an energy scale
given by the mean level spacing $d$. The unphysical properties of
random--matrix theory (i.e., the shape of the average spectrum) matter
on the scale $N d$.

The ensuing statistical theory is a complete theory that uses the
minimum number of input parameters and has predictive power. As is
typical for many applications of random--matrix theory, the
statistical theory predicts fluctuations in terms of mean values. The
latter comprise the values of the energy--averaged elements of the
scattering matrix and, in the case of correlation functions, the mean
level spacing $d$. It is here that system--specific features
enter. Examples are the optical model and the strength function which
reflect properties of the nuclear shell model. The theory predicts the
values of moments and correlation functions of $S$--matrix
elements. These determine mean values and fluctuation properties of
cross sections and other observables. The resulting expressions
vindicate the early conjectures, define the limits of their
applicability, and yield expressions that hold under more general
circumstances. The $S$--matrix autocorrelation function is a case in
point. In nuclei, the range of energies where the statistical theory
applies is limited by the underlying assumption that random--matrix
theory correctly describes the statistical properties of resonances.
That is true only when the nuclear equilibration time is shorter than
the average life time of the compound nucleus and holds for bombarding
energies up to 20 MeV or so. Beyond that range, precompound processes
modify the reaction dynamics.

The theory has been the object of stringent tests, largely performed
with the help of microwave billiards. These have been extremely
successful. There is no reason to doubt that the theory adequately
accounts for all aspects of chaotic scattering even though one aspect
of the theory has received little attention so far and has not been
seriously tested: The unitary transformation in channel space that
takes account of direct reactions. The reason is that direct reactions
and CN processes are almost mutually exclusive. When one is important,
the other one typically is not, and vice versa.

The theory has found numerous applications both within and outside the
field of nuclear physics. These have only partly been reviewed. We
have paid special attention to violations of parity, of isospin, and
of time--reversal invariance. In the first two cases, the theory
allows for the determination of the strength of the
symmetry--violating interaction. In the case of time reversal, it
yields an upper bound on that strength.

The calculation of moments and correlation functions of the
$S$--matrix from the statistical theory is not as complete as one may
wish. It would perhaps be unrealistic to expect complete knowledge of
all moments and all correlation functions. The theoretical effort
grows immensely with increasing order of such expressions. Moreover,
moments and correlation functions of low order only can reliably be
determined from the finite range of data typically available in
experiments. But it would be desirable to have theoretical expressions
for cross--section correlation functions even though approximate
expressions are available. That poses a continuing challenge for
theorists.

{\bf Acknowledgments.} We are indebted to many colleagues from whom we
have learned much about nuclear reactions and stochastic scattering.
We are especially grateful to Y. Alhassid, Y. Fyodorov, G. Garvey,
S. Grimes, H. L. Harney, S. Tomsovic, and V. Zelevinsky for a critical
reading of (parts of) the manuscript and for many valuable comments
and suggestions, and to T. Ericson, K. Jungmann, H.-J. St{\"o}ckmann,
W. Wilschut, and N. Severijns for discussions. We also thank
F. Sch{\"a}fer for his help in the preparation of the manuscript. AR
is much indebted to B. Dietz and the other members of the Quantum
Chaos Group in Darmstadt for the long and intense collaboration in the
experiments with microwave billiards. GEM and HAW thank the Institute
for Nuclear Theory for its hospitality during the completion of this
paper. This work was supported in part by the U.S. Department of
Energy Grant No. DE-FG02-97ER41042 and by the DFG within the SFB 634.

\end{document}